\documentclass[showpacs, superscriptaddress,prb]{revtex4-1}
\usepackage{amssymb}
\usepackage{amsmath}
\usepackage{array} 
\usepackage{bm,color}
\usepackage{url}
\usepackage{graphicx}

\makeatletter
\newif\if@restonecol
\makeatother

\usepackage[figure,longend]{algorithm2e}

\begin{document}

\title{Fully self-consistent $GW$ and
quasi-particle self-consistent $GW$ for molecules} 

\author{P.~Koval}
\email{koval.peter@gmail.com}
\affiliation{Centro de F\'{\i}sica de Materiales CFM-MPC, 
Centro Mixto CSIC-UPV/EHU, Paseo Manuel de Lardizabal 5, E-20018 San Sebasti\'an, Spain}
\affiliation{Donostia International Physics Center (DIPC), 
Paseo Manuel de Lardizabal 4, E-20018 San Sebasti\'an, Spain}

\author{D.~Foerster}
\email{d.foerster@loma.u-bordeaux1.fr}
\affiliation{CPMOH/LOMA, Universit\'e de Bordeaux 1, 
351 Cours de la Liberation, 33405 Talence, France}

\author{D. S\'anchez-Portal}
\email{sqbsapod@ehu.es}
\affiliation{Centro de F\'{\i}sica de Materiales CFM-MPC, 
Centro Mixto CSIC-UPV/EHU, Paseo Manuel de Lardizabal 5, E-20018 San Sebasti\'an, Spain}
\affiliation{Donostia International Physics Center (DIPC), 
Paseo Manuel de Lardizabal 4, E-20018 San Sebasti\'an, Spain}



\date{\today}

\begin{abstract}
Two self-consistent schemes involving Hedin's $GW$ approximation
are studied for a set of sixteen different atoms and small molecules.  
We compare results from the fully 
self-consistent $GW$ approximation (SC$GW$)
and the quasi-particle self-consistent $GW$ approximation (QS$GW$) within
the same numerical framework.  Core and valence electrons are treated
on an equal footing in all the steps of the calculation. We use basis sets of localized
functions to handle the space dependence of quantities and spectral functions
to deal with their frequency dependence. 
We compare SC$GW$ and QS$GW$ on a qualitative level by comparing the computed
densities of states (DOS). To judge their relative merit on a quantitative level,
we compare their vertical ionization potentials (IPs) with those obtained from
coupled-cluster calculations CCSD(T).
Our results are futher compared with ``one-shot''
$G_0W_0$ calculations starting from Hartree-Fock solutions ($G_0W_0$-HF).
Both self-consistent $GW$ approaches behave quite similarly. Averaging over all
the studied molecules, both methods
show only a small 
improvement (somewhat larger for SC$GW$) 
of the calculated IPs with respect to $G_0W_0$-HF results. 
Interestingly, SC$GW$ and QS$GW$ calculations tend to deviate in 
opposite directions with respect to CCSD(T) results. SC$GW$ systematically 
underestimates the IPs, while QS$GW$ tends to overestimate them. $G_0W_0$-HF
produces results which are surprisingly close to QS$GW$ calculations
both for the DOS and for the numerical values of the IPs.
\end{abstract}

\maketitle

\section{Introduction}

Self-consistent methods are commonly used to solve the non-linear equations
appearing in electronic structure theory.
For instance, in the Hartree-Fock (HF) method,~\cite{Fulde,Martin} one iteratively
determines the best single-determinant wave function,
starting from a reasonable initial guess, until the energy is minimized. 
In the Kohn-Sham framework of density-functional theory (DFT) one uses
self-consistency to find, for a given  
exchange-correlation functional, a set of single-particle orbitals that 
are used to  determine the electron density~\cite{PhysRev.136.B864,PhysRev.140.A1133,Martin}.
Self-consistency is, in principle,  also an essential ingredient to solve
Hedin's coupled equations to compute the interacting single-particle 
Green's function~\cite{Hedin:1965,Hedin:1999}. 
Unfortunately, the full system of Hedin's equations contains unknown functional
derivatives that prevent an exact solution. However, Hedin also proposed
a simpler approximation, the so-called $GW$ approximation, which is numerically
tractable and has proven to be a useful tool to study the electronic properties
of real materials~\cite{Hedin:1965,Strinati80,Pickett84,
AryasetiawanGunnarsson:1998,Hedin:1999,Aulbur19991,
OnigaReiningRubio:2002,SchilfgaardeKotaniFaleev:2006,
FriedrichSchindlmayr:2006,RinkeQteishNeugebauerScheffler:2008}.

In the $GW$ approximation, the self energy $\Sigma$ is obtained from the product
of the electron Green's functions ($G$) and 
the screened interaction ($W$) as  $\Sigma=\mathrm{i}GW$.
However, in spite of their apparent simplicity, $GW$ calculations can be 
numerically quite involved and  demanding for real materials. 
For this reason, a popular approach has been the so-called ``one-shot''
$GW$,~\cite{Strinati80,Pickett84,Hybertsen86}
where one computes the electron self energy directly
from the Green's function $G$ obtained from DFT or HF results
and the corresponding screened interaction $W$. As an alternative, 
one can iterate the process and feed back the electron self energy
into the computation of $G$ and try to achieve self consistency in
the relation $\Sigma=\mathrm{i}GW$.
This seems a good idea for several reasons. For example, it eliminates the 
undesired dependence of the results on the arbitrary starting point
that is inherent in the one-shot $GW$ scheme and is often quite large
\cite{Rinke-etal:2005,Fuchs:2007-HSE+G0W0,Bruneval12,Marom-etal:2012}.
Even more importantly, it has been shown that
self-consistent $GW$ (SC$GW$) is a conserving approximation, respecting the 
conservation of the number of particles, momentum and energy, 
among others.~\cite{BaymKadanoff:1961}
Unfortunately, it was demonstrated for the 
homogeneous electron gas~\cite{HolmBarth:1998} that 
SC$GW$ tends to worsen the agreement
of the band structure
with respect to experimental results for nearly-free-electron metals,
as compared to the simpler one-shot $GW$ scheme.
This has been a widely accepted conclusion for years. However, recent work
on small molecules and atoms~\cite{Stan06,Stan09,RostgaardJacobsenThygesen:2010,
CarusoRinkeRenSchefflerRubio:2012,Marom-etal:2012, Caruso2013} has reported some improvements,
although moderate,  with the use of SC$GW$.

There is an alternative self-consistent $GW$ procedure, 
the so-called ``quasi-particle self-consistent approximation'' (QS$GW$), 
that has been shown 
to be more accurate than the one-shot $GW$ approximation for
several solids and
molecules.~\cite{SchilfgaardeKotaniFaleev:2006,PhysRevB.84.205415}
Surprisingly, in spite of the conflicting claims of accuracy for 
the self-consistent SC$GW$ and QS$GW$, there are 
few direct comparisons of their respective performances.
Indeed, to the best of our knowledge, a comparison in which these two approaches
are treated using the same numerical approach and where their comparative merits
can be compared unambiguously, is still lacking.
The purpose of this article is to provide such a consistent comparison 
between SC$GW$ and QS$GW$ using the same numerical implementation. 

Our results do not indicate that any of the two self-consistent
$GW$ approaches is clearly superior to the other, 
at least for the description of the small molecules considered here. 
Indeed, averaging over the set of studied molecules, 
they give results quite close and only
slightly better than those of one-shot $G_0W_0$ calculations using
HF as a starting point, and SC$GW$ gives results only marginally closer
to our reference CCSD(T) calculations than QS$GW$. 
During the self-consistent iteration 
QS$GW$ only requires the evaluation of the self energy at the
quasiparticle energies obtained in the previous step. This is computationally 
much less demanding than SC$GW$, which needs the self energy at all frequencies. 
For this reason, QS$GW$ could be a more suitable method for
calculations in large systems.

The rest of the article is organized as follows. We briefly describe Hedin's
$GW$ approximation in Section~\ref{s:gw-theory}.
In Section~\ref{s:SC$GW$-and-QS$GW$}, 
the two self-consistent $GW$ approaches are presented.
In Section~\ref{s:domi-prod-sf} and \ref{s:conv}, we elaborate our numerical methods and
their particular usage for the present all-electron SC$GW$ and QS$GW$ 
calculations.
Section~\ref{s:results} contains our results and discussion.
We present our main conclusions in Section~\ref{s:conslusion}.

\section{Hedin's $GW$ approximation}
\label{s:gw-theory}

Green's functions have been a method of choice in solid state physics where
electron correlations play an important role. In particular the interacting
single-particle Green's function $G(\bm{r},\bm{r}',\omega)$ depends 
only on two spatial variables and frequency, but it directly accounts for
the electron density, electron removal and addition energies, and it also
allows the computation of the total energy.~\cite{Fetter-Walecka,GM}
The interacting single-particle Green's function can be found by solving
Dyson's equation~\cite{Fetter-Walecka}
\begin{equation}
G(\bm{r},\bm{r}',\omega) = G_0(\bm{r},\bm{r}',\omega) +
G_0(\bm{r},\bm{r}'',\omega)\Delta\Sigma(\bm{r}'',\bm{r}''',\omega)
G(\bm{r}''',\bm{r}',\omega).
\label{Dyson-eq-0}
\end{equation}
Please, notice that here we adopt the convention that an integral over spatial
variables is implied in any equation unless these variables appear on its
left-hand side. In Eq.~(\ref{Dyson-eq-0}), 
$G_0(\bm{r},\bm{r}',\omega)$ is the single-particle 
Green's function of a reference, artificial, system
of non-interacting electrons 
\begin{equation}
G_0(\bm{r},\bm{r}',\omega) = \left[\omega\delta(\bm{r}-\bm{r}') - H_{\mathrm{eff}}(\bm{r},\bm{r}')\right]^{-1},
\label{gf-0}
\end{equation}
described by an  effective one-electron Hamiltonian 
\begin{equation}
\label{Heff}
\hat{H}_{\mathrm{eff}}=\hat{T}+\hat{V}_{\text{ext}}+\hat{V}_H+\hat{V}_{\text{xc}}\equiv
\hat{H}_0+\hat{V}_H+\hat{V}_{\text{xc}}.
\end{equation}
Here, $\hat{H}_0$ includes the one-electron terms, i.e., the 
kinetic energy operator  $\hat{T}$ and the external potential $\hat{V}_{\text{ext}}$ (electrostatic field
of the nuclei). The Hartree term (electrostatic field of the electron density)
is $\hat{V}_H$,
and the exchange and correlation operator is denoted by
$\hat{V}_{\text{xc}}$. Finally, 
\begin{equation} 
\Delta\Sigma(\bm{r},\bm{r}',\omega)=
\Sigma(\bm{r},\bm{r}',\omega)-\hat{V}_{\text{xc}}(\bm{r},\bm{r}'), 
\end{equation}
where $\Sigma(\bm{r},\bm{r}',\omega)$
is the self energy that describes the effects of electron correlations. 
In order to avoid double counting, it is necessary to subtract
the approximate description of those effects already 
included in the effective one-electron Hamiltonian ($\hat{V}_{\text{xc}}$). 
Standard choices for the reference non-interacting system are given
by the Kohn-Sham and HF methods. The interacting Green's function is
then obtained by solving Dyson's equation
\begin{equation}
G(\bm{r},\bm{r}',\omega) = 
\left[\omega \delta(\bm{r}-\bm{r}')-H_{\mathrm{eff}}(\bm{r},\bm{r}')-
\Delta\Sigma(\bm{r},\bm{r}',\omega) \right]^{-1}=
\left[ (\omega -V_H(\bm{r}))\delta(\bm{r}-\bm{r}')-H_{0}(\bm{r},\bm{r}') - 
\Sigma(\bm{r},\bm{r}',\omega) \right]^{-1}.
\label{Dyson-eq-h}
\end{equation}

A closed set of exact equations for the Green's functions, the self energy
(and a vertex) was written down by Hedin.~\cite{Hedin:1965}
However, these equations have been solved so far only for model systems 
\cite{Molinari:2005,Lani:2012}.
Fortunately, Hedin~\cite{Hedin:1965} also proposed an expansion of the 
self energy in powers of the screened interaction $W(\bm{r},\bm{r}',\omega)$.
To the lowest order he obtained a simple expression for the self energy,
the so-called $GW$ approximation, 
where the self energy is given by the product of the Green's function
and the screened Coulomb interaction~\cite{Hedin:1965}
\begin{equation}
\Sigma(\bm{r},\bm{r}',\omega) =
\frac{\mathrm{i}}{2\pi} \int d\omega' G(\bm{r},\bm{r}',\omega+\omega')
W(\bm{r},\bm{r}',\omega') e^{i\eta \omega'},
\label{self energy}
\end{equation}
with $\eta$ being a positive infinitesimal. 
The screened Coulomb interaction $W(\bm{r},\bm{r}',\omega)$ takes into account 
that an electron repels other electrons and thereby effectively creates 
a cloud of positive charge around it that weakens or screens the bare
Coulomb potential. The screened interaction can be found as a solution 
of an integral equation 
\begin{equation}
W(\bm{r},\bm{r}',\omega)=v(\bm{r},\bm{r}') + v(\bm{r},\bm{r}'')
\chi(\bm{r}'',\bm{r}''',\omega) W(\bm{r}''',\bm{r}',\omega),
\label{W}
\end{equation}
where, to the lowest order in the electron-electron interaction, 
the polarization operator
can be evaluated as~\cite{Hedin:1965}
\begin{equation}
\chi(\bm{r},\bm{r}',\omega)=-\frac{\mathrm{i}}{2\pi}
\int d\omega' G(\bm{r},\bm{r}',\omega+\omega')
G(\bm{r}',\bm{r},\omega') e^{i\eta \omega'}.
\label{response}
\end{equation}
Equations (\ref{Dyson-eq-0}), (\ref{self energy}), (\ref{W}) and
(\ref{response}) constitute a closed set of equations that can be iteratively 
solved in order to find an approximation to the interacting one-electron 
Green's function $G(\bm{r},\bm{r}',\omega)$. This is usually known
as the self-consistent $GW$ approximation (SC$GW$). The corresponding
cycle is schematically depicted in Fig.~\ref{a:SCGW-principle}.
It is important to stress that, as already noted above, 
SC$GW$ is just an approximation to the exact set of Hedin's equations.
The exact set of equations involves the vertex function 
$\Gamma(\bm{r},\bm{r}',\omega;\bm{r}'',\omega')$,
which requires computing the functional 
derivative of the exact self energy. The $GW$ approximation replaces
the vertex function by $\delta(\bm{r}-\bm{r}')\delta(\bm{r}-\bm{r}'')$,
which is the zeroth order expression for the expansion 
of the vertex function in terms of the screened interaction $W$. Thus, 
the $GW$ approximation transforms Hedin's equations into 
a numerically tractable set of equations. 

In spite of their apparent simplicity, $GW$ calculations are still numerically 
demanding. This is one of the reasons why most studies of
real materials to date do not use the SC$GW$ approach, i.e. do not iterate
$GW$ equations until self-consistency, but rather use
the so-called $G_0W_0$ approximation.  
In this ``one-shot" calculation, 
the non-interacting Green's function $G_0(\bm{r},\bm{r}',\omega)$
is used instead of the interacting one in 
Eqs. (\ref{self energy}),
(\ref{W}) and (\ref{response}).
The screened Coulomb interaction obtained in this way
is referred to as $W_0$ in the following. A clear drawback of the $G_0W_0$ calculation is the 
dependence of the results on the approximation used
to compute the non-interacting Green's function
$G_0$.~\cite{Rinke-etal:2005,Fuchs:2007-HSE+G0W0,
Bruneval12,Marom-etal:2012,Marom12bis,Bruneval13}
This dependence gives rise to sizable differences, for example, starting from HF or 
DFT effective Hamiltonians. The SC$GW$ scheme can correct
this undesired feature of $G_0W_0$. Furthermore,
it can be shown~\cite{BaymKadanoff:1961} that the
self-consistent version of $GW$ is a conserving approximation, i.e., 
respects electron number, momentum and energy conservation.

\section{Self-consistent approaches involving Hedin's $GW$}
\label{s:SC$GW$-and-QS$GW$}

\begin{figure}[htbp]
\includegraphics[width=7cm, angle=0,clip]{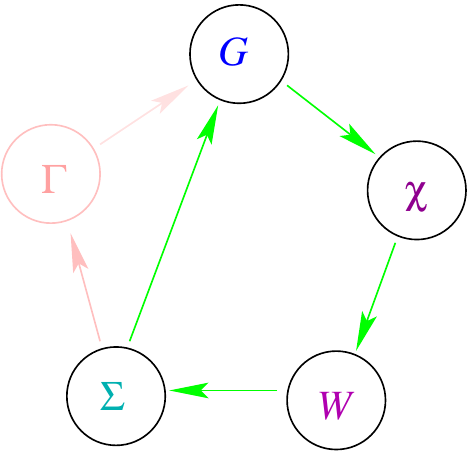}
\caption{
Schematic representation of the cycle in the self-consistent $GW$ (SC$GW$) 
approach versus exact Hedin's equations.
Exact  equations involve the vertex function $\Gamma$ for which, unfortunately,
there is not an explicit formula
available. Instead, the SC$GW$ method
approximates $\Gamma$ by its zeroth order term in an expansion as
function of $W$,
$\Gamma(\bm{r},\bm{r}',\omega;\bm{r}'',\omega')\approx\delta(\bm{r}-\bm{r}')\delta(\bm{r}-\bm{r}'')$,
giving rise to equations (\ref{self energy}), (\ref{W}) and
(\ref{response}) in the text. 
These equations, together with Dyson's equation~(\ref{Dyson-eq-0})
define a self-consistent procedure to compute the interacting Green's
function $G$. 
\label{a:SCGW-principle}}
\end{figure}

The formally simplest self-consistent $GW$ approximation
is illustrated in Fig.~\ref{a:SCGW-principle}. 
In this procedure, the self energy  at a given iteration 
is computed with the Green's function from the previous iteration
using the equations (\ref{self energy}), (\ref{W}) and
(\ref{response}) presented above.    This new self energy is 
then used to calculate a new Green's function, and the process
is iterated until a stable solution is found. 

In the first iteration, to start the self-consistent loop,
we need an initial approximation to the Green's function. This is 
typically obtained from the non-interacting 
Green's function $G_0(\bm{r}, \bm{r}',\omega)$
according to equation (\ref{gf-0}) using some suitable one-electron 
effective theory.  The non-interacting electron density
response $\chi_0(\bm{r},\bm{r}',\omega)$ and
the screened interaction $W_0(\bm{r},\bm{r}',\omega)$
are then obtained using equations (\ref{response}) and (\ref{W}). 
With the screened interaction, we can already calculate the 
self energy $\Sigma(\bm{r},\bm{r}',\omega)$ according to equation (\ref{self energy}).
So far the calculation is equivalent to a ``one-shot'' $G_0W_0$
calculation. However, inserting the calculated 
self energy into equation (\ref{Dyson-eq-h}) we can obtain 
our first approximation to the  interacting Green's function $G(\bm{r},\bm{r}',\omega)$.
 
We can now start the $GW$ calculation again, using the obtained interacting
Green's function $G(\bm{r},\bm{r}',\omega)$
(instead of the non-interacting one $G_0(\bm{r}, \bm{r}',\omega)$),
to compute $\chi(\bm{r},\bm{r}',\omega)$ and repeat the 
cycle until reaching self-consistency.  In such cycle, the 
Green's function in step $n$, $G^{(n)}$, is computed from the
self energy $\Sigma^{(n-1)}$ obtained using the information from the
previous step
\begin{equation}
G^{(n)}(\bm{r},\bm{r}',\omega) = 
\left[(\omega -V^{(n-1)}_H(\bm{r}))\delta(\bm{r}-\bm{r}') - H_{0}(\bm{r},\bm{r}') - 
\Sigma^{(n-1)}(\bm{r},\bm{r}',\omega) \right]^{-1}.
\label{Dyson-eq-h-n}
\end{equation}
The electron
density $n(\bm{r})$ has to be recalculated at the end of 
each iteration according to the relation 
\begin{equation}
n(\bm{r}) = -\frac{1}{\pi}\mathrm{Im}\left[
\int^{E_{\text{F}}}_{-\infty} G(\bm{r},\bm{r},\omega) d\omega\right]
\label{g2n}
\end{equation}
and, therefore, the Hartree potential $V_{H}(\bm{r})$ must be also updated after each iteration.
$E_{\text{F}}$ in Eq.~\ref{g2n} is the Fermi energy of the system, which is determined
by the number of electrons.

The most detailed studies on the performance of the SC$GW$ scheme
have been carried out for the homogeneous electron
gas.~\cite{PhysRevB.54.8411,PhysRevB.54.7758,HolmBarth:1998,
AryasetiawanGunnarsson:1998}
For this system it has been shown that SC$GW$ does not improve or
even worsens the description of the band structure, overestimating the 
bandwidth.~\cite{AryasetiawanGunnarsson:1998}
Furthermore, the weight of the plasmon satellite is reduced 
with respect to $G_0W_0$ and it
almost disappears in some cases. Part of these deficiencies seem
to be related to the use of the interacting Green's function in the
definition of the polarizability function $\chi$ (Eq.~\ref{response}). 
Due to the renormalization of the quasiparticle weight and the transfer
of spectral weight to higher energies (plasmon satellite), 
$\chi$ looses its clear physical meaning as a response function
and it no longer satisfies the $f$-sum rule.~\cite{AryasetiawanGunnarsson:1998}
As a consequence, the description of the screened interaction
$W$ is also affected and the plasmon resonance becomes very broad and ill-defined.
For  systems other than the homogeneous electron gas, the situation is
not so clear. Recent studies for atoms and small molecules seem
to reach conflicting conclusions about whether SC$GW$ improves the ionization
energies given by  $G_0W_0$ with suitable starting points, and whether these improvements
are sufficiently systematic to justify the use of the computationally more demanding
SC$GW$.~\cite{Stan06,Stan09,RostgaardJacobsenThygesen:2010,CarusoRinkeRenSchefflerRubio:2012,Marom-etal:2012,Marom12bis,Bruneval13,Caruso2013} 
In general,  the improvements, when present, seem to be small. 
In spite of  these deficiencies, the total energies obtained from 
SC$GW$ Green's functions, using either the Galitskii-Migdal formula~\cite{GM} or
the Luttiger-Ward functional~\cite{LW}, 
are quite accurate.~\cite{HolmBarth:1998,
AryasetiawanGunnarsson:1998,Stan06,Stan09,CarusoRinkeRenSchefflerRubio:2012,Caruso2013}
The good behavior of the total energy is probably related to the energy
conserving character of the SC$GW$
approximation.~\cite{BaymKadanoff:1961,AryasetiawanGunnarsson:1998}
Furthermore, the conserving character of SC$GW$ is an interesting
property that becomes useful in transport calculations.~\cite{PhysRevB.83.115108}

\begin{figure}[htb]
\includegraphics[width=7cm, angle=0,clip]{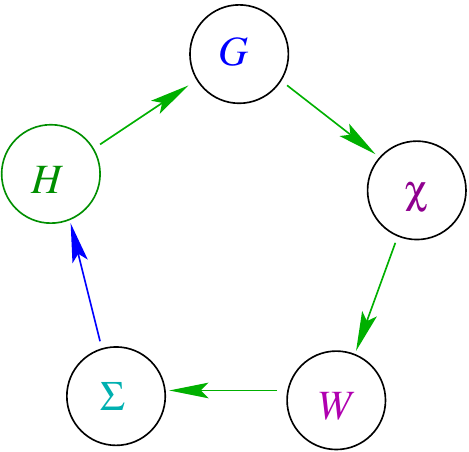}
\caption{Principle of the quasi-particle self-consistent $GW$ 
 approximation (QS$GW$). The calculated self energy at the $G_0W_0$
 level in one iteration is used to define 
 a new one-electron effective
 Hamiltonian. This new $\hat{H}_{\mathrm{eff}}$ provides the starting point for
 the next $G_0W_0$-like iteration. The procedure is repeated 
 until we get a stable
 $\hat{H}_{\mathrm{eff}}$. The method is based on a heuristic
 mapping $\Sigma_{G_0W_0}(\omega) \rightarrow
 \hat{H}_{\mathrm{eff}}$ as defined in Eq.~\ref{modeA}. 
}
\label{a:QSGW-principle}
\end{figure}

An alternative to this straightforward, self-consistent $GW$ approach
is given by the so-called ``quasi-particle self-consistent $GW$'' (QS$GW$) 
approximation recently proposed by Kotani,
Schilfgaarde and
Faleev.~\cite{SchilfgaardeKotaniFaleev:2006,PhysRevB.76.165106}
The rationale behind this approach is based on the perturbative character of the $GW$ approximation,
where the electron self energy is treated as a small perturbation.
Therefore, $GW$ should become a more accurate approximation if applied in
conjunction with a suitable effective one-electron Hamiltonian 
$\hat{H}_{\mathrm{eff}}$ that already 
provides a fair description of the one-electron-like excitations
of the many-electron system or quasiparticles (QP). The quasiparticles
can be obtained as solutions of the equation
\begin{equation}
\label{QPeq}
\{ \hat{H}_0+\hat{V}_H+\text{Re}
\left[\hat{\Sigma}(\epsilon_i)\right] -\epsilon_i \} |\psi_i\rangle = 0, 
\end{equation}
where $\text{Re}$ extracts the Hermitian part of the self-energy operator.
In QS$GW$, $\hat{H}_{\mathrm{eff}}$ is optimized such that its eigenfunctions
($\Psi_i$) and
eigenvalues ($E_i$) are good approximations to      
the QP wavefunctions ($\psi_i$) and 
energies ($\epsilon_i$) obtained using Eq.~\ref{QPeq} and 
a $G_0W_0$ self energy. This is done by defining a suitable mapping
$\Sigma_{G_0W_0}(\omega) \rightarrow \hat{H}_{\mathrm{eff}}$. 
Of course, as already described above, 
in order to compute the self energy $\Sigma_{G_0W_0}$ it 
is necessary to use a one-electron Hamiltonian as a starting point. 
Thus, in each iteration $n$ we obtain a new self energy 
$\Sigma^{(n)}_{G_0W_0}$, and 
a new effective Hamiltonian from it $\hat{H}^{(n)}_{\mathrm{eff}}$, that is then 
used to start the next iteration. 
The procedure finishes when $\Psi_i({\bf r})$ and $E_i$ do not change anymore
and, therefore, we have reached a self-consistent result for 
the ``optimum'' $\hat{H}_{\mathrm{eff}}$ (of course, the quality of these
results is determined
by the quality of the $\Sigma_{G_0W_0}(\omega) \rightarrow
\hat{H}_{\mathrm{eff}}$ mapping). 
Self-consistency in QS$GW$ is therefore not sought within the $GW$
calculation, but rather generating an optimal (in the sense
that minimizes the $\Delta\Sigma_{G_0W_0}(\epsilon_i)$=
$\Sigma_{G_0W_0}(\epsilon_i)-V_{\text{xc}}$ evaluated at the quasiparticle 
energies $\epsilon_i$~\cite{PhysRevB.76.165106}) 
non-interacting Green's function $G_0$ to perform a $G_0W_0$
calculation. The principle of this 
QS$GW$ approach is illustrated in Fig.~\ref{a:QSGW-principle}.

So far we have not specified the procedure to perform the mapping 
$\Sigma_{G_0W_0} \rightarrow \hat{H}_{\mathrm{eff}}$. This mapping is not unique
and Kotani {\it et al.} have actually proposed 
several ways to perform it. Here we have adopted the procedures called
``mode~A'' and ``mode~B'' in Ref.~\onlinecite{PhysRevB.76.165106},
which we recast in a single expression:
\begin{equation}
\label{modeA}
\hat{V}_{\text{xc}}= \frac{1}{2}(\hat{V}^{\dagger}_{\text{sfe}}+\hat{V}_{\text{sfe}}),
\end{equation}
where the operator $\hat{V}_{\text{sfe}}$ is given by
\begin{equation}
\label{vnsym}
\hat{V}_{\text{sfe}}=
\sum_{ij}|\Psi_i\rangle\text{Re}[\Sigma^{ij}(\omega_{ij})]\langle\Psi_j|.
\end{equation}
The frequency $\omega_{ij}$ is different for ``mode~A'' and ``mode~B''.
For ``mode~A'' $\omega_{ij}=E_j$, while for ``mode~B'' 
$\omega_{ij}=E_j, \text{ if } i=j, \text{ and } \omega_{ij}=E_{F} \text{ otherwise}$.
For the closed-shell molecules considered
here we take $E_{\text{F}}$ in the middle of the gap between
the highest occupied (HOMO) and lowest unoccupied (LUMO) molecular
orbitals.

Here $\text{Re}[\Sigma^{ij}(\omega)]$ denotes the Hermitian
part of the matrix elements of the self energy between the 
QP wavefunctions $\Psi_i({\bf r})$, and they are evaluated at the 
QP energies $E_i$.  
These QP wavefunctions $\Psi_i({\bf r})$ and  energies $E_i$ correspond to
the solutions of the QS$GW$ effective 
Hamiltonian at a given iteration and must be updated during the self-consistent loop.
Equation~(\ref{modeA}) is derived from the fact that
$\{\Psi_i\}$ forms a complete set and the requirement of having an Hermitian 
$\hat{V}_{\text{xc}}$ operator.~\cite{PhysRevB.76.165106} 
In Ref~\onlinecite{PhysRevB.76.165106} it was also shown that
Eq.~(\ref{modeA}) can be obtained from minimizing the norm of 
$\sum_{ij}
|\langle\Psi_i|\hat{\Sigma}(\epsilon_j)-\hat{V}_{\text{xc}}|\Psi_j\rangle|^2$.
However, the ultimate justification of the use of expression (\ref{modeA})
comes from the fact that it has been shown to provide accurate results for the
band structure of a large variety of semiconductors and transition metal
oxides.~\cite{SchilfgaardeKotaniFaleev:2006,PhysRevB.76.165106}

It is worth noting that in the present calculations we do not observe any
evidence of a starting-point dependence of the QS$GW$ results,
as recently suggested by calculations in oxides.~\cite{LiaoCarter:2011, IsseroffCarter:2012}
In the case of the small molecules studied here, HF and 
local density approximation DFT starting points converged always to the
same IPs and the same density of states. 
 
\section{Implementation of SC$GW$ and QS$GW$ schemes} 
\label{s:domi-prod-sf}

In the present work we compare the results of $G_0W_0$, SC$GW$ and
QS$GW$ calculations performed using the same numerical framework. 
Our numerical procedure is based on the use of a basis set of atomic orbitals
and a basis set of dominant products to express the products among 
those orbitals, as well as the use of spectral functions to treat the frequency
dependence of the functions involved in $GW$ calculations.~\cite{df-pk-dsp:2011}
In this Section, we focus on the main technical differences and describe
the additional procedures necessary to perform the present all-electron
self-consistent $GW$ calculations.

First, in our previous work~\cite{df-pk-dsp:2011} we presented
$G_0W_0$ results for several aromatic molecules starting from
DFT pseudopotential~\cite{Martin} calculations. In contrast, here we perform
all-electron calculations. This eliminates the important uncertainties
associated with the use of pseudopotentials, as discussed by several
authors.~\cite{Ku02,Delaney04,RostgaardJacobsenThygesen:2010,
CarusoRinkeRenSchefflerRubio:2012, Caruso2013, Gomez-AbalLiScheffler:2008} 
The basis of dominant products had to be improved
to adapt the basis for core-valence orbital products. The construction of 
the basis and the necessary improvements are described in subsection~\ref{ss:dp-basis}.

Second, in previous works we have used numerical orbitals with 
a finite spatial support.~\cite{SIESTA} However, here 
we use Gaussian basis sets to be able to carry out
consistent comparisons with coupled-cluster calculations performed using the
NWChem package.~\cite{Valiev20101477}

Third, for small molecules, HF solutions seem to be a better starting
point for $GW$ calculations than local or semilocal DFT functionals.~\cite{Bruneval13} 
For this reason, most of our calculations were initiated from a HF solution of the system.
The final results in the self-consistent schemes are independent
of the starting point as we will show explicitly.
For our HF calculations we have used a modified version of a code
originally due to James Talman.~\cite{PhysRevLett.84.855} In the present work,
the Hartree and exchange operators are computed using the dominant products basis. 

Fourth, some modifications are necessary in our {\it non-local compression}
scheme~\cite{df-pk-dsp:2011} of the dominant product basis to perform SC$GW$
calculations as explained in some detail in the subsection \ref{ss:nl}.

Fifth, both self-consistent methods, SC$GW$ and QS$GW$, need some 
mixing procedure to achive convergence. The mixing procedures are explained 
in the subsection \ref{ss:mix}.

Finally, we use spectral functions to deal with the frequency dependence 
of Green's function, response function, screened interaction and 
self energy. Although the method had not changed substantially 
since our publication~\cite{df-pk-dsp:2011}, we briefly describe
our method in subsection~\ref{ss:sf} for the sake of the readability of the manuscript.

\subsection{Expansions using orbital and dominant-products basis sets}
\label{ss:dp-basis}

We use linear combination of atomic orbitals (LCAO) approach \cite{Mulliken07071967}
and expand the eigenfunctions $\Psi_E(\bm{r})$ of the one-electron Hamiltonian
in terms of atom-centered localized functions $f^{a}(\bm{r})$
\begin{equation}
\Psi_E(\bm{r}) = \sum_{a} X^E_a f^{a}(\bm{r}).
\label{lcao}
\end{equation}
The atomic orbitals $f^{a}(\bm{r})$ have a predefined angular momentum
and radial shape, while the coefficients $X^E_a$ must be 
determined by solving the corresponding eigenvalue equation. 
In  this work we have used a basis set of atomic orbitals 
expanded in terms of Gaussian functions.~\cite{JCC:JCC9,doi:10.1021/ci600510j}
These basis sets are the same used by most of the Quantum Chemistry codes.
We have used NWChem code~\cite{Valiev20101477} to perform
the $\Delta$SCF coupled-cluster calculations that will be compared
with our $GW$ results. In particular, for most
calculations we have used two different sets of basis for
all our calculations: a correlation-consistent double-$\zeta$
(cc-pVDZ) and a triple-$\zeta$ (cc-pVTZ) basis. 
This choice represents a trade off between the computational cost of
our all-electron $GW$ calculations, their accuracy and our intent to perform
calculations for a relatively large set of molecules. 
Having results with two different basis sets allows estimating the
dependence of the observed behaviors on the size of the basis set. 
Furthermore, the smaller cc-pVDZ basis also allowed us to perform calculations with a higher
frequency resolution, which 
is instrumental to study the convergence with respect
to this computational parameter.
As commented in more detail in Section~\ref{ss:numerical-results},
several recent studies of the convergence of $GW$ calculations 
with respect to the size of the basis set
indicate that, for several small molecules and atoms, the cc-pVTZ basis
provides results for the IPs within few tenths of eV of the converged
values.~\cite{PhysRevB.84.205415,Bruneval12,Bruneval13} This is
further confirmed by a systematic convergence study as a function
of the basis set size that we have performed for two small systems,
He and H$_2$. For these two species we could explore the convergence
of the results using basis sets up to 
cc-pV5Z. As described in detail in Subsection~\ref{ss:mult-conv} and
Section~\ref{ss:numerical-results}, 
these highly converged results seem to confirm that 
the main conclusions of our comparison 
among different self-consistent $GW$ schemes remain valid in the limit
of saturated basis sets.  

In the case of the initial HF calculations, we must 
self-consistently solve the equation
\begin{equation}
\left(-\frac{1}{2}\nabla^2 + V_{\mathrm{ext}}(\bm{r})+V_{\mathrm{H}}(\bm{r})
\right)\Psi_E(\bm{r})  \\
 + \int \Sigma_{\mathrm{x}}(\bm{r},\bm{r}') \Psi_E(\bm{r}') d^3 r' 
 =  E \Psi_E(\bm{r}),
 \label{hf-eigen}
\end{equation}
where Hartree and exchange operators depend on the eigenfunctions $\Psi_E(\bm{r})$, with 
\begin{equation}
V_{\mathrm{H}}(\bm{r}) = 2 \sum_{E < E_{\text{F}}}
\int \frac{\Psi^*_E(\bm{r}')\Psi_E(\bm{r}')}{|\bm{r}-\bm{r}'|} d^3r'
\end{equation}
(we assume here a closed-shell system and the factor
of two stands for the two orientations of the spin),
 and
\begin{equation}
 \Sigma_{\mathrm{x}}(\bm{r},\bm{r}') 
 = \sum_{E < E_{\text{F}}}
 \frac{\Psi_E(\bm{r})\Psi^*_E(\bm{r}')}{|\bm{r}-\bm{r}'|}.
\label{hf-vh-x}
\end{equation}
Introducing (\ref{lcao}) in equations (\ref{hf-eigen}) and (\ref{hf-vh-x}),
we obtain the Hartree-Fock equations in a basis of atomic orbitals
\begin{equation}
H^{ab}X^E_b = ES^{ab}X^E_b,
\label{sp-equation}
\end{equation}
with
$H^{ab}\equiv T^{ab}+V_{\mathrm{ext}}^{ab}+V_{\mathrm{H}}^{ab}+\Sigma_{\mathrm{x}}^{ab}$
and $S^{ab}$, respectively, the matrix elements of the Fock operator and the overlap.
The exchange operator
$\Sigma_{\mathrm{x}}^{ab}$ is given by 
\begin{equation}
\Sigma_{\mathrm{x}}^{ab} = 
\sum_{E < E_{\text{F}}}  X^{E}_{a'}X^{E}_{b'} 
\iint 
\frac{f^{a}(\bm{r})f^{a'}(\bm{r})f^{b'}(\bm{r}')f^{b}(\bm{r}')}{|\bm{r}-\bm{r}'|}
d^3r d^3 r'.
\label{exchange-lcao}
\end{equation}

The appearance of products of atomic orbitals $f^{a}(\bm{r})f^{a'}(\bm{r})$ 
in this expression gives rise, in principle, 
to the need of computing cumbersome four-center integrals.
In practice, this can be avoided using an auxiliary basis set that spans
the space of orbital products and largely simplifies the
calculations.~\cite{PhysRevB.49.16214,PhysRevB.69.085111}.
Furthermore, the set of products of 
atomic orbitals usually comprise strong collinearities. 
Therefore, if properly defined, the number of elements in this 
auxiliary basis can be much smaller than the total number of orbital products,
making the calculations more efficient. In Ref.~\onlinecite{df:2008},
one of us presented a well-defined method to obtain such an auxiliary basis 
for an arbitrary set of atomic orbitals. In this work we use this set
of {\it dominant products} in all the operations involving products of
atomic orbitals. The dominant
products $F^{\mu}(\bm{r})$ are independently defined for each atom pair and
provide an optimal, orthogonal (with respect to the Coulomb metric)
basis to expand the products of orbitals within that pair of atoms, i.e.,
\begin{equation}
f^{a}(\bm{r})f^{b}(\bm{r}) = \sum_{\mu} V^{ab}_{\mu} F^{\mu}(\bm{r}).
\label{prod-vertex-identity}
\end{equation}
Therefore, the dominant products preserve the local character of the original
atomic orbitals and $V^{ab}_{\mu}$ is a sparse table by construction.

The dominant products $F^{\mu}(\bm{r})$ are expanded in terms of spherical
harmonics about a center. In the case of valence--valence 
and core--core bilocal products (i.e., involving two 
atoms at different locations and valence or core orbitals in both atoms), the midpoint along
the vector that joins both nuclei is chosen as the expansion center.
However, for pairs of orbitals involving 
core orbitals in one atom and valence orbitals in the other atom, 
we use an expansion center that is much
closer to the nucleus of the first atom. The center of
expansion for such core--valence
products is determined using information about the 
spatial extension of the core
and valence shells. As a measure of the spatial extension of a given shell,
we take an average of the square-root of the expectation values of $r^2$ among all the 
radial orbitals
belonging to that shell, 
$
R =  \frac{\sum_{s} (2 l_s+1) \sqrt{\int f_{s}(r) r^4 dr}}{\sum_{s} (2 l_s+1)},
$
where $2 l_s+1$ is the multiplicity of a given orbital with angular momentum $l_s$.
The coordinate of this core-valence bilocal dominant product is then 
calculated as a weighted sum of the positions of the two shells (atoms) involved,
$\bm{C}_{\text{core}}$ and $\bm{C}_{\text{val}}$,
$
\bm{C}_{\text{expand} }= \frac{\bm{C}_{\text{val}} R_{\text{core}} +
\bm{C}_{\text{core}} R_{\text{val}}}{R_{\text{val}}+R_{\text{core}}}$.
This adjustment of the expansion center significantly increased the 
accuracy of the expansion (Eq.~\ref{prod-vertex-identity}). For instance, the precision 
of the computed overlaps and dipoles improved by an order of magnitude.

The product expansion in Eq.~(\ref{prod-vertex-identity}) allows reducing
substantially the dimension of the space of orbital products.
For example,  using a cc-pVDZ basis we have 38 orbitals to 
describe acetylene (C$_2$H$_2$), leading  to 703 products.
However, they can be expressed in terms of 491 dominant products
with high precision (throwing away 
eigenfunctions of the local Coulomb metric
with eigenvalues
lower than $10^{-6}$).~\cite{df:2008} 
In general, we typically found a reduction in the number of products
by at least 30\% with this local compression scheme in these accurate
calculations. Still, as we will see in subsection \ref{ss:nl} it is 
generally possible to reduce further the dimension of the product
basis using a {\it non-local compression} scheme. We can now rewrite the 
exchange operator (\ref{exchange-lcao}) as
\begin{equation}
\Sigma_{\mathrm{x}}^{ab} = V^{aa'}_{\mu} D_{a'b'} v^{\mu\nu} V^{b'b}_{\nu},
\label{exchange-domi-prod}
\end{equation}
where $D_{ab}=\sum_{E < E_{\text{F}}} X^{E}_{a}X^{E}_{b}$ is a density matrix,
and $v^{\mu\nu}$ are matrix elements 
\begin{equation}
v^{\mu\nu}=\iint\frac{F^{\mu}(\bm{r})F^{\nu}(\bm{r}')}{|\bm{r}-\bm{r}'|}d^3r d^3 r'.
\label{coulomb-metric}
\end{equation}
Therefore, the exchange operator (\ref{exchange-domi-prod}) is 
efficiently calculated in terms of two-center integrals (\ref{coulomb-metric}).
The matrix elements of Hartree potential $V_{\mathrm{H}}(\bm{r})$
are also calculated in this basis of dominant products
$V_{\mathrm{H}}^{ab} = 2 V^{ab}_{\mu} v^{\mu\nu} D_{a'b'} V^{a'b'}_{\nu}$.

As shown in Ref.~\onlinecite{df-pk-dsp:2011}, the $GW$ equations 
(\ref{Dyson-eq-h}), (\ref{self energy}), (\ref{W}) and (\ref{response}) can also be
conveniently rewritten  within the basis sets of atomic orbitals $\{f^{a}(\bm{r})\}$
and dominant products $\{F^{\mu}(\bm{r})\}$. We state these equations without 
derivation for the sake of completeness
\begin{eqnarray}
G_{ab}(\omega) &= & 
\left[ \omega S^{ab} -V^{ab}_H-H^{ab}_{0} -\Sigma^{ab}(\omega) \right]^{-1}, 
\label{gf_tensor}\\
\Sigma^{ab}(\omega) &= &
\frac{\mathrm{i}}{2\pi} \int d\omega' V^{aa'}_{\mu}G_{a'b'}(\omega+\omega')
W^{\mu\nu}(\omega')V^{b'b}_{\nu} e^{i\eta \omega'}, \label{se_tensor} \\
W^{\mu\nu}(\omega) &= &
\left[\delta^{\mu}_{\nu'} - v^{\mu\mu'}\chi_{\mu'\nu'}(\omega)\right]^{-1} v^{\nu'\nu}, 
\label{si_tensor}\\
\chi^{\mu\nu}(\omega) & = & -\frac{\mathrm{i}}{2\pi}
\int d\omega' V^{ad}_{\mu} G_{ab}(\omega+\omega')G_{cd}(\omega')
V^{bc}_{\nu} e^{i\eta \omega'} \label{rf_tensor}.
\end{eqnarray}
The treatment of convolutions in the latter equations is done with 
spectral function technique as explained below.

\subsection{Spectral functions technique}
\label{ss:sf}

As customary, the screened interaction $W(\bm{r},\bm{r}',\omega)$ 
in our calculation is separated into the bare Coulomb 
interaction $v(\bm{r},\bm{r}')$ and a frequency-dependent component 
$W_{\text{c}}(\bm{r},\bm{r}',\omega)=W(\bm{r},\bm{r}',\omega)-v(\bm{r},\bm{r}')$.
The bare Coulomb interaction $v(\bm{r},\bm{r}')$ gives rise to the
HF exchange operator.~\cite{FriedrichSchindlmayr:2006}
It can be computed with the space of dominant products 
without much computational effort according to Eq.~(\ref{exchange-domi-prod}).
The $GW$ correlation operator $\Sigma_{\text{c}} = \mathrm{i}GW_{\text{c}}$ is more
demanding due to the frequency dependence combined with the rather
large dimension of the space of products.

Because of the discontinuities of the electronic Green's functions, a
straightforward convolution to obtain either response function (\ref{rf_tensor})
or the self-energy operator (\ref{se_tensor}) is practically impossible both 
in the time domain and in the frequency domain.
However, one can use an imaginary time technique \cite{Godby:1999} or spectral function
representations \cite{Shishkin-Kresse:2006,df-pk:2009,df-pk-dsp:2011}
to recover a computationally feasible approach. In this work,
we continue to use the spectral function technique and rewrite the time-ordered
operators as follows 
\begin{equation}
\begin{aligned}G_{ab}(t) &=
-\mathrm{i}\theta(t)\int_{0}^{\infty}ds\,\rho_{ab}^{+}(s)e^{-\mathrm{i}st}
+\mathrm{i}\theta(-t)\int_{-\infty}^{0}ds\,\rho_{ab}^{-}(s)e^{-\mathrm{i}st}; \\
\chi_{\mu\nu}(t) &=
-\mathrm{i}\theta(t)\int_{0}^{\infty}ds\,a_{\mu\nu}^{+}(s)e^{-\mathrm{i}st}
+\mathrm{i}\theta(-t)\int_{-\infty}^{0}ds\,a_{\mu\nu}^{-}(s)e^{-\mathrm{i}st}; \\
W_{\text{c}}^{\mu \nu }(t) &=
-\mathrm{i}\theta(t)\int_{0}^{\infty}ds\,\gamma_{+}^{\mu \nu}(s)e^{-\mathrm{i}st}
+\mathrm{i}\theta(-t)\int_{-\infty}^{0}ds\,\gamma_{-}^{\mu \nu}(s)e^{-\mathrm{i}st}; \\
\Sigma^{ab}_{\text{c}}(t) &=
-\mathrm{i}\theta(t)\int_{0}^{\infty}ds\,\sigma_{+}^{ab}(s)e^{-\mathrm{i}st}
+\mathrm{i}\theta(-t)\int_{-\infty}^{0}ds\,\sigma_{-}^{ab}(s)e^{-\mathrm{i}st}, \\
\end{aligned}\label{spectral_1}
\end{equation}
where ``positive'' and ``negative'' spectral functions define the whole spectral 
function by means of Heaviside functions $\theta(t)$. For instance, the spectral
function of the electronic Green's function reads
$\rho_{ab}(s)=\theta(s)\rho^{+}_{ab}(s)+\theta(-s)\rho^{-}_{ab}(s)$.
Transforming the first of equations (\ref{spectral_1}) to the frequency
domain, we obtain the familiar expression for the spectral representation of
a Green's function 
\begin{equation}
G_{ab}(\omega) = \int_{-\infty}^{\infty} \frac{\rho_{ab}(s) \, ds }{
\omega-s+\mathrm{i}\, \mathrm{sgn}(s) \varepsilon}.
\end{equation}%
Here $\varepsilon $ is a small line-broadening constant. 
In practice, the choice of $\varepsilon $ is related 
to the spectral resolution $\Delta \omega $ of the
numerical treatment and will be discussed below in section
\ref{s:conv}.

One can derive expression for spectral function of response 
$a_{\mu \nu }(s)$ using equations (\ref{rf_tensor}) and (\ref{spectral_1})
\begin{equation}
a_{\mu \nu }^{+}(s)=\iint V_{\mu }^{ad}
\rho _{ab}^{+}(s_{1})\rho _{cd}^{-}(-s_{2})V_{\nu}^{bc}
\delta (s_{1}+s_{2}-s)ds_{1}ds_{2}.
\label{sf_response_tensor}
\end{equation}%
Here, the convolution can be computed with fast Fourier methods and the
(time-ordered) response function $\chi_{\mu \nu }(\omega )$ can be
obtained with a Kramers-Kronig transformation 
\begin{equation}
\chi_{\mu \nu }(\omega )=\chi _{\mu \nu }^{+}(-\omega )+\chi _{\mu \nu
}^{+}(\omega ),\text{ where }\chi _{\mu \nu }^{+}(\omega )=\int_{0}^{\infty
}ds\,\frac{a_{\mu \nu }^{+}(s)}{\omega +\mathrm{i}\varepsilon -s}.
\label{sf2response}
\end{equation}

The calculation of the screened interaction $W_{\text{c}}^{\mu \nu }(\omega )$ must be
done with the response function, rather than with its spectral representation,
because of the
inversion in equation (\ref{si_tensor}). The spectral function of the
screened interaction $\displaystyle\gamma ^{\mu \nu }(\omega )$ 
can be easily recovered from the screened interaction
itself \cite{FriedrichSchindlmayr:2006}.
Deriving the spectral function $\sigma (\omega )$ of the self energy, we
arrive at 
\begin{align}
\sigma_{+}^{ab}(s)& =\int_{0}^{\infty }\,\int_{0}^{\infty }\delta
(s_{1}+s_{2}-s)\,V_{\mu }^{aa^{\prime }}\rho _{a^{\prime }b^{\prime
}}^{+}(s_{1})V_{\nu }^{b^{\prime }b}\gamma _{+}^{\mu \nu
}(s_{2})ds_{1}ds_{2},  \label{spectral_3} \\
\sigma_{-}^{ab}(s)& =-\int_{-\infty }^{0}\,\int_{-\infty }^{0}\delta
(s_{1}+s_{2}-s)V_{\mu }^{aa^{\prime }}\rho _{a^{\prime }b^{\prime
}}^{-}(s_{1})V_{\nu }^{b^{\prime }b}\gamma _{-}^{\mu \nu
}(s_{2})ds_{1}ds_{2}.  \notag
\end{align}%
These expressions show that the spectral function of a convolution is given
by a convolution of the corresponding spectral functions. As in the 
response functions, we compute these convolutions employing fast Fourier transforms.

\subsection{Frequency-dependent functions on the equidistant grid}

The spectral functions of the non-interacting Green's function
(\ref{gf-0}) are merely a set of poles at the eigenenergies $E$ 
\begin{equation}
\rho _{ab}^{+}(\omega )=\sum_{E>E_{\text{F}}}\delta (\omega -E)X_{a}^{E}X_{b}^{E},\
\rho _{ab}^{-}(\omega )=\sum_{E<E_{\text{F}}}\delta (\omega -E)X_{a}^{E}X_{b}^{E}.
\label{sf_0}
\end{equation}%

The use of fast Fourier techniques for convolution, for instance in equation
(\ref{sf_response_tensor}), requires that the spectral functions
$\rho _{bc}^{+}(\omega )$, $\rho_{da}^{-}(\omega )$ be known
at equidistant grid points 
$\omega _{j}=j\Delta\omega ,j=-N_{\omega }\ldots N_{\omega }$,
rather than at a set of energies resulting from a diagonalization procedure.
The solution to this problem (discretization of spike-like functions)
is known and well tested.~\cite{Shishkin-Kresse:2006,df-pk:2009,df-pk-dsp:2011}
We define a grid of points that covers the whole range of eigenenergies $E$.
Going through the poles $E$, we assign their spectral weight
$X_{a}^{E}X_{b}^{E}$ to the neighboring grid points $n$
and $n+1$ such that $\omega _{n}\leq E<\omega _{n+1}$ according to the
distance between the pole and the grid points 
$\displaystyle p_{n,\,ab}=\frac{\omega _{n+1}-E}{\Delta \omega }X_{a}^{E}X_{b}^{E},\
p_{n+1,\,ab}=1-p_{n,\,ab}.$
Such a discretization keeps both the spectral weight and the center of mass
of a pole. Convergence of discretization parameters is discussed below, in
section \ref{s:conv}.

As a result of our calculation, we obtain the density of states (DOS) directly 
from the imaginary part of the converged Green's function 
\begin{equation}
\mathrm{DOS}(\omega)=-\frac{1}{\pi} \mathrm{Im} 
\left[G_{ab}(\omega)S^{ab}\right],
\end{equation}
where $G^{ab}(\omega)$ is obtained by solving Dyson's equation (\ref{gf_tensor}).
In our approach, the ionization potential IP is found directly from the 
density of states $\mathrm{DOS}(\omega)$ on a uniform frequency grid.
We find the IP by fitting the density of states locally by a third order
polynomial and by finding the maximum of this fit.

The convergence of both SC$GW$ and QS$GW$ loops
is determined by the $\mathrm{DOS}(\omega)$
\begin{equation}
\label{conv}
\mathrm{Conv} = 
\frac{1}{N_{\text{orbs}}}\int 
\left|\mathrm{DOS}_i(\omega) -\mathrm{DOS}_{i-1}(\omega)\right| d\omega,
\end{equation}
where $N_{\text{orbs}}$ is total number of orbitals in the molecule ---
the $\text{DOS}_{i}(\omega)$ is normalized to this number and $i$ is 
the iteration number.
We have chosen a small threshold on this convergence  parameter
$\mathrm{Conv}<10^{-5}$ in order to stop $GW$ the iteration of
both self-consistency schemes.
In general we observe that this criterium translates to
an even larger accuracy in the convergence of IP
(better than $10^{-5}$ relative error).

\subsection{Non-local compression of the dominant-products basis }
\label{ss:nl}

The calculation of screened interaction $W_{\text{c}}(\bm{r},\bm{r}',\omega)$
should have been performed in the space of orbital products, thus
requiring the inversion of matrices of large dimensions.
The basis of dominant products partially alleviates this problem
by eliminating the collinearities between products of 
orbitals corresponding to the same pair of atoms. However, there 
are still strong linear dependencies between products of orbitals
corresponding to neighboring pairs of atoms. Thus, the 
number of elements in the auxiliary basis set for the orbital product
expansion can be further reduced with important savings 
in the required memory and run time. 
In order to address this problem, we perform an additional 
non-local compression:
the new product basis is formed by linear combinations of the  
dominant products of all the pairs of atoms in the
molecule. As described in detail in Ref~\onlinecite{df-pk-dsp:2011},
these linear combinations are obtained by first constructing
the Coulomb metric projected into a relevant 
function manifold, and second keeping only the eigenfunctions of
this projected metric with eigenvalues larger than a threshold
value $\lambda_{\mathrm{thrs}}$. Thus, the   
elements of this new basis are orthogonal to each other with
respect to Coulomb metric. The relevant manifold is determined
by low-energy {\it electron-hole} pair excitations according to:
$\{V_{\mu }^{EF}\equiv X^{E}_a V^{ab}_{\mu} X^{F}_b\}$, where 
$X^{E}_a$ are the eigenvectors of the effective Hamiltonian (\ref{sp-equation}),
and $V^{ab}_{\mu}$ is the product ``vertex'' (\ref{prod-vertex-identity}).
In the construction of the metric only low-energy excitations are
included according to the criterium:
\begin{equation}
|E-F|<E_{\mathrm{thrs}} \,\text{and} \,E-E_{\text{F}}<0, F-E_{\text{F}}>0.
\label{subsetof_vef}
\end{equation}
Using Eq.~(\ref{subsetof_vef}) to select the relevant
electron-hole pair excitations to describe the dynamics
provides good results for one-shot $G_0W_0$ calculations 
if $E_{\mathrm{thrs}}$ is sufficiently large. 
However, for  SC$GW$ one has to reconsider this point more carefully.
During the iteration process, the restriction that the relevant subspace
to represent the polarization function $\chi_{\mu\nu}(\omega)$ necessarily
corresponds to pairs of occupied--unoccupied eigenstates of the initial
one-electron Hamiltonian $\hat{H}_{\mathrm{eff}}$ is \textit{relaxed}. 
With each iteration we are loosing the information about the initial
$\hat{H}_{\mathrm{eff}}$ and its sharp division of the Hilbert space
into one occupied and one unoccupied manifolds.
Therefore, we have used a more general subset of vectors 
$V_{\mu }^{EF}=X_{a}^{E}V_{\mu }^{ab}X_{b}^{F}$
in which more general \textit {low-energy pairs} $EF$ were included according to 
\begin{equation}
|E-F|<E_{\mathrm{thrs}}.
\label{subsetof_vef_gen}
\end{equation}
So we consider products of occupied/occupied, unoccupied/unoccupied
and occupied/unoccupied pairs of eigenfunctions of $\hat{H}_{\mathrm{eff}}$, 
provided that their energies are sufficiently close. 

In our calculations $E_{\mathrm{thrs}}$ and $\lambda_{\mathrm{thrs}}$ are treated
as convergence parameters, which are refined until convergence is reached
in the self energy for the range of frequencies 
under exploration. Here we consider small molecules 
with a relatively small basis set. Therefore it was actually 
possible to include all possible pairs of eigenvectors in the 
compression procedure, while $\lambda_{\mathrm{thrs}}$ was taken $10^{-3}$ for all molecules.
With this choice, we could get a significant reduction in the
size of the product basis. For example, for the acetylene molecule with
a cc-pVDZ basis, from 
the 703 initial products of orbitals, we made a first local compression to 
491 dominant products and, with the non-local compression, this was
reduced to 128 basis elements.

\subsection{ $\Sigma(\omega) \rightarrow \hat{V}_{\text{xc}}$ mapping
in a basis of atomic orbitals}

The map of the self energy to an exchange-correlation operator 
(\ref{modeA}) is made separately for the frequency-independent
(exchange) self energy $\Sigma_{\text{x}}=\mathrm{i}Gv=\Sigma^{\mathrm{HF}}_{\text{x}}$,
and for the frequency-dependent correlation self energy
$\Sigma_{\text{c}}(\omega)=\mathrm{i}GW_{\text{c}}$.
Obviously, the exchange operator $\hat{V}_{\mathrm{x}}$ is identical
to the exchange part of the self energy 
$V^{ab}_{\mathrm{x}} = \Sigma_{\mathrm{x}}^{ab}$ (i.e. to the HF exchange operator 
\ref{exchange-domi-prod}).

The correlation operator $\hat{V}_{\mathrm{c}}$ is found by using 
equation (\ref{modeA}) and inserting the LCAO expansion (\ref{lcao})
into equation (\ref{vnsym}) 
\begin{equation}
V_{\text{sfe},\mathrm{c}}^{ab} = \sum_{ij}S^{aa'}X^{i}_{a'} X^{i}_{a''}
\text{Re}[\Sigma_{\mathrm{c}}^{a''b''}(\omega_{ij})]
X^{j}_{b''}X^{j}_{b'}S^{b'b}.
 \label{self energy2vxc-map-lcao}
\end{equation}
Because we use real-valued basis functions $f^{a}(\bm{r})$, the 
Hermitian part of operator reduces to the real part.
In our approach, we obtain the self energy $\Sigma_{\mathrm{c}}^{ab}(\omega)$
on an equidistant frequency grid, which allows the calculation of
convolutions by means of fast Fourier transforms. The eigenvalues $E$ of
the QP equation
do not necessarily fit with any equidistant grid, but we have found that a
linear interpolation procedure
provides a reliably converging
approximation to the self energy in an arbitrary energy $\Sigma_{\mathrm{c}}^{ab}(E)$.

\subsection{Mixing schemes for SC$GW$ and QS$GW$}
\label{ss:mix}

Mixing of successive iterations is often necessary to achieve convergence in iterative 
approaches to nonlinear equations.
Mixing is needed to solve the Hartree-Fock equations and the same is true for 
the self-consistent equations of SC$GW$ and QS$GW$. 

In the SC$GW$ scheme (Fig.~\ref{a:SCGW-principle}) we have to mix
frequency-dependent operators, which unfortunately leads to large memory
requirements. Therefore, we resorted to the simplest linear mixing scheme. 
Initially, we tried to mix the Green's functions calculated in sucessive steps 
as suggested in Ref.~\onlinecite{PhysRevB.83.115108}. However,  
we found examples where the convergence was unstable and the results unreliable. 
By contrast, a linear mixing of the self energy 
\begin{equation}
\label{slfemixing}
\Sigma^{i}(\omega) = (1-\alpha) \Sigma_{\mathrm{in}}^{i-1}(\omega) + 
\alpha\Sigma_{\mathrm{out}}^{i-1}(\omega)
\end{equation}
always worked in the case of SC$GW$ and it was possible to use a mixing weight 
as large as $\alpha=0.35$.

In the case of QS$GW$ calculations (Fig.~\ref{a:QSGW-principle}) 
the self energy mixing sometimes failed to achieve convergence.
A convenient solution was to mix the correlation operator (\ref{self energy2vxc-map-lcao})
rather than the self energy. This mixing of correlation operator has been also used in
the MOLGW code by Bruneval.~\cite{Bruneval12}
For the molecules considered here, the linear mixing of the
correlation operator has been used with $\alpha=0.25$.

\subsection{Independence of SC$GW$ and QS$GW$ on their starting points}

\begin{figure*}[htb]
\begin{tabular}{p{6cm}p{5.4cm}p{6cm}}
\includegraphics[height=4.2cm, angle=0,viewport=35 50 445 320,clip]{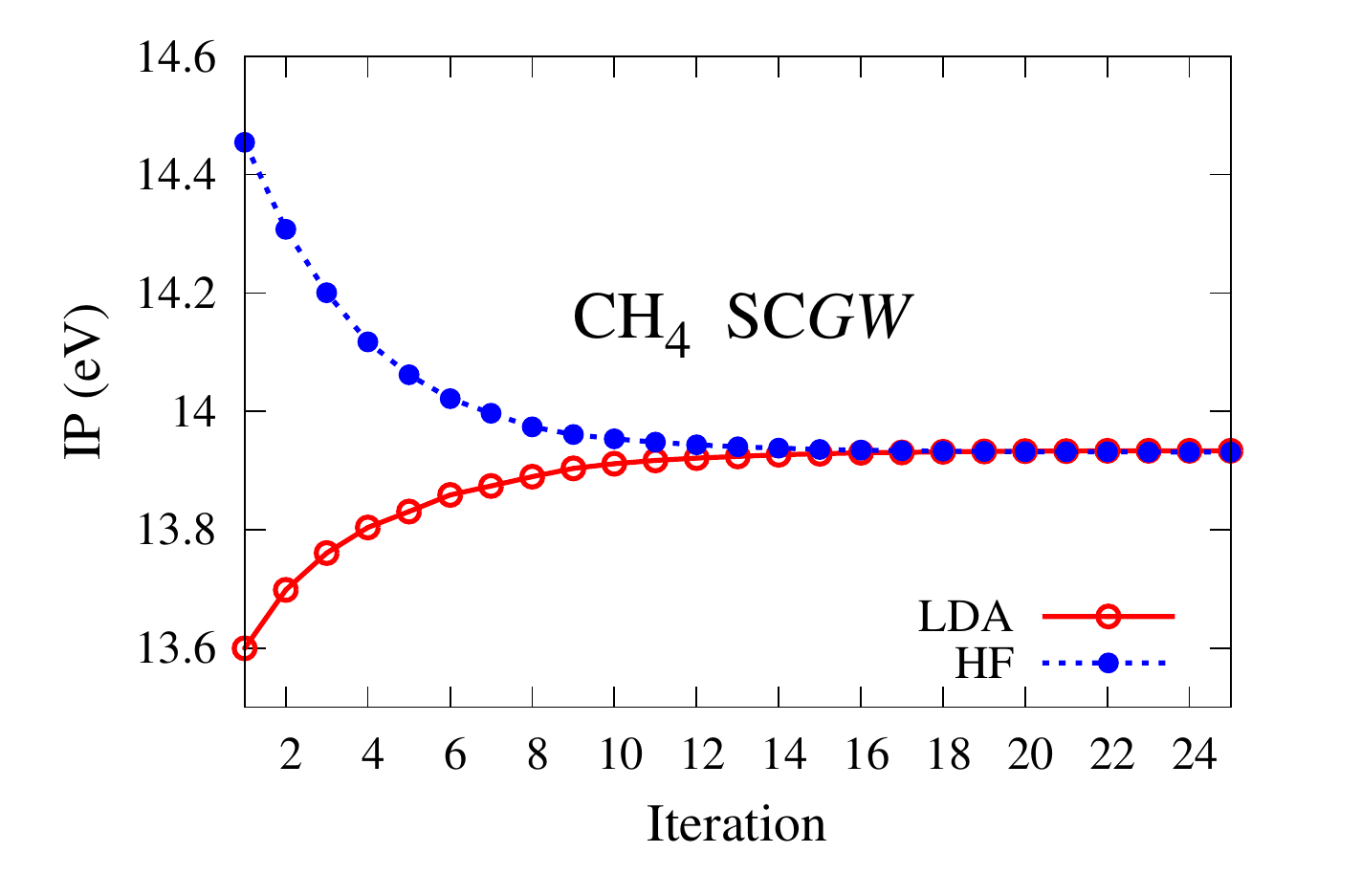} &
\includegraphics[height=4.2cm, angle=0,viewport=75 50 445 320,clip]{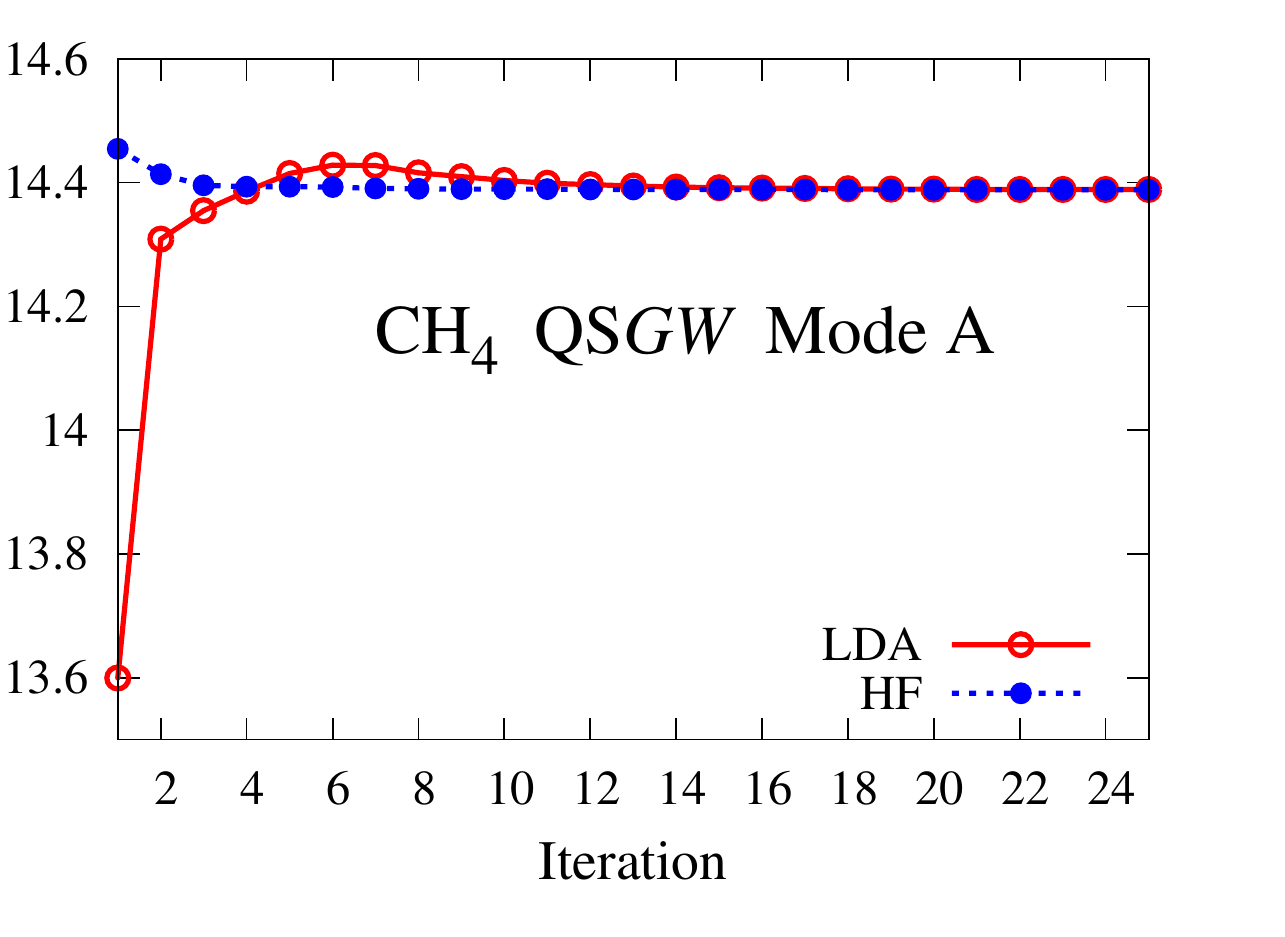} &
\includegraphics[height=4.2cm, angle=0,viewport=75 50 445 320,clip]{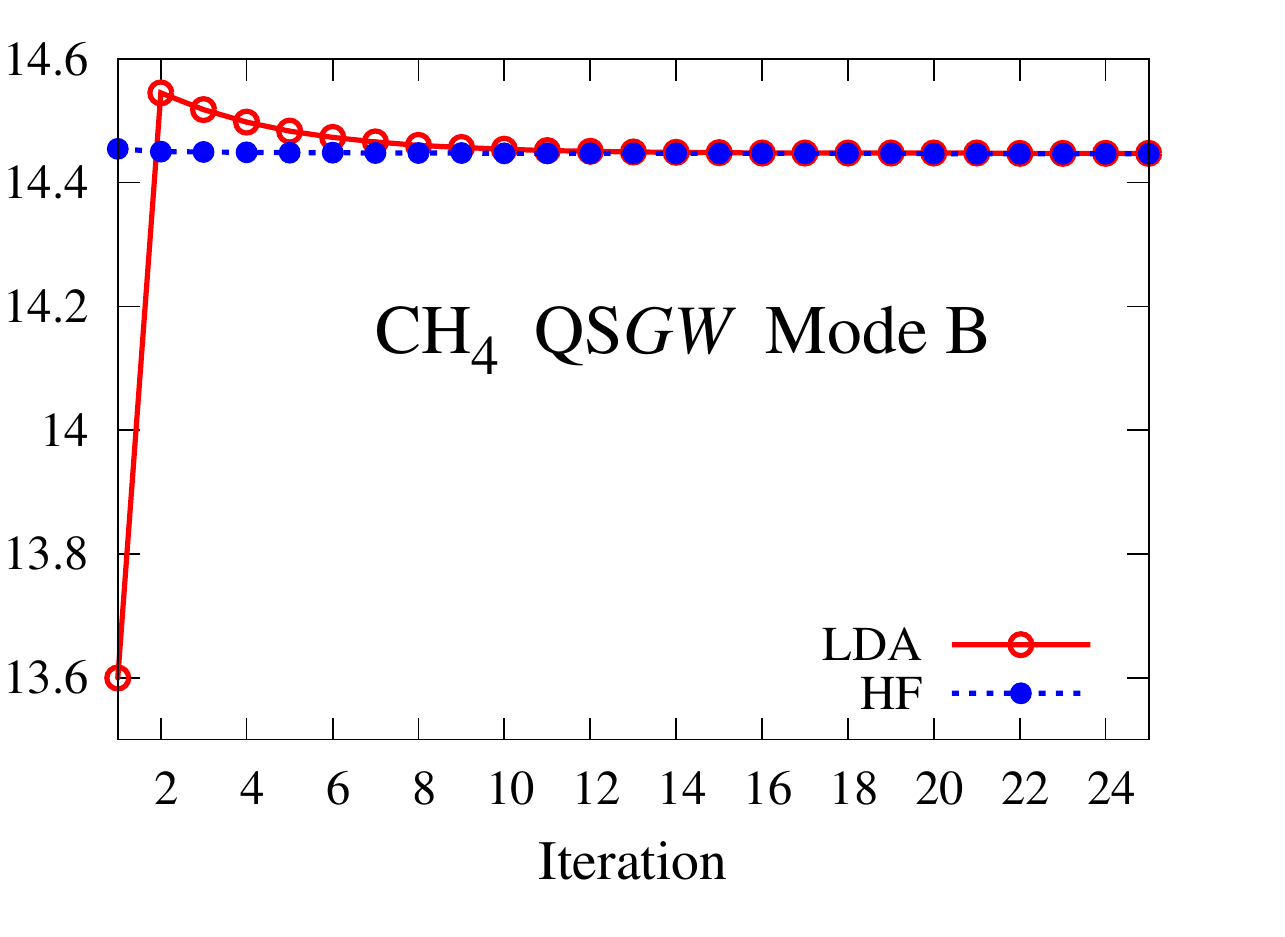} \\
\centerline{a)} & \centerline{b)} & \centerline{c)}\\
\end{tabular}
\caption{\label{f:independence} Evolution with the iteration number of
the ionization potential of methane CH$_4$ during SC$GW$ and QS$GW$ calculations.
A Hartree-Fock calculation and a density functional theory calculation using the local density approximation were used
as starting points. The first iteration corresponds to a $G_0W_0$ calculation.}
\end{figure*}

In both methods, SC$GW$ and QS$GW$, the Hartree potential $V_{\mathrm{H}}$,
as well as the exchange $\Sigma_{\mathrm{x}}$
and correlation $\Sigma_{\mathrm{c}}(\omega)$ components of the 
self energy are recomputed in every iteration.
Only the matrix elements of the 
kinetic energy $\hat{T}$ and the nuclear attraction $V_{\mathrm{ext}}$
are kept fixed. In such self-consistent loop, we expect that any reasonable starting
Green's function will converge to the same interacting Green's function,
but this expectation has to be confirmed by actual
calculations~\cite{CarusoRinkeRenSchefflerRubio:2012}.
Such a test also provides a measure of the achievable accuracy in the numerical procedure.
We present such test in Fig.~\ref{f:independence} for the methane molecule,
where the convergence of the IP is
accomplished using HF and the local density approximation (LDA)
to DFT as starting points. For these calculations we have used a  
frequency resolution $\Delta\omega=0.05$ eV and a broadening constant 
$\varepsilon=0.1$~eV
for both SC$GW$ and QS$GW$. This 
choice of frequency resolution and broadening constant will
be justified in section \ref{s:conv}. The frequency grid covers a range of
[$-$1228.8 eV, 1228.8 eV] for both starting points: HF and LDA, which 
is sufficient to obtain converged SCGW calculations.
The non-local compression was done with all possible pairs of molecular orbitals
(i.e. $E_{\text{thrs}}$ is chosen higher than maximal difference of eigenvalues)
and threshold for eigenvalues $\lambda_{\text{thrs}}$ is set to
$\lambda_{\text{thrs}}=10^{-5}$.

We can see that the convergence behavior of SC$GW$ is monotonic and, in this case,
almost symmetric with respect to the LDA/HF starting points.
After 25 iterations, both starting points converge to the same IP within 3~meV
for the SC$GW$ calculation, which is well within the used frequency resolution 
of 50 meV.

QS$GW$ converges rather fast at the beginning of the self-consistent loop,
but the convergence behavior is not monotonic in general. However, the ``mode B''
converges somewhat more reliably because a monotonic convergence sets in earlier
than for the ``mode A'', as shown in Fig.~\ref{f:independence} 
Moreover, QS$GW$ ``mode B'' can achieve a better and faster
convergence of the DOS (Eq.~\ref{conv}) than with ``mode A''. For instance, in the present case, 
we reached $\text{Conv} \sim 2\cdot 10^{-3}$ for ``mode A'' after 150 iterations
both with HF and LDA starting points, while for ``mode B'' we found
$\text{Conv}\sim 10^{-6}$ after 31 iterations for HF and 40 iterations
for LDA starting points. In both cases we used mixing parameter $\alpha=0.25$.
These indications of better convergence properties of ``mode B'' comparing 
to ``mode A'' will be further discussed below, in subsection \ref{ss:mult-conv},
in relation to the convergence with respect to the basis set size. 

The negligible starting point dependence of the IP seems to indicate that we are indeed reaching
the same self-consistent solution either starting from HF or LDA, both for SC$GW$ and QS$GW$
self-consistent schemes. This is further confirmed by the direct comparison of the iterated DOSs.
For all the cases examined we have found that LDA and HF starting points always arrive to indistinguishable
DOSs.

\section{Convergence studies}
\label{s:conv}

Here we discuss the dependence of our results on different technical parameters.
The set of convergence parameters is 
rather large. Namely, we should explore the convergence with respect
to the extension of the frequency grid $[\omega_{\min},\omega_{\max}]$,
the frequency resolution of the grid $\Delta \omega$,
the broadening constant $\varepsilon$ and 
the parameters defining the non-local compression ($E_{\text{thrs}}$, $\lambda_{\text{thrs}}$),
for the three self-consistent schemes 
SC$GW$, QS$GW$ ``mode A'' and QS$GW$ ``mode B''. We have chosen
to study these parameters for two systems: helium and methane with cc-pVDZ basis set.
A full range-covering convergence study is practically impossible with
such a large set of convergence parameters. However, it is possible to show
the convergence with respect to each parameter separately, keeping the other parameters fixed.
Additionally we explore the convergence with respect to the basis set size for two small 
systems, He and H$_2$, using basis sets up to cc-pV5Z basis. As we will see,  
this study will unveil the poor convergence properties of QS$GW$ ``mode A''
with respect to the size of the basis. 

Notice that in our previous publication,~\cite{df-pk-dsp:2011}
we proposed the use of two grids with different resolution: a 
finer grid covering the low energies of interest, and a coarser
grid with larger extension. However, here we do not use this so-called second
window technique. We prefer to converge the results with respect to a single
frequency grid and, thus, eliminate this additional source of uncertainties.

\subsection{Frequency grid extension}
\label{ss:fm-conv}

Here we consider the convergence with respect to frequency grid extension.
Analyzing the changes in the DOS as a function of the 
self-consistency iteration, we have clearly seen the  
appearance of satellite structures besides the main peaks. 
The satellites at the $G_0W_0$ level
can reach approximately twice $\Delta E$, where $\Delta E=|E_1-E_N|$ and
$E_1$ and $E_N$ are, respectively, the lowest and highest eigenvalues of the
starting point Hamiltonian.
The subsequent iterations in the SC$GW$ loop lead to the appearance
of even larger frequencies in the self energy and, consequently, in the DOS.
However, the higher-order satellites are weak and do not significantly contribute
to the numerical value of the ionization potential. We discuss the satellite 
structure of SC$GW$ in more detail in the Supplementary Material.~\cite{supplementary-material}
We take into account the above mentioned facts and parametrize
the range of the frequency grid as a function of $\Delta E$, defining a new parameter
$f_{\omega}$,
$[-f_{\omega} \Delta E,f_{\omega}\Delta E]$. 
The other parameters were chosen as following:
$\varepsilon=0.2$~eV, $\Delta \omega=0.1$~eV, $E_{\text{thrs}}=\Delta E$,
$\lambda_{\text{thrs}}=10^{-3}$; this choice will be justified later in
this section.

Table \ref{t:fm-conv} shows the IPs for several extensions of
the frequency grid for helium and methane.
\begin{table}[htb]
\begin{tabular}{|c|c|c|c||c|c|c|}
\hline
& \multicolumn{3}{c||}{Helium} & \multicolumn{3}{c|}{Methane} \\
Prefactor $f_{\omega}$ 
    & QS$GW$~A&QS$GW$~B& SC$GW$    &QS$GW$~A&QS$GW$~B& SC$GW$ \\
\hline
1.0 & 24.852  & 24.852 & 24.738    & 14.379 & 14.420 & 13.742 \\
1.5 & 23.689  & 23.683 & 23.685    & 14.379 & 14.420 & 13.736 \\
2.0 & 24.349  & 24.345 & 24.140    & 14.380 & 14.420 & 13.735 \\
2.5 & 24.350  & 24.346 & 24.120    & 14.380 & 14.420 & 13.735 \\
3.0 & 24.350  & 24.346 & 24.116    & 14.380 & 14.420 & 13.735 \\
\hline
\end{tabular}
\caption{Ionization potential of helium and methane as a function of
the frequency grid extension. One can see that results converge after 
$f_{\omega}=2.0$ both for SC$GW$ and QS$GW$ self-consistency schemes.
The values of $\Delta E$ using a cc-pVDZ basis for He and CH$_{4}$
are, respectively, 93.6 and 381.5~eV. \label{t:fm-conv}}
\end{table}
The inspection of the data shows that results converge for 
large enough grid extensions. Incidentally, the convergence is
much faster for CH$_4$ than for He.  According to these data, 
$f_{\omega}=2$ seems to set the smallest frequency grid extension  
after which the results become reliable.
In the rest of the calculations presented here, we will use 
$f_{\omega}=2.5$ to ensure
a good convergence of the obtained IP (now within  a few meV).

\subsection{Frequency grid resolution}
\label{ss:fr-conv}

\begin{figure}[htb]
\begin{tabular}{p{7cm}p{7cm}}
\includegraphics[height=5.5cm, angle=0,viewport=60 30 405 305,clip]{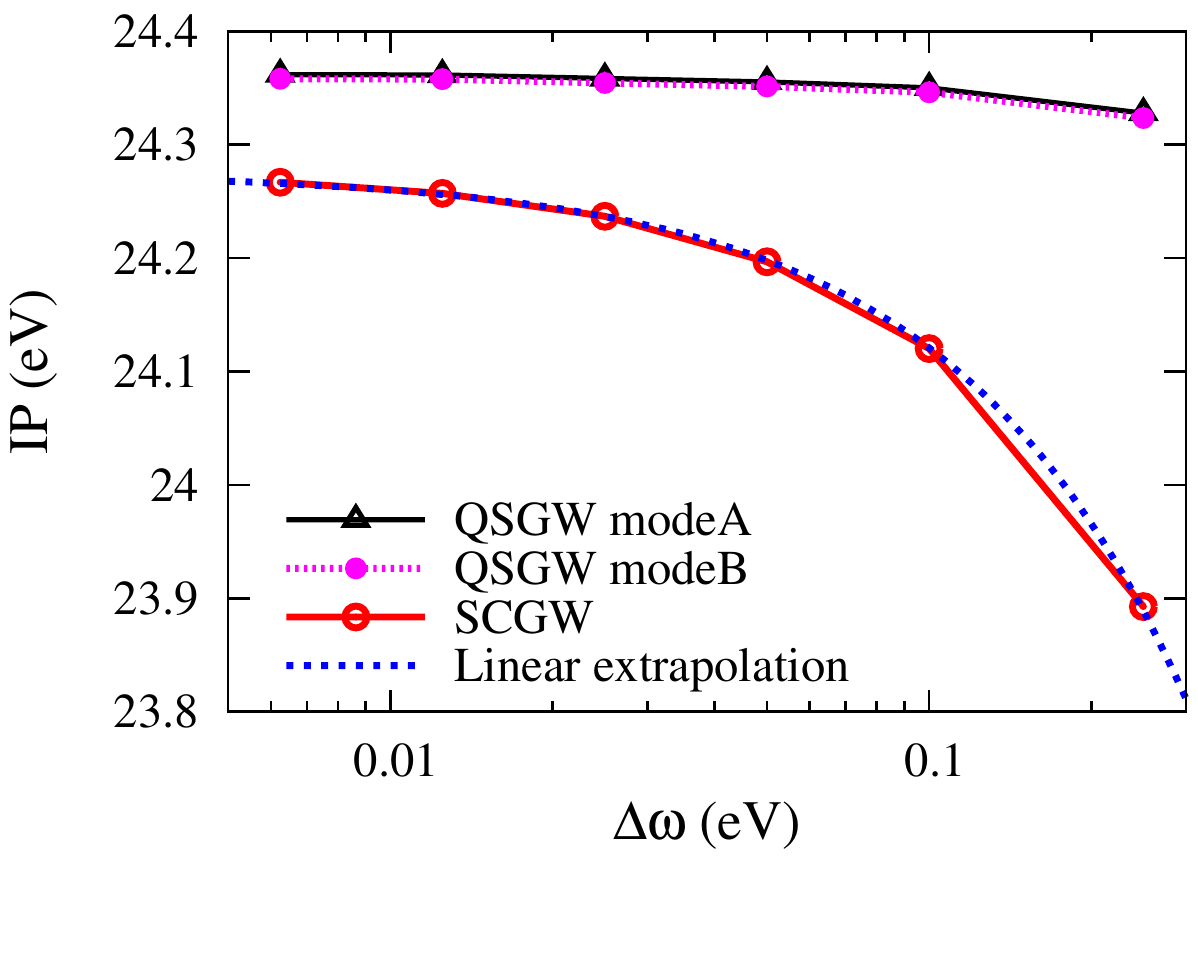} &
\includegraphics[height=5.5cm, angle=0,viewport=80 30 405 305,clip]{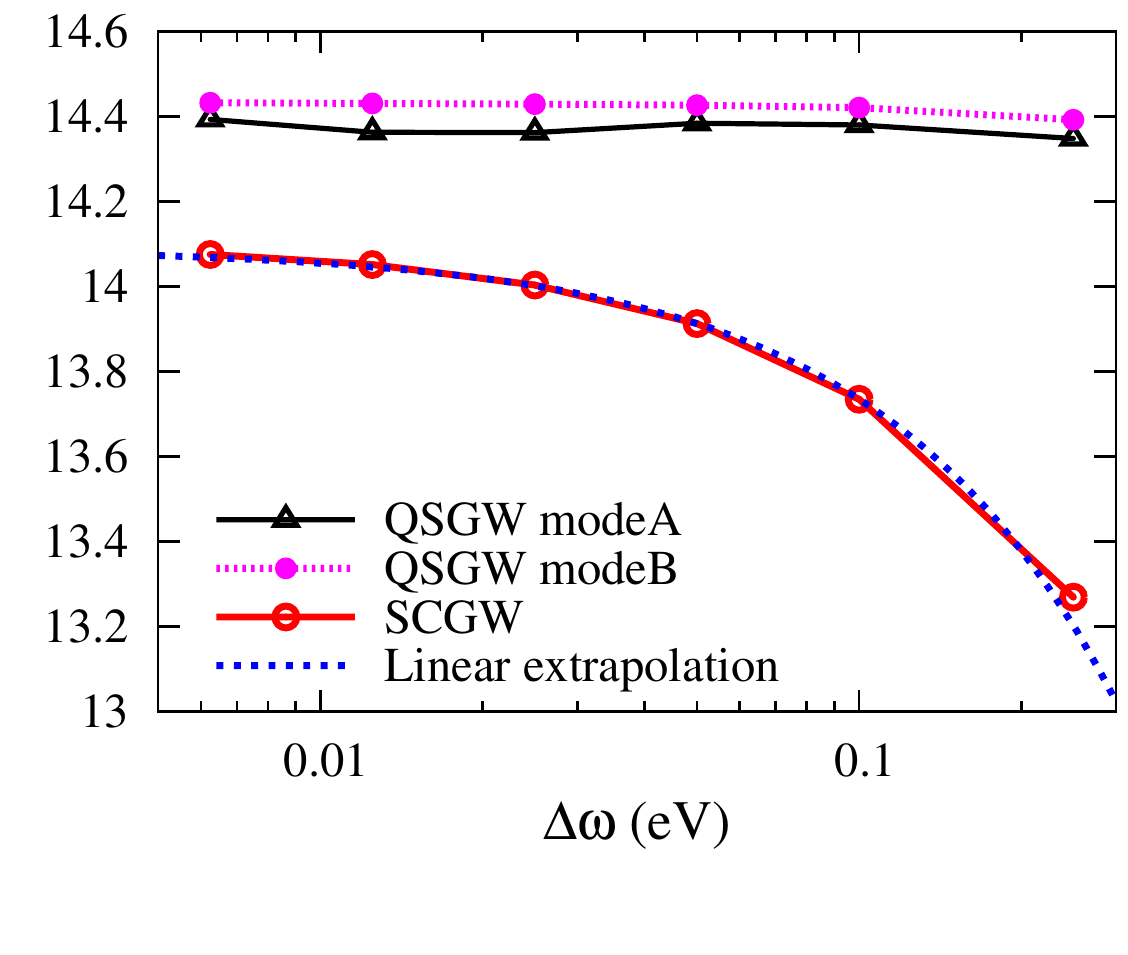} \\
\centerline{a)} & \centerline{b)}
\end{tabular}
\caption{\label{f:fr-conv}
Ionization potential of helium (panel a) and methane (panel b) as
functions of the frequency grid resolution.
The IPs are essentially independent on frequency resolution in the  QS$GW$ procedures.
The SC$GW$ procedure shows an almost linear dependence of the IP on $\Delta\omega$.
The linear extrapolation for SC$GW$ (dotted line)  is computed from two
IPs calculated using frequency resolutions $0.1$ and $0.05$~eV.}
\end{figure}

We turn now to the role of the frequency resolution. In this study, we fixed
the extension of the grid to $[-2.5\Delta E, 2.5\Delta E]$ as discussed above,
varied the frequency resolution $\Delta\omega$, and compared the calculated IPs.
The broadening constant is $\varepsilon=2\Delta\omega$. The parameters of non-local compression are
chosen as in the previous subsection. 
The results for helium and methane are presented in Fig.~\ref{f:fr-conv}.

Both QS$GW$ ``modes'' give results largely independent on the frequency resolution
$\Delta \omega$.
This is a welcome feature because a relatively coarse frequency grid can be used
with QS$GW$. It is interesting to note that a similar behavior is generally found
for one-shot $G_0W_0$ calculations. 
In contrast, the SC$GW$ procedure exhibits a stronger dependence
on the frequency resolution. We observe an almost linear dependence 
of the calculated IP on $\Delta\omega$. This (less welcome) feature
has its roots in the computation of the density matrix from the Green's
function (Eq.~\ref{g2n}). The spectral function treatment using a coarse grid
results in rather broad resonances of Lorentzian shape, and their width
deteriorates the quality of the density matrix. This convergence behavior
can be seen already in a self-consistent loop without any correlation
self energy $\Sigma_{\text{c}}(\omega)$, i.~e. performing the Hartree-Fock
calculation with Green's functions. Regarding this point it is interesting 
to note that, although the deviations
of the electron number are usually rather small in 
the present $GW$ calculations, typically not larger than 1\%, we renormalize
the density matrix to right number of electrons 
after each iteration to avoid uncontrolled variations
of the Hartree potential.
Notice that this consequence of the spectral function representation
does not affect the QS$GW$ calculations, because the density matrix
in QS$GW$ is obtained directly from the eigenvectors of the
QS$GW$ effective Hamiltonian 
$\hat{H}_{\mathrm{eff}}$.

The approximate linear dependence of the SC$GW$ IP (Fig.~\ref{f:fr-conv})
for small values of $\Delta \omega$ is seen in all the examples we have considered.
For most atoms and molecules the calculated IP increases as $\Delta\omega$
decreases, with the sole exception of LiF that shows the opposite behavior.
Therefore, we will estimate the results in the limit of infinite resolution
($\Delta \omega \rightarrow 0$) from two calculations with different 
frequency resolutions.
The SC$GW$ results presented in subsection \ref{ss:numerical-results}
have been obtained using this linear extrapolation to infinite resolution.

\subsection{Broadening constant}
\label{ss:eps-conv}

The choice of broadening constant $\varepsilon$ in our calculations
with equidistant frequency grid is rather intuitive.
If the broadening constant is smaller than frequency resolution $\Delta\omega$,
then a resonance may ``squeeze'' unnoticed between two neighboring frequency
points and become missed. Therefore, the broadening constant $\varepsilon$ must
be necessarily larger than the frequency spacing $\Delta\omega$.

In this work, we will parametrize the broadening constant as
$\varepsilon=f_{\varepsilon}\Delta\omega$, where $f_{\varepsilon}>1$ is a new parameter.
We are interested to keep the number of frequencies in the grid as small as possible
to minimize the computational cost connected to the size of the frequency grid.
Here the frequency grid extension is set using $f_{\omega}=2.5$.
The frequency resolution is chosen to be $\Delta \omega=0.1$ eV for QS$GW$,
while for SC$GW$ the data presented correspond to a 
linear extrapolation of the IPs from the data computed 
for $\Delta \omega=0.1$ and $\Delta \omega=0.05$~eV as described
in subsection \ref{ss:fr-conv}. The parameters of non-local compression 
are chosen as in subsection \ref{ss:fm-conv}. In Table~\ref{t:eps-conv}
we show the IPs computed with different broadening constants 
$f_{\varepsilon}\Delta\omega$.
One can see that the IPs change steadily with decreasing of parameter $f_{\varepsilon}$
from $3.0$ to $2.0$ in all calculations, while between $f_{\varepsilon}=2.0$ to 
$f_{\varepsilon}=1.0$ there is no clear trend. Moreover, the SC$GW$ calculation
for methane failed to converge to our target Conv accuracy 
with $f_{\varepsilon}=1.0$. Therefore, we regard $f_{\varepsilon}=2.0$ as an optimal
parametrization for broadening constant $\varepsilon$. 
\begin{table}[htb]
\begin{tabular}{|c|c|c|c||c|c|c|}
\hline
& \multicolumn{3}{c||}{Helium} & \multicolumn{3}{c|}{Methane} \\
Prefactor $f_{\varepsilon}$ 
    & QS$GW$~A & QS$GW$~B & SC$GW$ & QS$GW$~A & QS$GW$~B & SC$GW$ \\
\hline
1.0 & 24.370   & 24.366 & 24.274   & 14.385   & 14.431   & 14.093 \\ 
1.5 & 24.355   & 24.351 & 24.286   & 14.383   & 14.425   & 14.103 \\
2.0 & 24.350   & 24.346 & 24.273   & 14.380   & 14.420   & 14.090 \\
2.5 & 24.347   & 24.343 & 24.274   & 14.376   & 14.416   & 14.081 \\
3.0 & 24.344   & 24.340 & 24.279   & 14.372   & 14.413   & 14.073 \\
\hline
\end{tabular}
\caption{Ionization potential of helium and methane as function of
the broadening parameter $\varepsilon$=$f_{\varepsilon}\Delta\omega$. \label{t:eps-conv}}
\end{table}

\subsection{Non-local compression}
\label{ss:lbd-conv}

The choice of non-local compression parameters was studied 
in Ref.~\onlinecite{df-pk-dsp:2011} for pseudo-potential based, LDA-$G_0W_0$
calculations. In the present work, we found the behavior of non-local compression
to be similar to that found in our previous study. However, here we prefer 
not to limit the number of molecular orbitals by the energy criterium
$E_{\mathrm{thrs}}$ (see section \ref{ss:nl}). This decision does not significantly
contributes to the runtime of any of our examples, while it removes one 
technical parameter to converge our calculations with respect to. 
Table \ref{t:lbd-conv} shows the dependence 
of the IPs on the threshold eigenvalue $\lambda_{\text{thrs}}$ of the Coulomb metric.  
The other calculation parameters has been chosen as in the previous subsection.

\begin{table}[htb]
\begin{tabular}{|c|c|c|c||c|c|c|}
\hline
& \multicolumn{3}{c||}{Helium} & \multicolumn{3}{c|}{Methane} \\
$\lambda_{\text{thrs}}$ 
          & QS$GW$~A & QS$GW$~B & SC$GW$ & QS$GW$~A & QS$GW$~B & SC$GW$ \\
\hline
$0.1$     & 23.404   & 23.403   & 23.456 & 13.776   & 13.821 & 13.667 \\
$10^{-2}$ & 24.350   & 24.346   & 24.273 & 14.350   & 14.386 & 14.065 \\
$10^{-3}$ & 24.350   & 24.346   & 24.273 & 14.380   & 14.420 & 14.090 \\
$10^{-4}$ & 24.350   & 24.346   & 24.273 & 14.385   & 14.425 & 14.093 \\
$10^{-5}$ & 24.350   & 24.346   & 24.273 & 14.385   & 14.425 & 14.093 \\
\hline
\end{tabular}
\caption{Ionization potential of helium and methane as function of
non-local compression threshold $\lambda_{\text{thrs}}$. \label{t:lbd-conv}}
\end{table}
From the table one can see that a large threshold for the eigenvalues 
of the Coulomb metric 
$\lambda_{\text{thrs}}=0.1$ leads to sizable changes of the 
computed IPs.
However, the non-local compression becomes reliable with 
thresholds $\lambda_{\text{thrs}}<10^{-3}$. 
The values of the IP with $\lambda_{\text{thrs}}=10^{-3}$
and $\lambda_{\text{thrs}}=10^{-4}$ vary less than $6$ meV.
Because a stronger reduction of the number of products positively impacts
the computational performance, we have chosen $\lambda_{\text{thrs}}=10^{-3}$
for the main calculations in section \ref{s:results}.

\subsection{Size of the cc-pV$\zeta$Z basis sets and failure of QS$GW$ ``mode A'' to converge}
\label{ss:mult-conv}

The correlation consistent basis sets cc-pV$\zeta$Z are supposed to provide
increasingly better results in terms of the convergence to the complete basis
set (CBS) limit as the cardinal number $\zeta$ of the basis set  is increased. We intent to study
this convergence for SC$GW$ and QS$GW$ schemes. The computational cost of 
using high-$\zeta$ basis grows very steeply. Therefore, we are limited
in this test to small systems and, as already mentioned, for larger
molecules we restrict to cc-pVDZ and cc-pVTZ basis. 
The covergence test as a function of the size of the basis is important to determine
whether a meaningful comparison between SC$GW$ and QS$GW$ can be done using
those smaller basis set.  The results presented here seem to 
indicate that this is the case because, although the convergence of
the IPs is quite slow with the size of the basis set,
both $GW$ schemes show a rather similar convergence behavior.

We focus in the helium atom and the hydrogen dimer. The frequency grid extension is fixed
by $f_{\omega}=2.5$. The frequency resolution is $\Delta\omega=0.1$ eV for both QS$GW$ ``modes''.
For SC$GW$, we report linearly extrapolated IPs  from data calculated using
$\Delta\omega=0.1$ and $\Delta\omega=0.05$~eV, following our discussion
in subsection~\ref{ss:fr-conv}.
The broadening constant  is set to $\varepsilon=2\Delta\omega$, and 
the non-local compression is performed with $\lambda_{\text{thrs}}=10^{-3}$.
These choices are justified by the tests presented in the subsections 
\ref{ss:fm-conv}, \ref{ss:fr-conv}, \ref{ss:eps-conv} and \ref{ss:lbd-conv}.
The data for the IPs as a function of the basis size are collected in the Table~\ref{t:bsc-conv}.
We present results obtained with our code 
for ``mode A'' and``mode B'' of QS$GW$ (henceforth QS$GW$~A and QS$GW$~B), 
and SC$GW$.  Table~\ref{t:bsc-conv} also presents the data computed with the MOLGW code
developed by F.~Bruneval \cite{MOLGW} as well as our reference ionization energies
from the CCSD calculations with the NWChem code \cite{Valiev20101477}.
Notice that for systems containing two electrons CCSD and CCSD(T) are identical,
due to the absence of triple excitations, and become equivalent to full-CI.~\cite{Szabo-Ostlund:MQC}
MOLGW implements (among other methods) the QS$GW$~A algorithm.~\cite{Bruneval12}
It is important to stress here that the MOLGW code employs other algorithms than 
used in this work and its implementation is independent on our implementation.
Therefore, the close agreement (maximal deviation of 0.03~eV) between the 
QS$GW$ IPs computed with our code and MOLGW  is an important cross-check.

\begin{table}[htb]
\begin{tabular}{|c|c|c|c|c|c||c|c|c|c|c|}
\hline
& \multicolumn{5}{c||}{Helium} & \multicolumn{5}{c|}{Hydrogen dimer} \\
\hline
Basis set
        & QS$GW$~A 
                 & QS$GW$~A$^{\star}$ 
                          & QS$GW$~B 
                                   & SC$GW$ & CCSD & 
                                                      QS$GW$~A & QS$GW$~A$^{\star}$ 
                                                                        & QS$GW$~B 
                                                                                 & SC$GW$ & CCSD\\
\hline
cc-pVDZ & 24.350 & 24.359 & 24.346 & 24.273 & 24.326  & 16.148 & 16.141 & 16.232 & 16.000 & 16.257 \\
cc-pVTZ & 24.340 & 24.320 & 24.554 & 24.409 & 24.528  & 16.378 & 16.357 & 16.455 & 16.171 & 16.394 \\
cc-pVQZ & 24.751 & 24.766 & 24.668 & 24.490 & 24.564  & 16.569 & 16.562 & 16.526 & 16.216 & 16.422 \\
cc-pV5Z & 24.799 & 24.825 & 24.705 & 24.522 & 24.580  & 16.538 & 16.519 & 16.553 & 16.232 & 16.430 \\
\hline
CBS     &  -     &  -     & 24.744 & 24.555 & 24.597  &      - &  -     & 16.581 & 16.250 & 16.438 \\
\hline
\end{tabular}
\caption{Ionization potential of helium atom and hydrogen dimer as function of
basis set size for different methods. Columns marked with $\star$
indicate results obtained with the MOLGW code~\cite{Bruneval12} 
for QS$GW$~A. CBS stands for the complete basis set extrapolation (see the text).
\label{t:bsc-conv}}
\end{table}

In agreement with previous studies,~\cite{PhysRevB.84.205415,Bruneval12} 
the data in Table~\ref{t:bsc-conv} illustrate the very slow convergence of the $GW$ results
with the basis set size. 
A more noticeable and unexpected finding is the non-monotonous convergence
of the QS$GW$~A method for the two systems considered here. 
This is in clear contrast with the behavior observed for both SC$GW$ and QS$GW$~B
and, to the best of our knowledge, it had not been reported previously.
Notice that the same irregular behavior is produced by our code and by MOLGW. 
According to our analysis,  this poor convergence can be traced back to the 
combination of two issues, one inherent to the QS$GW$~A scheme, and the other
related to the use of atomic orbitals as a basis set. 
The difficulties arise from the fact that in QS$GW$~A the non-diagonal
elements (in the basis set of QP wavefunctions) of 
the correlation operator (Eq.~\ref{modeA}) contain 
contributions from the self energy evaluated at two different QP energies.
Therefore, e.g., the calculation of the 
HOMO
is influenced by the self energy calculated 
at all other energies, including energies corresponding to the highest molecular states.
In spite of the lack of justification for having this mixing of information evaluated
at different energies (other than defining an Hermitian operator in Eq.~\ref{modeA}), 
this should not necessarily cause difficulties for the convergence if those
self-energy cross-terms would be small or would have a smooth dependence
on frequency. Unfortunately this is not always the case. In particular, 
using a basis set of atomic orbitals (even a quite complete one), 
the self energy is very spiky even 
at high energies. This reflects the fact that the continuum of states,
that one should find above the vacuum level, is replaced by a discrete collection
of states. Therefore, when one of the eigenvalues of the
QS$GW$ QP equation lies in a region
where the self energy is large, this might have a large influence on 
the results at low energies through the self-energy cross-terms. 
In this situation, self-consistency might be difficult to achieve (due to changes
in the sign of the self-energy contribution during the self-consistent process), and even 
if self-consistency is reached the results do not show a steady trend with the 
basis set size (since increasing the basis set modifies strongly the 
structure of the self energy at high energies).

The bad convergence properties of QS$GW$~A in combination 
with basis set of atomic orbitals is a serious draw back for the applicability 
of this scheme in our case. Fortunately, this property is not shared by 
QS$GW$~B, that shows a slow but steady convergence with 
the basis set size for both He and H$_2$. 
The reason is that, in ``mode B'', all the non-diagonal components of the 
correlation operator are computed at the Fermi energy, and the difficulties
described above disappear. Therefore, in the rest of the paper we will concentrate
in the QS$GW$~B method. 

The steady convergence of the QS$GW$~B and SC$GW$ methods with respect to 
the basis set allows extrapolating to the CBS limit. This extrapolation is
performed using an inverse cubic function on the cardinal number $\zeta$
of the cc-pV$\zeta$Z basis, IP=IP$_{\text{CBS}}$ + A$\zeta ^{-3}$, with $\zeta=4$ and 
$5$. This formula is frequently used to extrapolate the 
correlation energy~\cite{CBS,CBS2} and we
have found that perfectly fits the dependence of our IPs calculated with $\zeta \geq 3$.
It is interesting to note  that our CBS-limit IPs using 
SC$GW$ 24.56 and 16.25~eV, respectively for He and H$_2$, are in excellent
agreement with the values, 24.56 and 16.22, given by Stan {\it et al.} using large bases
of Slater orbitals.~\cite{Stan06,Stan09} Interestingly, if we use our CCSD
results as a reference in the CBS limit, in the case of He we find that the SC$GW$ IP is much
closer to the reference value than 
the QS$GW$~B one, while for H$_2$ we have the opposite behavior and QS$GW$~B performs
somewhat better than SC$GW$. 

The slow convergence of the self-consistent $GW$ schemes with the basis set is certainly
an undesirable feature. The IPs calculated with a cc-pVTZ basis are still $0.1$--$0.2$~eV from the 
CBS limit. However, a very interesting feature is that the covergence
behavior is very similar for both methods, and the differences between 
the calculated IPs converges much faster with the basis set size. 
In particular, we observed that the IPs obtained with the 
QS$GW$ scheme are always higher than those obtained with SC$GW$.
For example, the IPs
calculated with QS$GW$ and SC$GW$ for He (H$_2$) using a TZ basis 
differ by 0.15 (0.29) eV, while the CBS-limit difference is 0.19 (0.33) eV.
So, at least for these two systems, the qualitative differences between QS$GW$
and SC$GW$
IPs obtained with a cc-pVTZ basis seem to be maintained all the way to 
the CBS limit.

Table~\ref{t:bsc-conv} also shows that CCSD results converge somewhat faster with
the basis set than the $GW$ ones. The IPs of He and H$_2$ calculated with a cc-pVTZ basis
are within 0.07~eV of our CBS limits. This different rate of convergence 
makes difficult the comparison of the performance of the self-consistent
$GW$ schemes against CCSD results using non-saturated basis sets.
Still for basis sets larger than DZ we see than the CCSD IPs always lie somewhere
in between the SC$GW$ lower bound and the QS$GW$ upper bound.  
One should keep in mind the different rate of convergence between the
$GW$ schemes and the CCSD when examining the results in Table~\ref{t:ip-ccsd_vs_gw}.
In particular, since the IPs tend to increase with the quality of the basis set, using 
basis sets which are not fully converged QS$GW$ could tend to  outperform SC$GW$.
However, as we will see below, we find the opposite trend and SC$GW$ is, on the average, 
marginally better than QS$GW$~B at the cc-pVTZ level. This is probably a robust
result which holds for larger basis sets. 

\section{Results}
\label{s:results}

The methods presented above allow realizing both SC$GW$ and QS$GW$ 
calculations within the same numerical framework. In subsection~\ref{ss:dos-examples} 
we present the densities of states (DOS) obtained with different $GW$ schemes.
The quantitative merit of the $GW$ methods is studied in subsection \ref{ss:numerical-results},
using the calculated IPs as a measure of such performance. 

\subsection{Densities of states for CH$_4$ and N$_2$}
\label{ss:dos-examples}

\begin{figure*}[htb]
\begin{tabular}{p{7cm}p{7cm}}
\includegraphics[width=7cm, angle=0,viewport=85 80 408 320,clip]{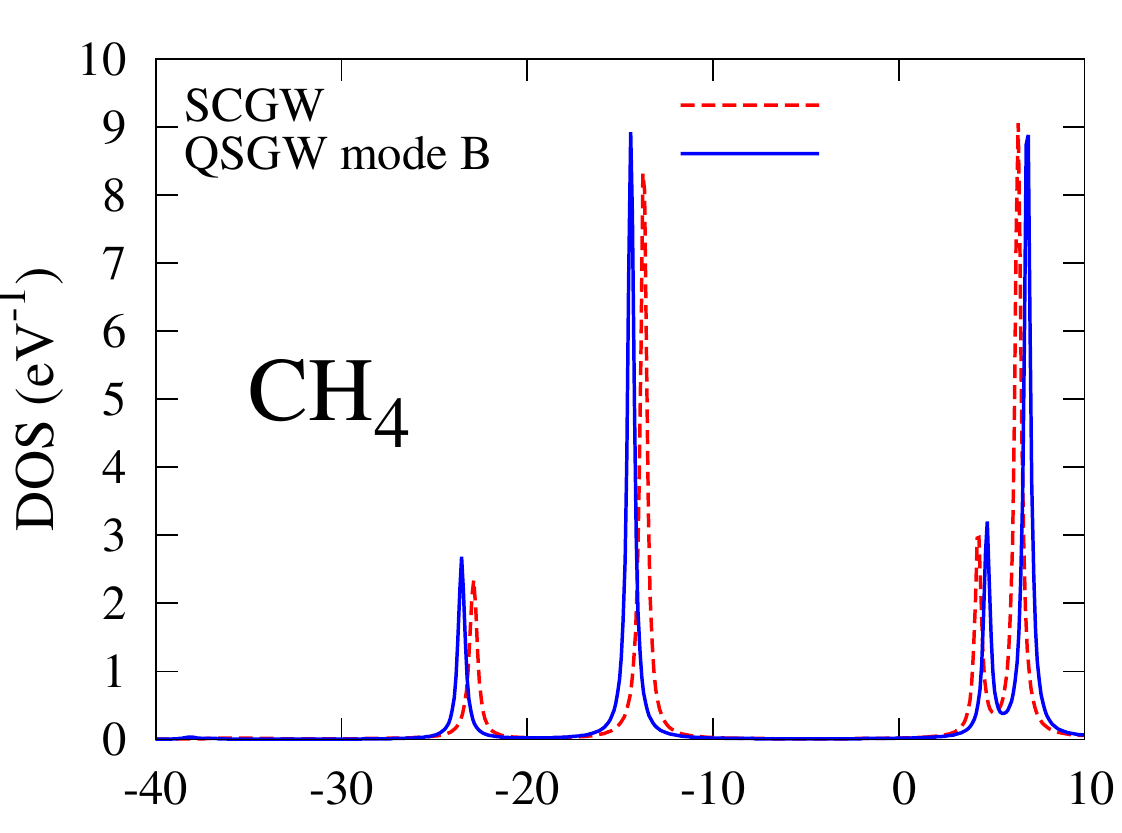} &
\includegraphics[width=7cm, angle=0,viewport=85 80 408 320,clip]{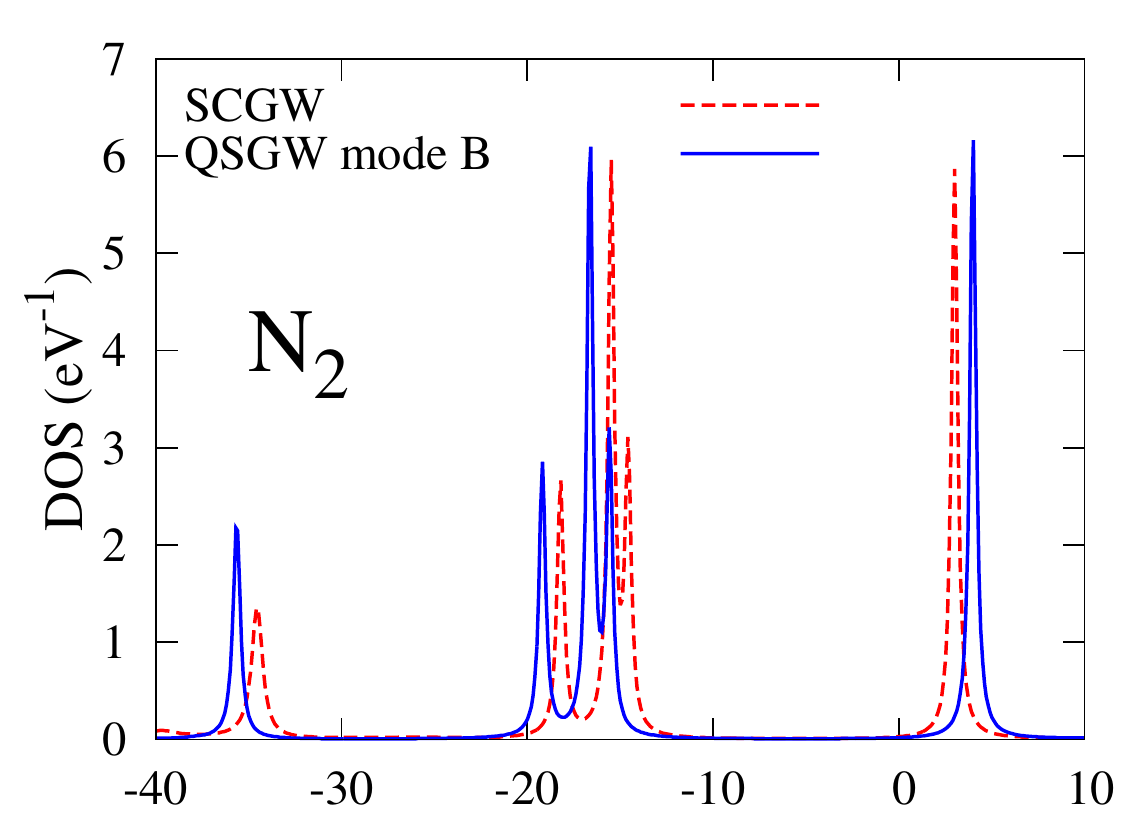} \\
\centerline{a)} & \centerline{b)} \\
\includegraphics[width=7cm, angle=0,viewport=85 80 408 311,clip]{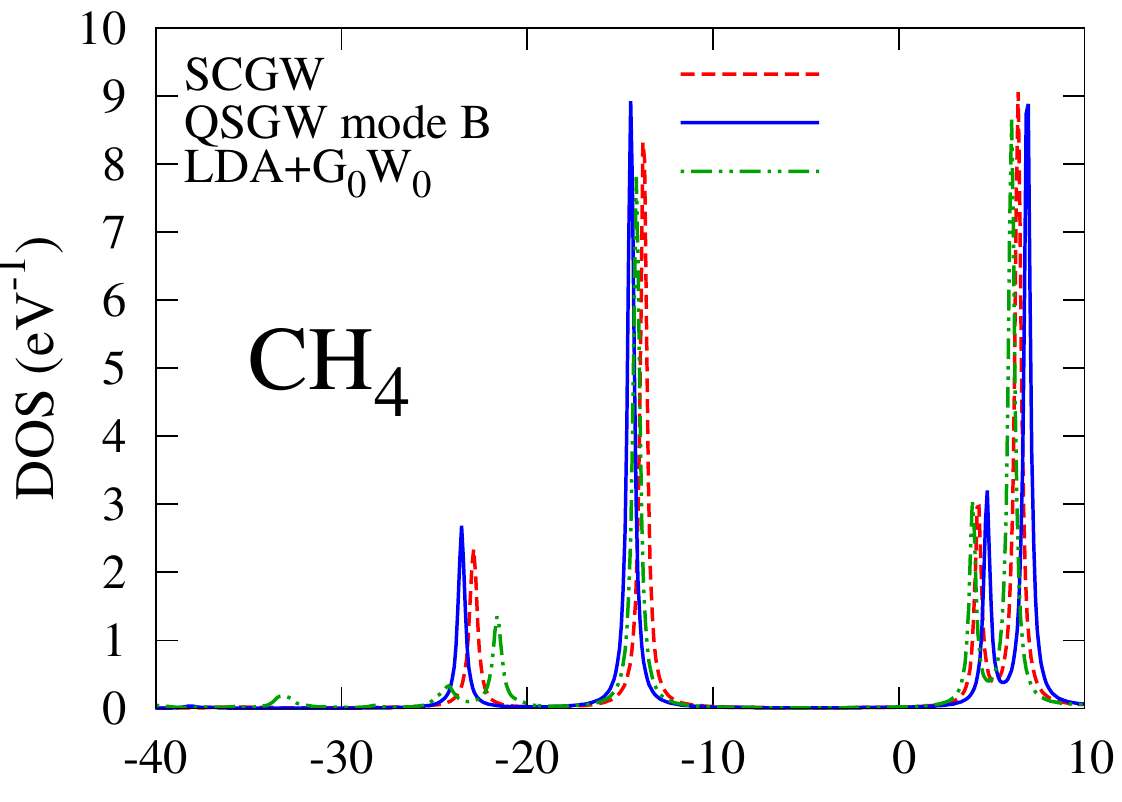} &
\includegraphics[width=7cm, angle=0,viewport=85 80 408 311,clip]{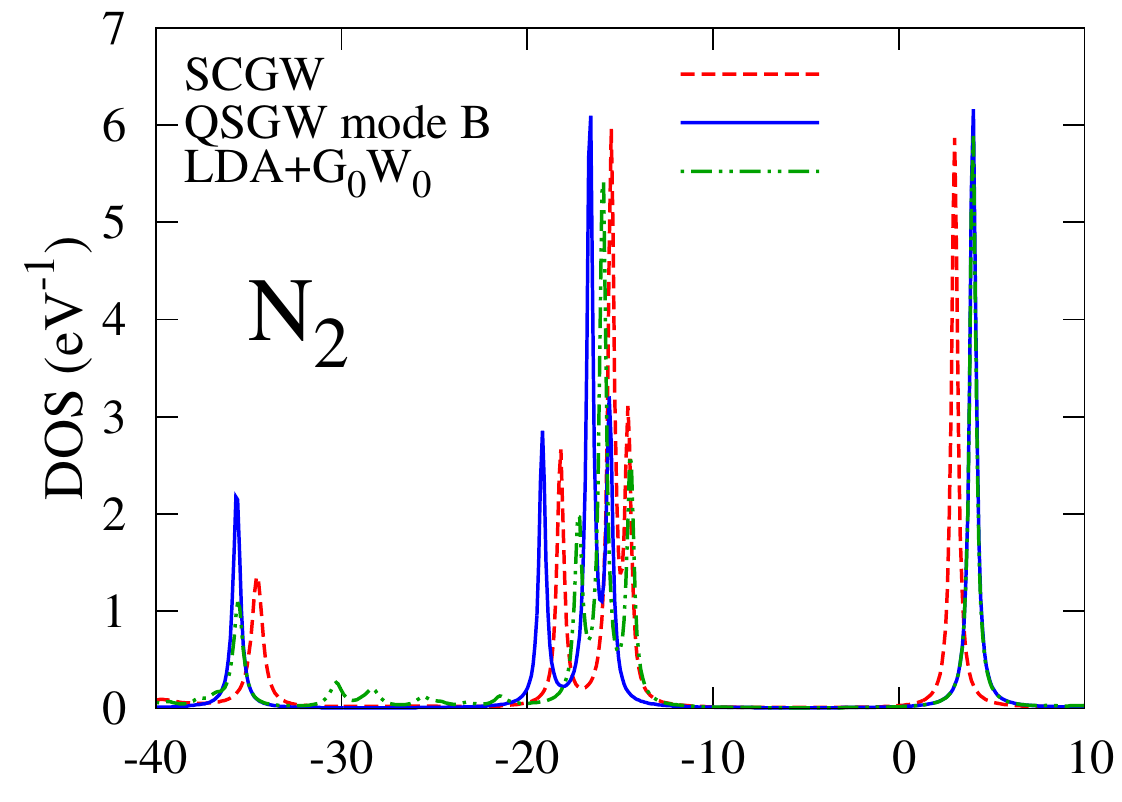} \\
\centerline{c)} & \centerline{d)} \\
\includegraphics[width=7cm, angle=0,viewport=85 50 408 311,clip]{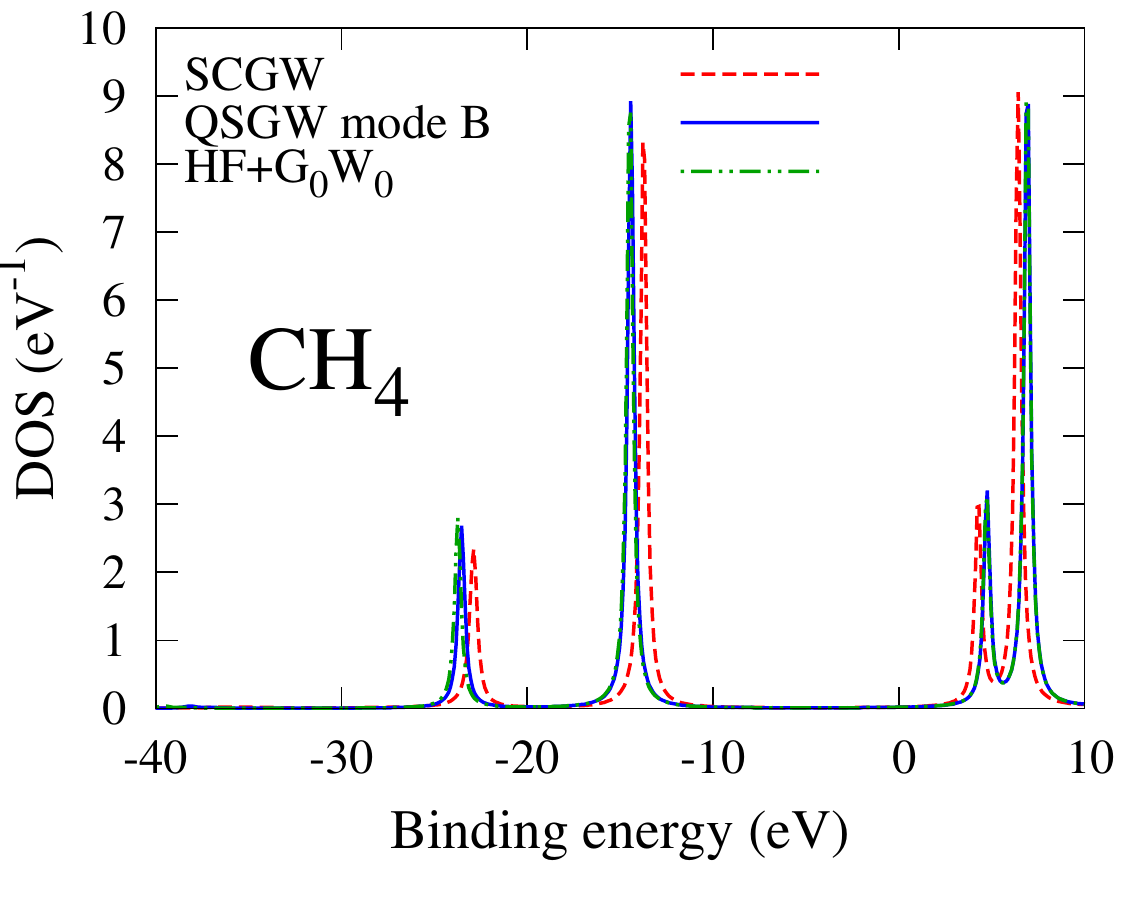} &
\includegraphics[width=7cm, angle=0,viewport=85 50 408 311,clip]{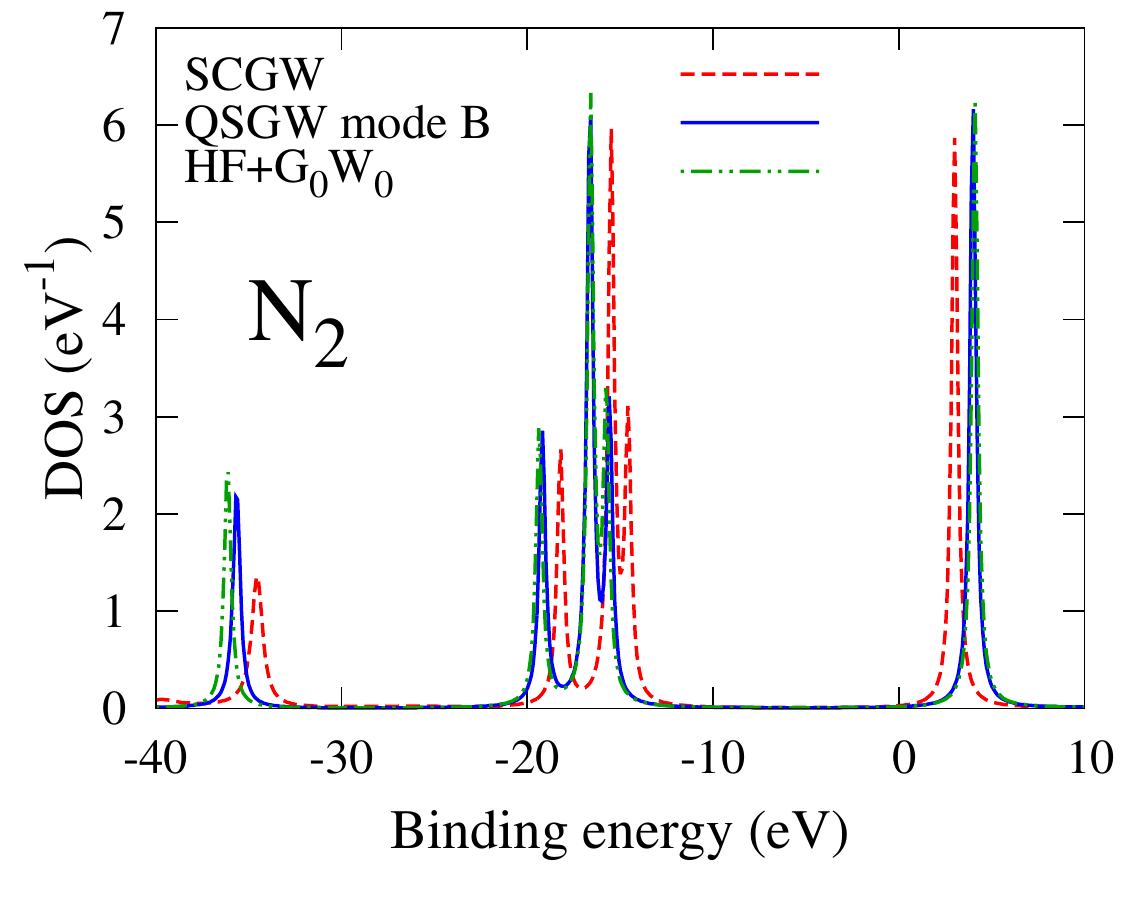} \\
\centerline{e)} & \centerline{f)} \\
\end{tabular}
\caption{\label{f:dos-methane-nitrogen-linear}
Densities of states of methane CH$_4$ [panels (a), (c) and (e)] and 
nitrogen N$_2$ [panels (b), (d), (f)] molecules
calculated using different approximations. 
The frequency resolution in these calculations is 
$\Delta \omega=0.1$ eV, and the broadening parameter $\varepsilon=0.2$ eV.}
\end{figure*}

Information about the effect of different self-consistent procedures can be obtained
from the DOS they provide. Figure~\ref{f:dos-methane-nitrogen-linear} compares the DOS 
of the methane molecule and the nitrogen dimer using different schemes.
Panels (a) and (b) demonstrate that SC$GW$ and QS$GW$~B behave quite similarly 
although the positions of the peaks are slightly shifted. 

Panels (c), (d), (e) and (f) illustrate the dependence of one-shot 
$G_0W_0$ on the starting point and its comparison with QS$GW$~And SC$GW$ B results.
The Hartree-Fock starting point ($G_0W_0$-HF) produces a DOS very close to that of
the self-consistent QS$GW$ solution (panels (e) and (f)).
In contrast, calculations using the Perdew-Zunger~\cite{PhysRevB.23.5048} 
local density exchange-correlation functional as a starting point ($G_0W_0$-LDA)
produce DOSs that depart more 
from those of both (SC$GW$ and QS$GW$) self-consistent
approaches. In particular, several satellite peaks can be seen in the frequency 
range below $-20$~eV for both, CH$_4$ and N$_2$.
Self-consistency tends to eliminate these features (see panels (c) and (d)).
However, weak satellite peaks also appear in both 
SC$GW$ and QS$GW$ approaches. For example, for methane we can find 
satellite peaks around $\pm$35~eV, although they are barely 
visible in Fig.~\ref{f:dos-methane-nitrogen-linear}.
To clearly visualize these structures it is necessary to plot the DOS
in logarithmic scale. This kind of analysis is presented in the Supplementary 
information.~\cite{supplementary-material} 

In agreement with previous observations,~\cite{PhysRevB.83.115103} we find that the
Hartree-Fock starting point in combination with the one-shot $G_0W_0$
approach tends to provide excellent estimations of one-electron
excitation energies in small molecules, see the example of methane 
in Fig.~\ref{f:dos-methane-nitrogen-linear}~(e) 
and Table~\ref{t:ip-ccsd_vs_gw}. For this reason we use HF as a starting point
in our calculations of ionization potentials in the next subsection.

\subsection{Ionization potential of atoms and small molecules}
\label{ss:numerical-results}

In order to assess the quality of the self-consistent $GW$ method for
atoms and small molecules at a quantitative level, we compare the performance
of SC$GW$ and QS$GW$ ``mode B'' with that of quantum chemistry methods, 
in particular with coupled-cluster (CC) calculations.
Here we focus in the first vertical IP. 
Although we further compare our results against experimental
data, a reliable study would require considering effects due to structural relaxations
in the final state and corrections related to the finite nuclear masses
for light elements, among others.
These effects are not taken into account in the present $GW$ calculations. 
Moreover, a comparison with other well-established 
theoretical methods using the {\it same} basis set also eliminates,
at least partially, the ambiguities related to the use
of a finite, necessarily incomplete, basis set of atomic orbitals 
(see the comments Sec.~\ref{ss:mult-conv}). 
This is an important point since, due to the use of all-electron 
calculations in the self-consistent $GW$ calculations (therefore
requiring the evaluation of the self energy in a very extended
frequency grid), even with the small molecules
considered here, we are limited to relatively modest basis sets
that might not provide fully converged results. 

We have chosen the coupled-cluster method with single, double and perturbative triple
excitations (CCSD(T)) as a reference theory to compare our $GW$ results with. 
This choice is motivated by the usefulness of CCSD(T) in many other applications
requiring to estimate the contribution of electron correlations in quantum
chemical calculations.~\cite{Rezac-Hobza:2013-ccsdt-gold}
We performed our CC calculations using the open-source NWChem package,~\cite{Valiev20101477}
and two different Gaussian basis sets~\cite{JCC:JCC9,doi:10.1021/ci600510j}
that we also adopted in our $GW$ calculations for consistency. 
We have used both, correlation-consistent double-$\zeta$ polarized (cc-pVDZ),
and triple-$\zeta$ polarized (cc-pVTZ) basis sets for all of our calculations. Comparison
of these two sets of results  
provides a rough estimation of the effect of the basis set incompleteness.
A systematic study of the convergence with respect to the basis set
size was presented in Sec.~\ref{ss:mult-conv} for two small 
systems, He and H$_2$. The basic conclusions obtained from these two systems are:
{\it i}) The convergence of the $GW$ results is rather slow; {\it ii}) Fortunately
the convergence of SC$GW$ B and QS$GW$ is very similar and differences
between IPs calculated with these two schemes are converged within 
0.05~eV already for cc-pVTZ basis sets; {\it iii}) The convergence
of CCSD(T) is somewhat faster than that of $GW$, which should be
taken into account when analyzing the data presented here. 

The molecular geometries were optimized at the level of CCSD(T) using 
the cc-pVTZ basis set \cite{supplementary-material}. 
These geometries were later used in all the other calculations, including
the self-consistent $GW$. In addition to the CCSD(T) calculations,
we have also performed calculations without perturbative triples (CCSD)
with the cc-pVTZ basis as a way to estimate the convergence of 
the description of correlations as provided by  
CCSD(T). Due to the use of relatively small basis sets in our calculations,
we limit our study to the IPs. An accurate calculation of electron affinities would 
require more complete augmented basis sets.

At the level of CC calculations, the vertical IPs were obtained from 
$\Delta$SCF-CC calculations, i.e., the IP is taken as the difference between
the total energy calculated for the neutral molecule and a singly-charged 
positive ion keeping the ground-state CCSD(T)/cc-pVTZ geometry. 
For the positive ions, unrestricted Hartree-Fock was used to produce
the starting point for the CC calculations.~\cite{DSCF-HF} Our calculations 
compare well with the literature. We checked our CCSD(T)/cc-pVTZ 
calculations against the data from NIST database CCCBDB.~\cite{NIST}
Ionization potential of atoms is the same as provided by NIST. Unfortunately,
there are only adiabatic IPs available from NIST for the small molecules we consider.
However, we compared the total energies of neutral molecules with the
corresponding NIST values and found a good agreement within a few meV.
Moreover, our ionization energies of N$_2$, CO, F$_2$
C$_2$H$_2$ and H$_2$CO agree well with some recent quantum chemical
calculations.~\cite{CCSDT-N2-CO-F2:1999,EOM-N2-CO-F2:2003,EOM-C2H2-H2CO-Musia2004210:2004,
EOM-CO-C2H2:2009}

In the $GW$ calculations, the IPs were obtained from the position of
the first peak below Fermi level in the DOS of each molecule.
The frequency grid resolution $\Delta\omega$ used with
the QS$GW$ approach was 0.05 eV for cc-pVDZ and 0.1~eV for cc-pVTZ basis sets.
In the case of SC$GW$, a linear extrapolation to the limit of infinite frequency
resolution was applied as discussed in subsection \ref{ss:fr-conv}.
Therefore, $\Delta\omega=0.05$ and $0.025$ eV were used in the calculations with
cc-pVDZ basis set, and $\Delta\omega=0.1$ and $0.05$ eV for those using a
cc-pVTZ basis set.

The convergence with the number of dominant products, used here to
express the products of basis functions, was monitored
comparing the energies of the HOMO
of the different molecules calculated at the Hartree-Fock level
with our code and with NWChem. Our code uses the basis of dominant products to compute
Hartree and exchange contributions to the energy and Hamiltonian.
We found maximal differences of at most 6 meV (for nitrogen containing molecules),
while the mean absolute error (MAE) of the HF-HOMO position is only 1.6 meV
for our set of sixteen atoms and molecules.

\begin{table*}[htb]
\begin{tabular}{|r||r|r||r|r||r|r||r|r||r||r|}
\hline 
& \multicolumn{10}{c|}{IP (eV)} \\
\hline
Method  
          &\multicolumn{2}{c||}{ $G_0W_0$-HF} 
                          &\multicolumn{2}{c||}{ SC$GW$}  
                                          & \multicolumn{2}{c||}{ QS$GW$~B} 
                                                             &\multicolumn{2}{c||}{ CCSD(T)} 
                                                                               & CCSD   & Exp.\\
Basis    &cc-pVDZ&cc-pVTZ &cc-pVDZ&cc-pVTZ&cc-pVDZ&cc-pVTZ   & cc-pVDZ&cc-pVTZ& cc-pVTZ & \\
\hline
He        &24.36 & 24.57  & 24.28 & 24.41 & 24.35 & 24.55    & 24.33  & 24.53  & 24.53 & 24.59\\
Be        & 8.98 &  9.05  &  8.46 &  8.53 &  8.95 &  9.03    &  9.29  &  9.29  &  9.28 &  9.32\\
Ne        &20.87 & 21.40  & 20.98 & 21.38 & 21.00 & 21.50    & 20.89  & 21.31  & 21.26 & 21.56 \\
H$_2$     &16.23 & 16.46  & 16.00 & 16.17 & 16.24 & 16.45    & 16.26  & 16.39  & 16.39 & 15.43$^*$\\
CH$_4$    &14.43 & 14.74  & 14.09 & 14.26 & 14.43 & 14.65    & 14.21  & 14.38  & 14.34 & 13.60  \\
H$_2$CO   &10.74 & 11.25  & 10.44 & 10.78 & 10.84 & 11.24    & 10.46  & 10.82  & 10.76 & 10.89 \\
C$_2$H$_2$&11.23 & 11.54  & 10.67 & 10.85 & 11.21 & 11.43    & 11.22  & 11.42  & 11.26 & 11.49 \\
HCN       &13.48 & 13.81  & 12.89 & 13.08 & 13.48 & 13.73    & 13.48  & 13.70  & 13.55 & 13.61 \\
CO        &14.39 & 14.74  & 13.53 & 13.81 & 14.03 & 14.34    & 13.62  & 13.93  & 13.93 & 14.01 \\
N$_2$     &15.84 & 16.30  & 15.05 & 15.38 & 15.57 & 15.95    & 15.10  & 15.46  & 15.59 & 15.58 \\
Li$_2$    & 5.23 &  5.34  &  4.88 &  4.98 &  5.28 &  5.35    &  5.19  &  5.23  &  5.22 &  5.11$^*$\\
LiH       & 7.96 &  8.15  &  7.74 &  7.84 &  7.97 &  8.15    &  7.85  &  7.98  &  7.98 &  7.90$^*$ \\
LiF       &10.72 & 11.32  & 10.85 & 11.13 & 11.27 & 11.77    & 10.90  & 11.34  & 11.24 & 11.30$^*$\\
HF        &15.55 & 16.17  & 15.54 & 16.05 & 15.89 & 16.43    & 15.44  & 15.97  & 15.90 & 16.12 \\
F$_2$     &15.93 & 16.30  & 15.46 & 15.74 & 16.06 & 16.36    & 15.38  & 15.69  & 15.91 & 15.70 \\
H$_2$O    &12.17 & 12.80  & 12.03 & 12.52 & 12.34 & 12.88    & 11.96  & 12.50  & 12.42 & 12.62$^*$ \\
\hline
MAE       & 0.22 & 0.28   & 0.21  & 0.22  & 0.25  & 0.27     & 0.00   & 0.00   & 0.069 & 0.19 \\ 
\hline
\end{tabular}







\caption{\label{t:ip-ccsd_vs_gw}
Vertical ionization potentials (in electon-volts) of the sixteen 
species studied in this work calculated using different
computational methods (see the text for more details) and two different basis sets,
a correlation-consistent double-$\zeta$ polarized (cc-pVDZ) and 
a triple-$\zeta$ polarized (cc-pVTZ) basis.~\cite{JCC:JCC9,doi:10.1021/ci600510j}  
The mean absolute error (MAE) is calculated with respect to the 
CCSD(T) results for each basis separately.  
Experimental data are taken from the NIST
CCCBDB database.~\cite{NIST}
The numbers marked with asterisks $^{*}$ are experimental adiabatic IPs.
}
\end{table*}

\begin{figure}[htb]
\begin{tabular}{p{8cm}p{8cm}}
\includegraphics[width=7.8cm, angle=0,viewport=75 75 410 310,clip]{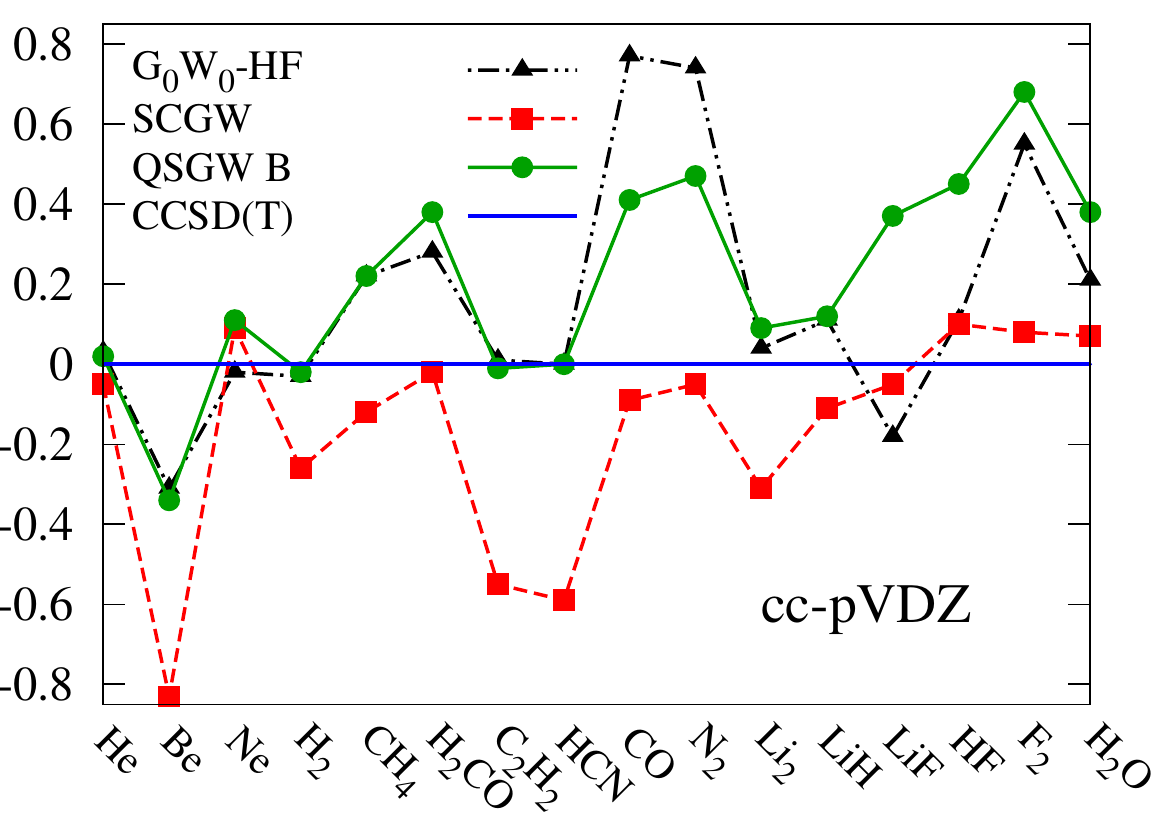} &
\includegraphics[width=7.8cm, angle=0,viewport=75 75 410 310,clip]{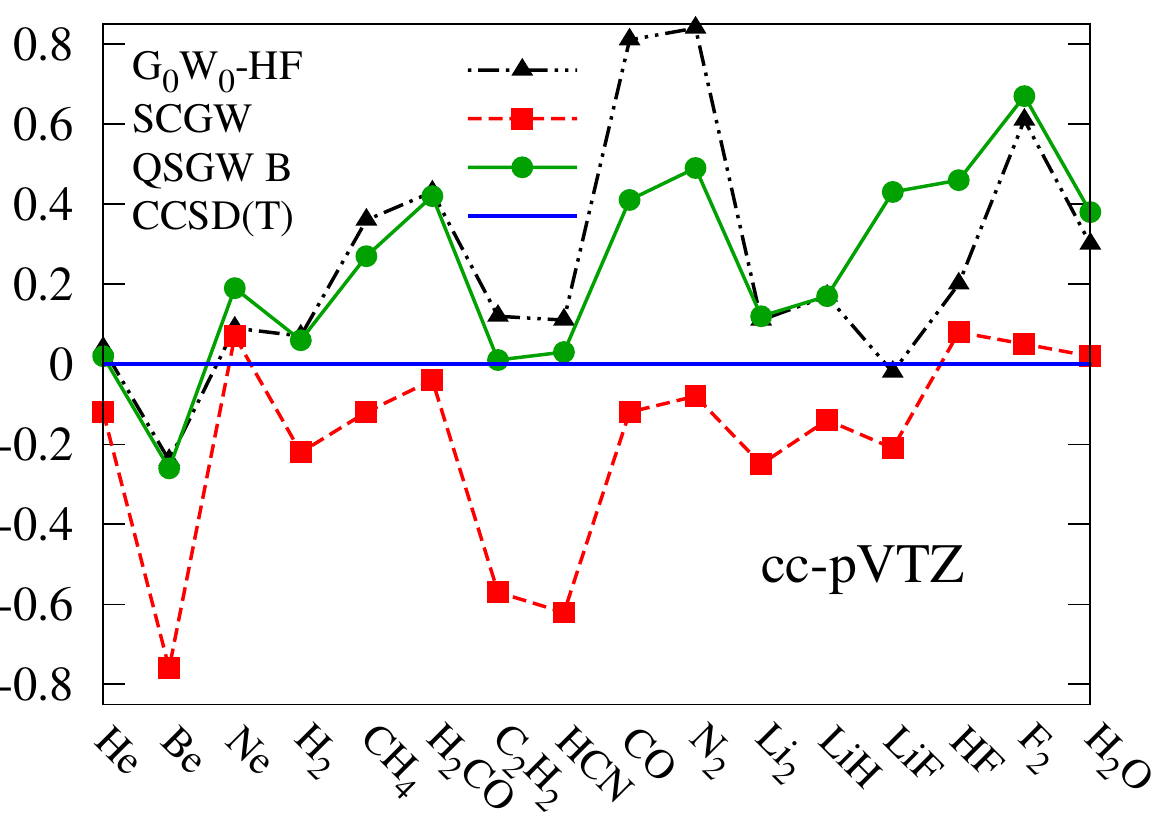} 
\end{tabular}
\caption{\label{f:DeltaIP}
Differences between vertical IPs calculated at the $G_0W_0$-HF, SC$GW$ and QS$GW$ levels,
and those obtained from coupled-cluster calculations. 
Panels (a) and (b) show calculations performed using cc-pVDZ and 
cc-pVTZ basis sets respctively. The data for the IPs can be found in Table~\ref{t:ip-ccsd_vs_gw}. 
}
\end{figure}

The results for the IPs of all the studied systems
are presented in Table~\ref{t:ip-ccsd_vs_gw}.
Before analyzing  the $GW$ results, it will be instructive to make
some comments about our CC reference calculations. 
Comparison between CCSD(T) and CCSD results (both using the cc-pVTZ basis)
indicates that the inclusion of triple excitations does not substantially
modify the calculated IPs on the average: 69~meV MAE and a maximal difference
of 0.22~eV for the F$_2$ molecule.
These differences are significantly smaller than those obtained when
comparing the CCSD(T) results with those of the different $GW$ methods.
This confirms that, at least for the systems considered here, CCSD(T) is 
a reasonable choice as a reference theory.

The convergence of the results with respect to the basis set is slower as
we could anticipate from our systematic study for He and H$_2$. 
Comparing CCSD(T) results calculated with cc-pVDZ and cc-pVTZ bases,
we find a  MAE of 0.27~eV and a maximal difference of 0.54~eV for the IP
of the water molecule. These larger variations are a clear
indication of the rather slow convergence of correlation effects
with respect to the basis size. The present results also confirm
the observation, made in Sec.~\ref{ss:mult-conv} for He and H$_2$, that the IPs
increase with the use of the more complete basis, 
with the exception of beryllium atom whose IP is unchanged when 
moving from a cc-pVDZ basis to a cc-pVTZ basis. 

The observed dependence of the IP on the basis set size
also agrees with the results of two recent convergence studies
of $G_0W_0$-HF IPs for light atoms as a function of the basis set
size.~\cite{PhysRevB.84.205415,Bruneval12} According to these studies,
$G_0W_0$-HF calculations using a cc-pVTZ basis set already produce
IPs converged within $\sim$0.15~eV for He and Be as compared with 
calculations using much larger bases. This agrees well with our observation
for He and H$_2$ IPs of a convergence with respect to the CBS limit 
within $\sim$0.2~eV using the TZ basis. However, for Ne, Bruneval~\cite{Bruneval12} 
has shown that this error can grow considerably ($\sim$0.4~eV) and it is necessary to
use a much larger basis, up to cc-pV5Z, in order to converge the results within
a range of $\sim$0.1 eV. 
Another convergence study at the $G_0W_0$ level was performed 
by Ren {\it et al.}~\cite{Ren-2012-RI}. It also shows the increase and slow convergence of the  
IPs of atomic and molecular systems with the basis set size. 
Unfortunately, the use of aug-cc-pV6Z bases, proposed in Ref.~\onlinecite{Ren-2012-RI}
as an appropriate reference basis set, 
is prohibitively expensive for the molecular study of self-consistent $GW$ schemes
presented here.
Thus, following Ke~\cite{PhysRevB.84.205415}, we use cc-pVTZ basis in our calculations.
We stress here that the main purpose of the present paper is not to provide fully
converged IPs, but to study how different self-consistent $GW$ schemes perform
for several representative molecules while keeping all other technical details identical.
As shown in detail below, the cc-pVTZ basis seems to be sufficient for this purpose.
This is indicated by the fact that the qualitative and quantitative deviations of
the different $GW$ IPs with respect to the CCSD(T) results, and among them, are rather
similar with the two basis sets used in this study (cc-pVDZ and cc-pVTZ).
In any case, Table~\ref{t:ip-ccsd_vs_gw} provides a consistent comparison, using the 
same basis sets and the same numerical implementation, between different schemes
to include correlation. 

Comparing our CCSD(T)/cc-pVTZ results with the experimental data 
in Table~\ref{t:ip-ccsd_vs_gw} we can find some significant deviations.
The larger deviation (0.96~eV) takes place for H$_2$. This is probably related
to the lack of corrections due to the finite mass of nuclei and the structural
relaxations in the final state in our calculations.  
The second largest difference (0.78~eV) happens for CH$_4$.
Relaxations in the final state are known to play a crucial 
role for methane~\cite{Grossman01} (the adiabatic IP is 12.61~eV~\cite{NIST}), and
this might be behind the poor comparison with the nominal experimental vertical
IP (13.60~eV~\cite{NIST}). 
In spite of the uncertainties about the comparison of our calculated vertical 
IPs with available experimental data, the overall agreement is good and the
MAE of the CCSD(T)/cc-pVTZ calculations with respect to the experimental results
in Table~\ref{t:ip-ccsd_vs_gw} is 0.19~eV, smaller than those
of most of the self-consistent $GW$ methods.

We now turn to the analysis of our $GW$ results. 
Both  self-consistent $GW$ approaches, SC$GW$ and 
QS$GW$~B, give results that are relatively close to the CC numbers
obtained using the same basis. Figure~\ref{f:DeltaIP} depicts
the differences between $GW$ and CC IPs.
We can see that the overall behavior of SC$GW$ and QS$GW$ IPs
is quite similar. However, QS$GW$ tends to overestimate the IPs as compared
to CC results, whereas SC$GW$ underestimates the IP in most cases.
In the case of He and H$_2$ such behavior is also observed for
IPs calculated using more complete basis sets. 
The $G_0W_0$ results starting from HF solutions are closer to those
of QS$GW$~B. Indeed the MAE with respect to CCSD(T) results using the 
cc-pVTZ basis is very similar for both methods.

QS$GW$ and SC$GW$ deviate from CC results in different directions. However,  
the mean absolute value of such deviation is quite similar
in both cases. The MAEs with respect to the 
CCSD reference can be found in Table~\ref{t:ip-ccsd_vs_gw}: 
0.21 and  0.25~eV, respectively for 
SC$GW$ and QS$GW$~B calculations using the cc-pVDZ basis,  
which increase to 0.22 and 0.27~eV when the larger
cc-pVTZ basis is used. 
It is interesting to note, following our discussion Sec.~\ref{ss:mult-conv}, 
that the MAE of QS$GW$~B IPs with respect to the CCSD(T) data is slightly 
larger than that of SC$GW$. 
If the observed differences were solely determined by the faster
convergence of CCSD(T) results with respect to the basis set size, 
we would expect the opposite behavior. Therefore, we can speculate that, for the set
of sixteen molecules considered here, it is likely that SC$GW$ will provide
better IPs (in average) than those given by QS$GW$~B. 
However, coming back to Table~\ref{t:ip-ccsd_vs_gw}, we can say
that using cc-pVTZ basis sets on average QS$GW$ and SC$GW$ perform very similarly.
The maximal discrepancies are somewhat larger for SC$GW$:
0.76~eV for the Be atom using the cc-pVTZ basis, to be compared with
the 0.67~eV deviation for F$_2$ in the case of QS$GW$.
The $G_0W_0$-HF is on average only slightly worse than SC$GW$ and quite
comparable to QS$GW$~B, 
with a MAE of 0.28 (0.22)~eV  and a maximal error of 0.86 (0.77)~eV for
the N$_2$ (CO) molecule using the cc-pVTZ (cc-pVDZ) basis.

We can now compare our results with previously published data for the
IPs of small molecules computed with self-consistent $GW$ schemes.
For this purpose we will use the results obtained with the more
complete cc-pVTZ basis.  
Most of the existing data for molecules correspond to the SC$GW$ 
method.~\cite{Delaney04,Stan06,Stan09,RostgaardJacobsenThygesen:2010,
CarusoRinkeRenSchefflerRubio:2012,Marom-etal:2012,Caruso2013}
We are only aware of three very recent studies using the QS$GW$ method 
for small molecules: one dealing with 
small sodium clusters up to five atoms~\cite{Bruneval09}, one studying 
small conjugated molecules~\cite{PhysRevB.84.205415} and 
one for first row atoms.~\cite{Bruneval12}

We start with the SC$GW$ results.
Stan {\it et al.}~\cite{Stan06,Stan09} performed all-electron SC$GW$ calculations
using large bases of Slater orbitals. They presented results for the IPs of 
the same atoms that
we have considered (He, Be and Ne), as well as for H$_2$ and LiH. In general 
we find good agreement with their data. Our IPs  are always somewhat smaller, although
differences stay within 0.15~eV, except for Ne, for which the difference grows
up to 0.39~eV.  Most of the differences are probably due to the basis set. 
As mentioned above, in the cases of He and H$_2$ in which we could
use larger basis sets, our IPs extrapolated to the complete basis set limit 
and those reported by Stan {\it et al.} agree within 0.03~eV. 
The large deviation for Ne seems to indicate some particular difficulty 
of the cc-pVTZ basis set to describe  the IP of this element.~\cite{Bruneval12} 
The MAE, over the five species mentioned above, 
of our SC$GW$ IPs with respect to those of Stan {\it et al.} is 0.15~eV (which grows up 
to 0.19~eV when we compare the $G_0W_0$-HF results).   
Delaney {\it et al.}~\cite{Delaney04} reported an all-electron SC$GW$ IP for Be of 
8.47~eV. Our SC$GW$/cc-pVTZ IP for Be (8.53~eV) lies in 
between this value and that given by Stan {\it et al.} (8.66~eV). 

More extensive sets of molecules have been studied by
Rostgaard {\it et al.}~\cite{RostgaardJacobsenThygesen:2010}
and Caruso {\it et al.}~\cite{CarusoRinkeRenSchefflerRubio:2012}. 
Rostgard {\it et al.} presented data for the all-electron SC$GW$ IPs of 34 different molecules,
including all the molecules considered here except H$_2$.
Their calculations used a double-$\zeta$ polarized basis set of augmented Wannier functions
(Wannier functions obtained from projector augmented wave calculations of the molecules,
supplemented with suitably chosen numerical atomic orbitals). 
Core states were taken into account 
in the calculation of the matrix elements of the exchange self energy. 
However, the contribution
of core states to the correlation  self energy of valence electrons was disregarded,
since it was assumed to be small due to the large
energy difference and small spatial overlap between valence and core states.
We find that the SC$GW$ IPs in Table~\ref{t:ip-ccsd_vs_gw} are larger 
(except for LiF and LiH) than those 
reported by Rostgard {\it et al.}. The maximal differences take place
for F$_2$ and LiF, where our calculated 
IPs are 0.54~eV larger and 0.67~eV smaller, respectively.
The average deviation between our SC$GW$ results
and those of  Rostgard {\it et al.} 
(MAE=0.32~eV, which grows up to 0.57~eV for the $G_0W_0$-HF results) 
is somewhat larger, although comparable,  to that between 
our SC$GW$ and CCSD(T) results. This seems to indicate that numerical 
and methodological aspects behind
each implementation still hinder the comparison of results obtained with different codes
using, formally, the same self-consistent $GW$ scheme. The use of different
basis is probably one of the most important causes of discrepancies, as recently pointed
out by Bruneval and Marques for $G_0W_0$ calculations.~\cite{Bruneval13}
However, part of the discrepancies might be related to two factors: {\it i}) 
the use of  MP2/6-31G(d)  geometries by Rostgard {\it et al.},
while we use CCSD(T)/cc-pVTZ relaxed geometries and, {\it ii}) 
the lack of core-valence correlations in their calculations. 
The better agreement of our results with the full all-electron SC$GW$ calculations 
in Ref.~\onlinecite{CarusoRinkeRenSchefflerRubio:2012} could support
this last conclusion on the influence of core-valence correlations.

Caruso {\it et al.}~\cite{CarusoRinkeRenSchefflerRubio:2012} report the values of 
the SC$GW$ IPs for the same set of molecules used by Rostgard {\it et al.}. 
Their all-electron calculations use a basis set of numerical atomic 
orbitals and the resolution of the identity technique to express the products
of those orbitals.
Their IPs are systematically larger than those reported here, although the differences
are relatively small, lower than 0.19~eV for all the molecules except for LiF,
for which the difference grows up to 0.46~eV. 
The MAE over the 12 molecules is only 0.14~eV for SC$GW$ and 0.15~eV for
$G_0W_0$-HF calculations.
Therefore, the overall agreement between our SC$GW$/cc-pVTZ results and
those of Caruso {\it et al.} is rather good.

Now we compare our QS$GW$ with the very scarce data available in the literature.
Ke has recently studied the IPs and electron affinities of 
a number of conjugated molecules using the QS$GW$ ``mode A'' method.~\cite{PhysRevB.84.205415}
Ke uses a cc-pVTZ basis, similar to that utilized here,   
and reports 11.31~eV and 11.44~eV for the IP 
of C$_2$H$_2$ calculated at the level
of QS$GW$~A and $G_0W_0$-HF, respectively. This is in excellent agreement with 
our corresponding results of 11.43~eV and 11.54~eV and indicates that, 
at least for this
molecule and the cc-pVTZ basis set, the calculated IP is rather stable 
against the use either QS$GW$~A or B schemes. Bruneval~\cite{Bruneval12}
reported 24.46 (24.72), 9.11 (9.16) and 21.62 (21.79) eV, respectively,
for the IPs of He, Be and Ne calculated using the QS$GW$~A ($G_0W_0$-HF) approach
and a very complete cc-pV5Z basis (of Cartesian kind). These values are in good agreement
with our results although they are always somewhat larger. 
This is  due to the use of a smaller cc-pVTZ basis
set in our case, as clearly demonstrated by the excellent agreement 
between data calculated using the MOLGW program~\cite{Bruneval12} and
our code when the same basis set are used (Table~\ref{t:bsc-conv}).
Furthermore, focusing on the results published by Bruneval 
in Ref.~\onlinecite{Bruneval12}, 
comparing our $G_0W_0$-HF with those reported
in Figure~1 of that paper, we find that the results reported there for
the cc-pVTZ basis are almost identical to those presented here. This again indicates
a very welcome consistency between both sets of calculations.

Finally, we can compare our $GW$ vertical IPs with the experimental data
in Table~\ref{t:ip-ccsd_vs_gw}. 
This comparison should be taken with some caution: as 
commented above, the comparison might be affected by other factors different 
from the ability of the $GW$ schemes to capture electron correlations.
In any case, it is interesting to obtain a quantitative measure of the deviation.
The MAE with respect to the experimental data are similar for the SC$GW$ and QS$GW$~B
results obtained using the cc-pVTZ basis, 0.26 and 0.35~eV, respectively.
It increases to 0.5~eV for the $G_0W_0$-HF approach.
These deviations of the $GW$ results with respect to the experiments are somewhat
larger than those with respect to the CCSD(T)/cc-pVTZ theoretical reference.
They seem to confirm a very similar degree of accuracy for the QS$GW$ and SC$GW$ methods,
as well as their moderate improvement as compared to the $G_0W_0$-HF approach.

\section{Conclusions and Outlook}
\label{s:conslusion}

In this article we studied two self-consistent $GW$ approaches,
the self-consistent $GW$ (SC$GW$) and the quasi-particle self-consistent $GW$ (QS$GW$),
within a single numerical framework. We explored two possible realizations of the 
QS$GW$ algorithm, the so-called ``mode A'' and ``mode B''.  A systematic 
study for He and H$_2$ indicated
that, for QS$GW$~A, the IPs do not show a monotonic convergence as a function
of the basis set size. This unexpected
results was traced back to the peculiar dependence on two different reference
energies of the cross-terms of the correlation operator in QS$GW$~A, in combination
with the use of basis sets of atomic orbitals that confers the self energy 
a complex and abrupt frequency dependence in the high frequency limit. 
Motivated by this observation, we concentrate our study of different molecules
in a comparison between standard self-consistent SC$GW$ and QS$GW$ ``mode B'' .

We focused on light atoms and small
molecules as examples of finite electronic systems and performed
all-electron $GW$ calculations for them. 
We have studied the density of states (or spectral function) given
by both approaches and, from a qualitative point of view and at low and moderate energies,
we did not find significant differences between both approaches. In both cases
the number and intensity of satellite structures is reduced 
with respect to one-shot $G_0W_0$ calculations. This is in agreement
with previous observations, for example, for the homogeneous electron
gas.~\cite{HolmBarth:1998} We have also compared both approaches
quantitatively by calculating the ionization potentials and comparing
them against coupled-cluster calculations. The comparison shows similar
qualities for both self-consistent $GW$ approaches, which are
only slightly better that one-shot $G_0W_0$ calculations starting from Hartree-Fock. 
Interestingly, SC$GW$ and QS$GW$ calculations tend to deviate in 
opposite directions with respect to CCSD(T) results. 
SC$GW$ systematically produces too low IPs, while QS$GW$ tends
to overestimate the IPs. We do not have a clear explanation 
for this different behavior of SC$GW$ and QS$GW$.
It is interesting to note, however, that the
behavior observed for QS$GW$ here seems to be consistent with the known tendency of
QS$GW$ to overestimate the band gaps of solids.~\cite{PhysRevB.76.165106,Shishkin07}
For the small molecules considered here, $G_0W_0$-HF produces results which are
surprisingly close to QS$GW$ calculations both for the DOS and for the numerical
values of the IPs. In spite of the similarities, 
SC$GW$ produces results somewhat closer to the CCSD(T) reference.

We chose to compare our results against CCSD(T) calculations, rather than against
experimental results for several reasons. One of them is the 
difficulty to converge the self-consistent $GW$ results with respect to the basis
set in our all-electron calculations.
Performing converged calculations with respect to the frequency grid
and size of the auxiliary basis of dominant products proved to
be computationally intensive  and, therefore, we are limited to cc-pVTZ
basis sets in most cases. 
However, comparison between CCSD(T) and $GW$ results obtained
with both the cc-pVDZ or cc-pVTZ bases, leads to very similar observations.
Furthermore, a systematic convergence test as a function of the basis set size
performed for He and H$_2$ indicates that our observation that QS$GW$ tends to overestimate,
while SC$GW$ tends to underestimate, the ionization potential of CCSD(T)
is very likely to remain valid using more complete basis sets. 
Regarding the observation that SC$GW$ is marginally closer to the 
CCSD(T) results than QS$GW$, we also believe that it will remain valid
with more complete basis sets. The reason for this suspicion is 
the steeper increase of the $GW$ IPs with the basis size
as compared to those calculated using CCSD(T) (that show a faster convergence).
We argue that this will tend to improve the agreement between SC$GW$ and
CCSD(T), and degrade that of QS$GW$, as the basis set size increases.  
Another interesting point is that the exclusion of triple excitations in the CC
calculations, i.~e. performing CCSD calculation,
produced only minor differences for most systems.  With all these ingredients,
we expect that the comparison presented here
among different self-consistent $GW$ methods, and
of those with CCSD(T),  reflects the ability of these 
schemes to deal with the effects of correlations in small molecules. 

Regarding the applicability of self-consistent $GW$ methods:
On the one hand, our results could not prove that any of the explored self-consistent $GW$
approaches is clearly superior to one-shot $G_0W_0$ calculations 
using an appropriate starting point (e.g., Hartree-Fock and certain
hybrid functionals have been shown to provide an excellent starting point
for one-shot GW calculations~\cite{Fuchs:2007-HSE+G0W0,Marom12bis,
Marom-etal:2012,Koerzdoerfer:2012,Atalla:2013,Bruneval13});
On the other hand, at least for the IPs of the set of atoms and molecules considered here,
the self-consistent results seems to improve, although slightly, the $G_0W_0$-HF
and we did not observe any clear signature that the self-consistent $GW$ results
were pathological.  This is interesting because there are situation where
one would like to improve the one-particle DFT spectra using a charge or
energy conserving scheme.  Transport calculations in molecular junctions are a clear
example.~\cite{PhysRevB.83.115108} In this context, it is also worth noting that our
calculations indicate that SC$GW$ shows a more stable convergence pattern of the 
self-consistent loop. The QS$GW$ method can be advantageous in many applications
because it generates an effective one-electron Hamiltonian with an improved spectrum.

\section*{Acknowledgments}

The authors want to thank James Talman for constant support and providing 
essential algorithms and programs at the initial stages of this work.
Eric Shirley and Russell Johnson are acknowledged for essential information about 
the computational procedures used in the NIST CCCBDB database.
Mathias Ljungberg and R\'emi Avriller made many useful 
comments that improved this manuscript. Fabien Bruneval shared with us information
about his recent QS$GW$ calculations and the latest version of his program.
Computing resources were provided by Donostia International Physics Center 
(Donostia-San Sebasti\'an, Spain), 
Centro de F\'{\i}sica de Materiales CFM-MPC Centro Mixto CSIC-UPV/EHU
(Donostia-San Sebasti\'an, Spain).
Part of the computer time for this study was provided by the computing facilities MCIA
(M\`esocentre de Calcul Intensif Aquitain) of the Universit\'e de Bordeaux and
of the Universit\'e de Pau et des Pays de l'Adour.
PK acknowledges support from the CSIC JAE-doc program,
co-financed by the European Science Foundation, and 
the Diputaci\'on Foral de Gipuzkoa. DSP and PK acknowledge
financial support from the Consejo Superior de Investigaciones Cient\'{\i}ficas
(CSIC), the Basque Departamento de Educaci\'on, UPV/EHU (Grant No. IT-366-07), 
the Spanish Ministerio de Ciencia e Innovaci\'on (Grant No. FIS2010-19609-C02-02), 
the ETORTEK program funded by the Basque Departamento de Industria and the
Diputaci\'on Foral de Gipuzkoa, and the German DFG through the SFB 1083.
DF acknowledges support from the
ORGAVOLT-ANR project and the Euror\'egion Aquitaine-Euskadi program. 

\bibliography{scgw-article}

\begin{thebibliography}{10}
\bibitem{DSCF-HF}%
  \BibitemOpen
  \bibinfo {note} {From these HF calculations we can also compute an
  approximation to the vertical IPs of the molecules. However, such
  $\Delta$SCF-HF IPs deviate severely from the experimental and CCSD results.
  The MAE of the $\Delta$SCF-HF IPs with respect to the CCSD calculations is
  1.0~eV, with a maximal error of $\sim$1.5~eV for the LiF molecule. These
  errors are substantially larger that those of the different $GW$ schemes
  described in the paper, and indicate the importance of correlation effects
  for the estimation of molecular IPs. Indeed, $\Delta$SCF-HF IPs are known to
  be significantly worser than those obtained from $\Delta$SCF-DFT
  calculations~\cite{Zevallos03,Zhan03}}%
  \bibAnnoteFile{NoStop}{DSCF-HF}%
\end{thebibliography}

\end{document}


\title{
Supplemental material for ``Fully self-consistent $GW$ and
quasi-particle self-consistent $GW$ for molecules''} 

\author{P.~Koval}
\email{koval.peter@gmail.com}

\affiliation{Centro de F\'{\i}sica de Materiales CFM-MPC, 
Centro Mixto CSIC-UPV/EHU, Paseo Manuel de Lardizabal 5, E-20018 San Sebasti\'an, Spain}
\affiliation{Donostia International Physics Center (DIPC), 
Paseo Manuel de Lardizabal 4, E-20018 San Sebasti\'an, Spain}

\author{D.~Foerster}%
\email{d.foerster@loma.u-bordeaux1.fr}
\affiliation{CPMOH/LOMA, Universit\'e de Bordeaux 1, 
351 Cours de la Liberation, 33405 Talence, France}

\author{D. S\'anchez-Portal}
\email{sqbsapod@sq.ehu.es}

\affiliation{Centro de F\'{\i}sica de Materiales CFM-MPC, 
Centro Mixto CSIC-UPV/EHU, Paseo Manuel de Lardizabal 5, E-20018 San Sebasti\'an, Spain}
\affiliation{Donostia International Physics Center (DIPC), 
Paseo Manuel de Lardizabal 4, E-20018 San Sebasti\'an, Spain}

\maketitle



In this supplemental material we present in detail
the differences between the density of states
(DOS) of atoms and molecules calculated using  the SC$GW$
and QS$GW$ approaches. In particular, we focus in the appearance,
energy distribution and intensities of satellite structures. 
The qualitative differences between the satellite structures obtained with 
these two self-consistent approaches and HF-$G_0W_0$ is also discussed.
It is important to notice that these results are certainly influenced by the
relatively small size of the basis set utilized in the present calculations.
Some of the structures that we discuss here appear in the DOS at high energies, 
well outside the window defined by the one-electron states that can be
represented with our basis sets. 
The high energy region of the spectrum should be covered 
by the continuum of unbound states that, however, cannot be represented
with our localized orbitals. We can expect that, in reality, the coupling to 
and overlap with this continuum of states should 
make these satellite structures less apparent than in the present calculations. 
However, even taking into account these limitations, 
we think that it is instructive to study how these satellite 
structures develop in order to better understand some of the different behaviors
of the two self-consistent approaches cosidered in our paper.

\section{Satellite structures in the calculated DOS }

\label{ss:qualitative-results}

\begin{figure}
\begin{tabular}{p{7.5cm}p{7.5cm}}
\includegraphics[width=7.5cm, angle=0,viewport=10 50 410 320,clip]{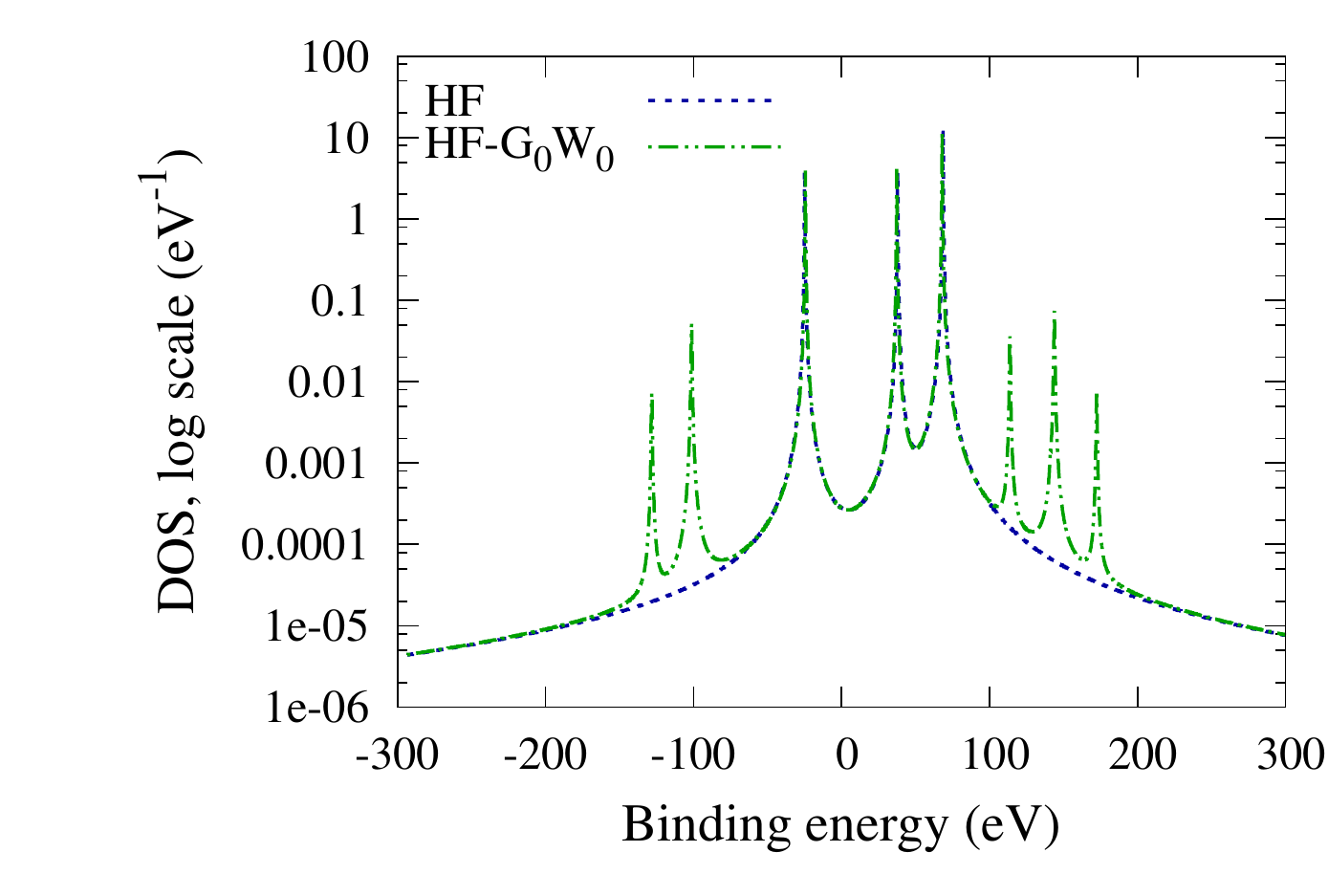} &
\includegraphics[width=7.5cm, angle=0,viewport=10 50 410 320,clip]{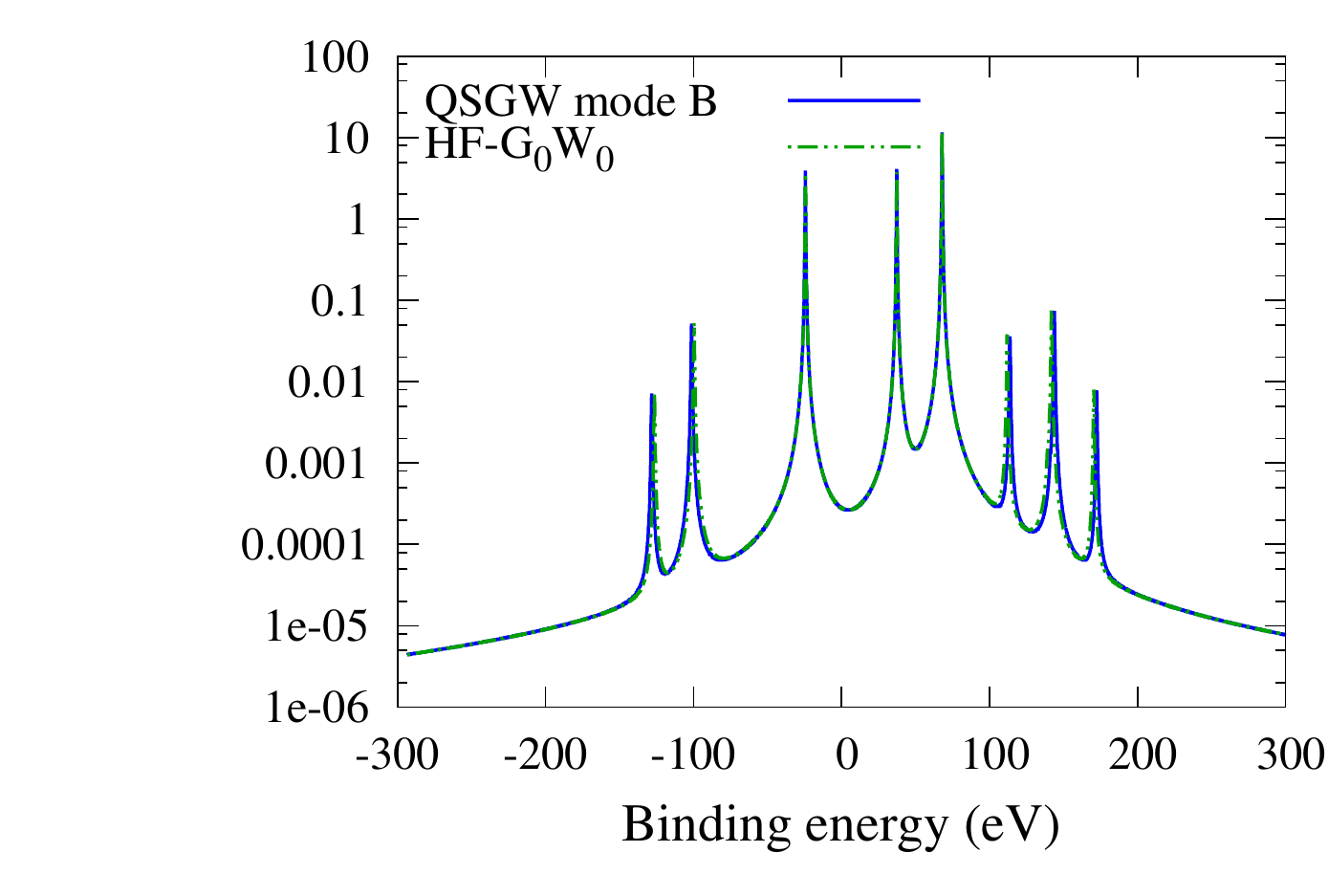} \\
\centerline{a)} & \centerline{b)} \\
\includegraphics[width=7.5cm, angle=0,viewport=10 50 410 311,clip]{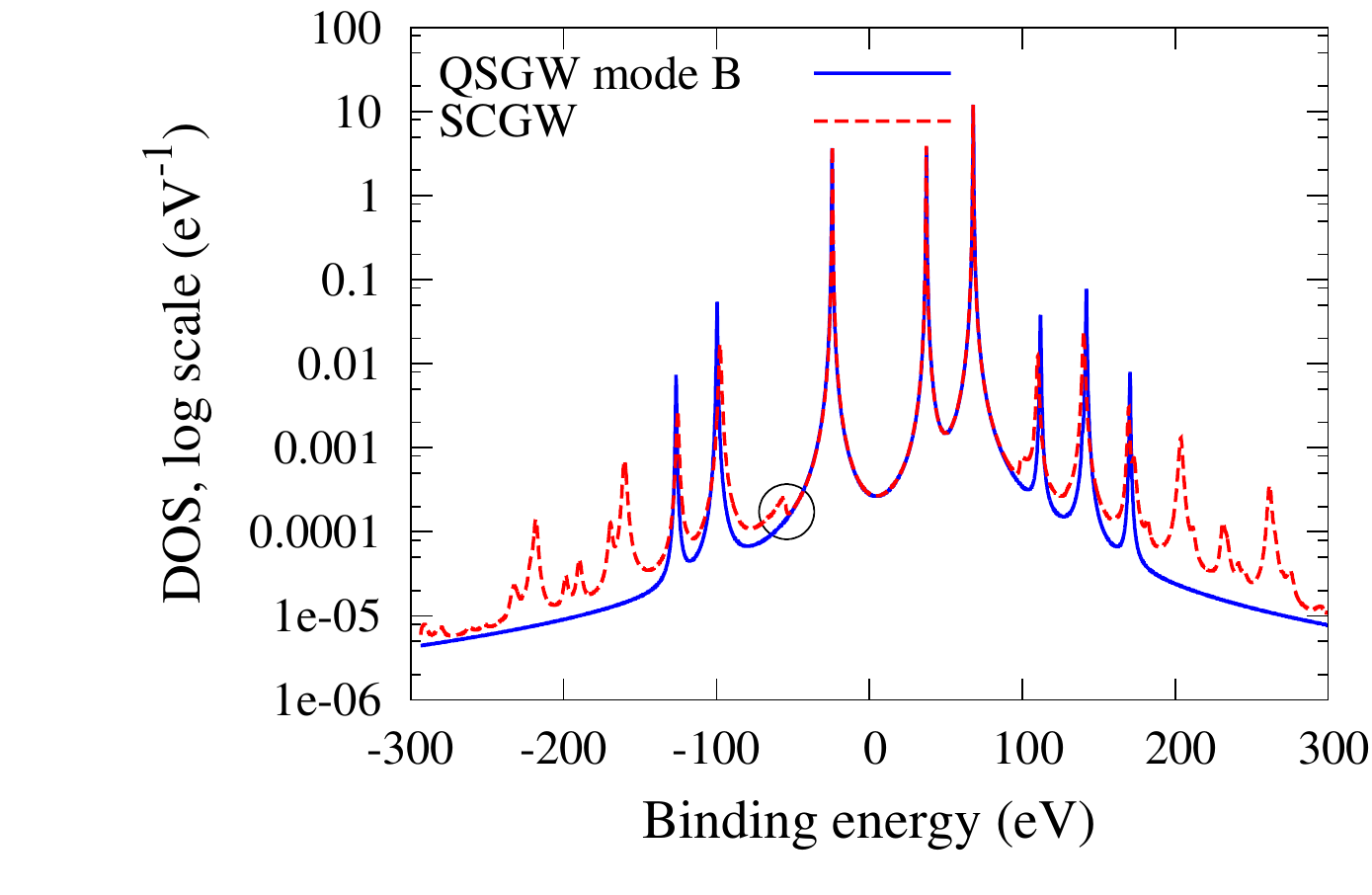} &
\includegraphics[width=7.5cm, angle=0,viewport=10 50 410 311,clip]{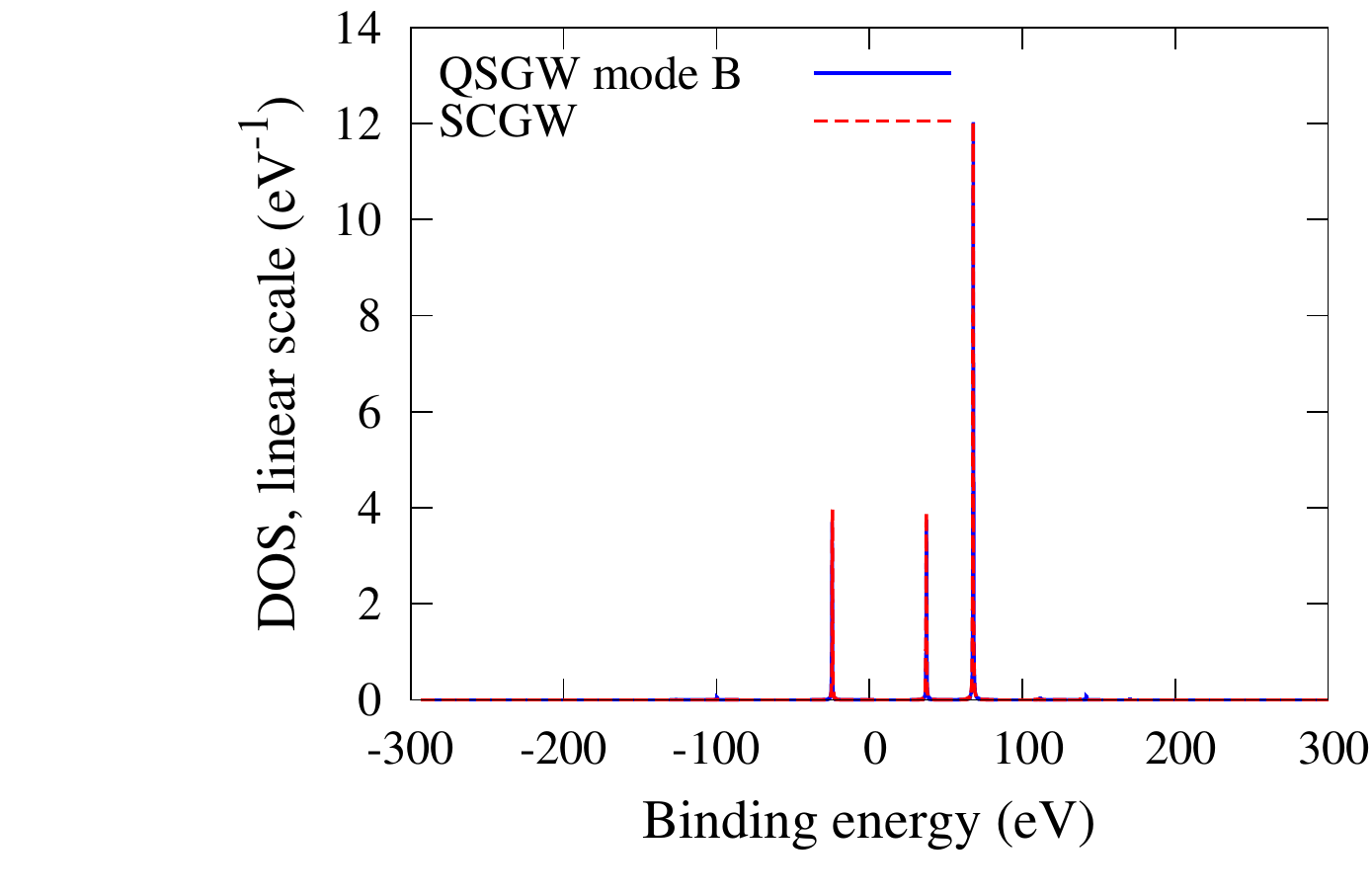} \\
\centerline{c)} & \centerline{d)} \\
\includegraphics[width=7.5cm, angle=0,viewport=10 50 410 311,clip]{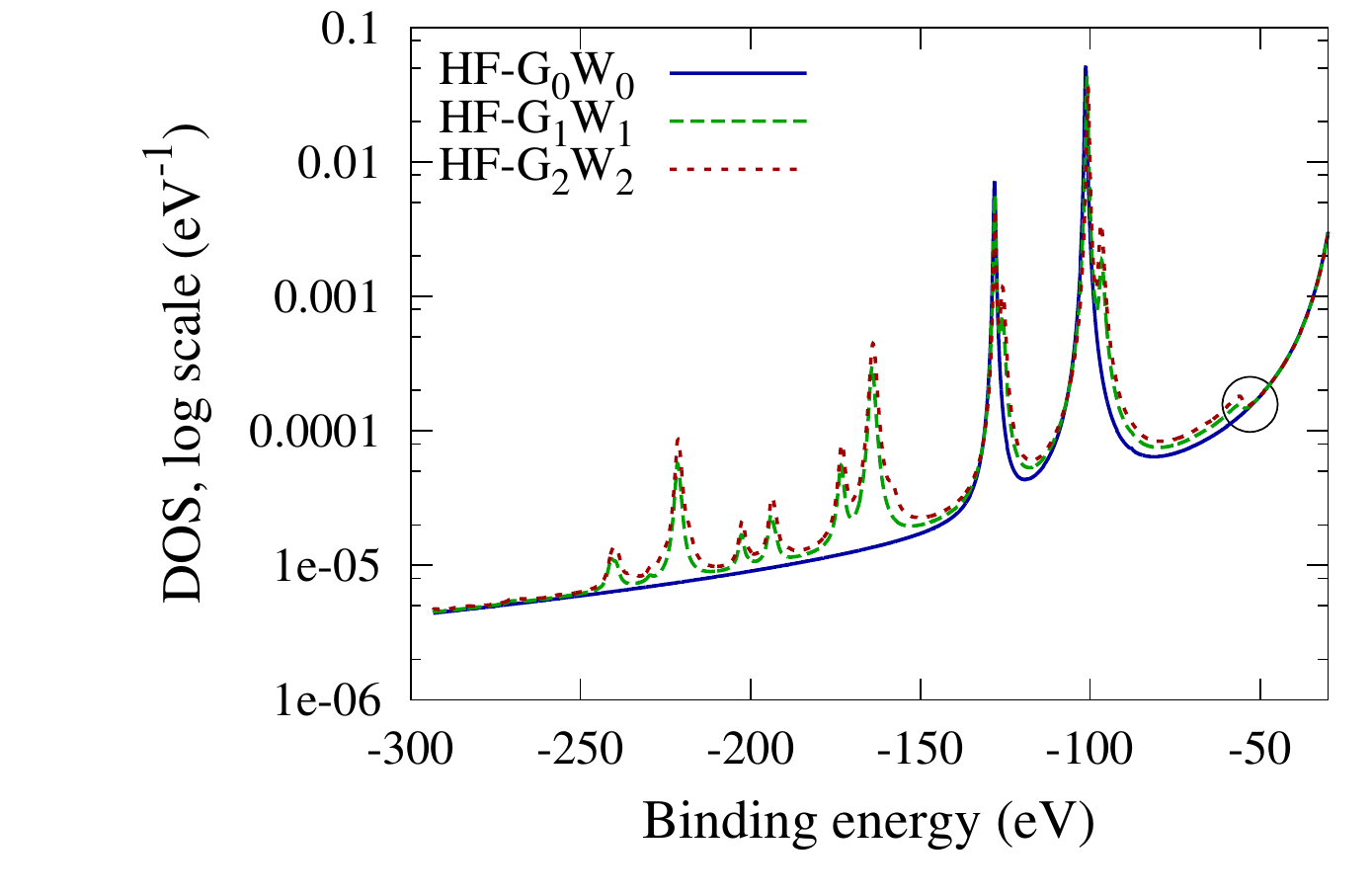}  &
\includegraphics[width=7.5cm, angle=0,viewport=10 50 410 311,clip]{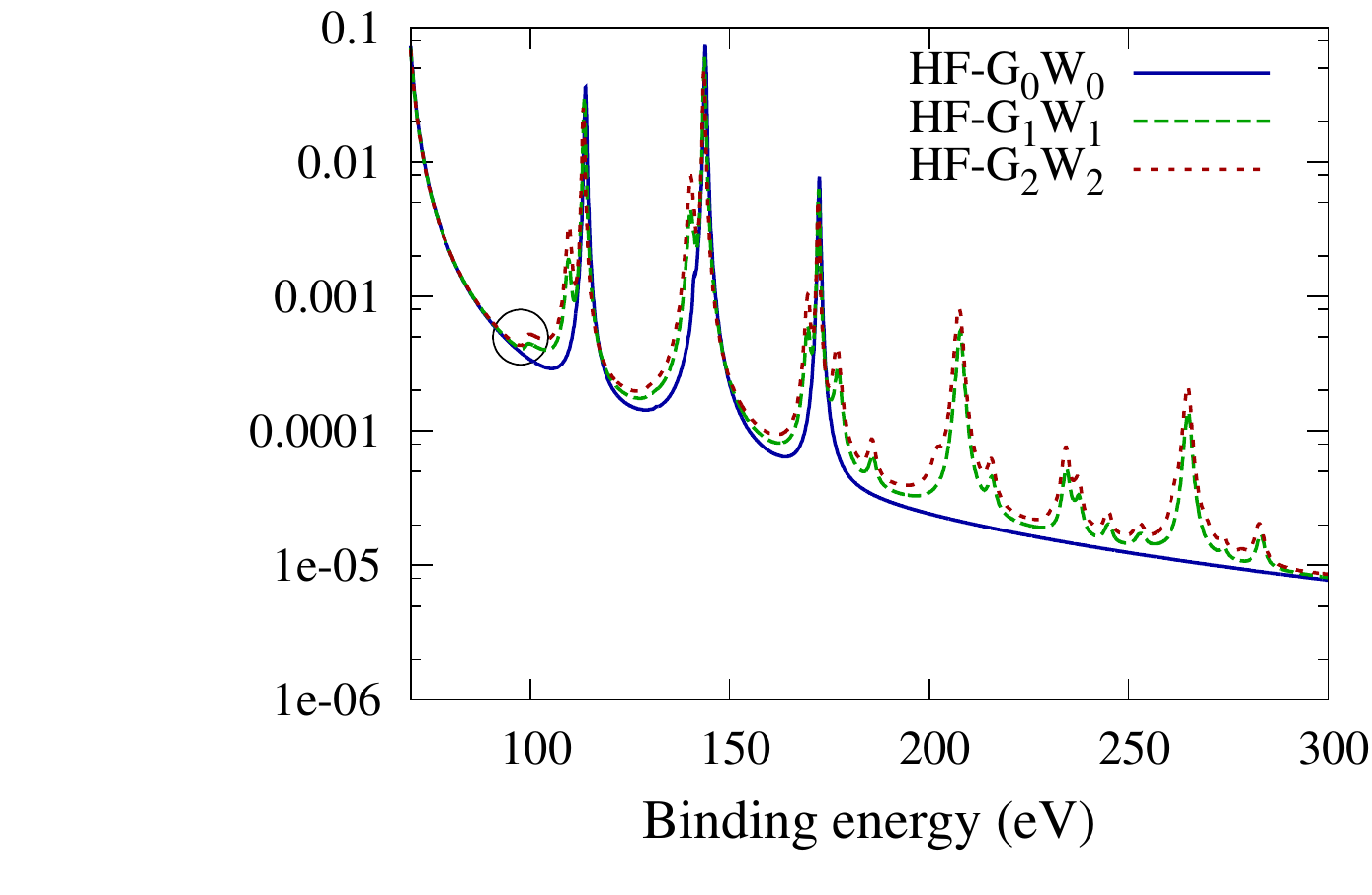} \\
\centerline{e)} & \centerline{f)} \\
\end{tabular}
\caption{\label{f:dos-helium}
DOS of Helium atom computed within HF, HF-$G_0W_0$, QS$GW$ and  
SC$GW$ approximations. A cc-pVDZ basis set has been used in these 
calculations.  Panels e) and f) demonstrate the
emergence of a more complex satellite structure in the SC$GW$ calculation as
a function of the number of iterations. 
The linear scale plot of the DOS for QS$GW$ and SC$GW$ is shown on panel d).
Using a linear scale only the main features
are clearly visible. This demonstrates the real size of the satellites, which 
indeed carry a small spectral weight.}
\end{figure}

We want to illustrate the implications of the
SC$GW$ and QS$GW$ approximations for the calculated densities of states.
For the sake of simplicity, we have chosen the helium atom for this
discussion, we will later show results for a larger molecule (methane). 
The DOSs obtained within the HF, HF-$G_0W_0$, QS$GW$ and  
SC$GW$ approximations are plotted in the Fig.~\ref{f:dos-helium}. A
logarithmic scale has been chosen in all the panels except (d)  
to emphasize the details of the satellite structures. As can be seen,
the satellite structures produced by the different approximations are clearly distinct.
The DOS obtained from a Hartree-Fock (HF) calculation is constituted 
by a few peaks at the positions 
of the eigenvalues of the Fock operator (using a cc-pVDZ basis set).
Of course, within HF all the peaks correspond 
to normalized one-particle excitations and no satellite structures are found. 
 However, a one-shot 
$G_0W_0$ calculation starting from the HF solution (HF-$G_0W_0$) 
exhibits satellite peaks within an energy window approximately 
twice as large as the range occupied by the HF eigenvalues [see Fig.~\ref{f:dos-helium}~(a)].  
The satellite structure obtained with a QS$GW$ calculation 
[Fig.~\ref{f:dos-helium}~(b)] is very similar
to that of HF-$G_0W_0$. In contrast, as shown in Fig.~\ref{f:dos-helium}~(c), a
fully self-consistent SC$GW$ calculation generates a more complex satellite structure, 
with peaks extending to much higher energies.  These additional peaks (secondary satellites)
do appear at the expense of the first satellites (those already present at the $G_0W_0$ level)
that get somewhat less intense.

As mentioned above, in the present calculations we 
used a small cc-pVDZ Gaussian basis, however the features we are aiming
to reveal are common for larger basis sets too. 
The calculations are fully converged respect to the number
of elements in the basis of dominant products. 
The frequency grid spacing is set to $\Delta \omega=0.1$ eV.
The other parameters were chosen as discussed in the main paper.
The SC$GW$ calculation for He converged after 35 iterations, while QS$GW$
required 13 iterations. The criterium for self-cosistency involves two 
subsequent DOSs at all the calculated frequency points:
$\mathrm{Conv} = 
\frac{1}{N_{\text{orbs}}}\int 
\left[\mathrm{DOS}_i(\omega) -\mathrm{DOS}_{i-1}(\omega)\right] d\omega$.
We have set the threshold for $\text{Conv}$ at 10$^{-5}$ and
always obtained very similar DOSs. This clearly shows that the complex
satellite structures of the SC$GW$ calculations described here are not spurious,
but remain in very well converged calculations.

The fact that the QS$GW$ approach
delivers qualitatively the same picture as HF-$G_0W_0$ can be easily understood. 
The QS$GW$ method is effectively a series
of one-shot $G_0W_0$ calculations (although the reference one-electron effective
Hamiltonian is modified along the calculation). 
More interesting is the large difference between QS$GW$ and SC$GW$ results,
with a far more complex structure in the case of the SC$GW$ calculations. 
This structure can be understood from the emergence of 
succesive satellite peaks during the SC$GW$ iteration. The emergence of 
such secondary satellites is illustrated in panels (e) and (f) of 
Fig.~\ref{f:dos-helium}. The second iteration in the SC$GW$ loop results
in a DOS reaching approximately twice as far as one-shot HF-$G_0W_0$, while
the third iteration exibits even higher frequency features which 
get accordingly less intense (i.e., they carry a smaller spectral weight
as we move further away in energy). This rich satellite structure can be revealed 
thanks to the spectral function treatment used in our calculations, which allows
a direct computation of the Green's function close to the real axis (at the expense
of using very fine frequency grids). 
The general smallness of the 
satellite structures for the particular case of He considered here can 
be seen in panel (d), which compares the DOS
obtained with QS$GW$ and SC$GW$ in a linear scale. Here one can only recognize the main 
resonances similar to those obtained with a HF calculation (although, of course,
appearing at shifted energies), i.e. some of the most important qualitative 
differences between the SC$GW$ and QS$GW$ DOS cannot be seen using a linear scale.

In Fig.~\ref{f:dos-helium}~(e) and (f) we can observe the appearance of 
step-like features in the DOS after the second iteration during the
SC$GW$ loop. These step-like features also appear 
in the converged SC$GW$ DOS and have been highlighted 
with circles in Fig.\ref{f:dos-helium}~(c), (e) and (f). 
These small features are connected to the broadening: their size 
can be reduced, together with Lorentzian-shape background,
by diminishing the broadening parameter $\varepsilon$ used
to compute the Green's functions.

\begin{figure}
\begin{tabular}{p{7cm}p{7cm}}
\includegraphics[width=7cm, angle=0,viewport=10 50 408 320,clip]{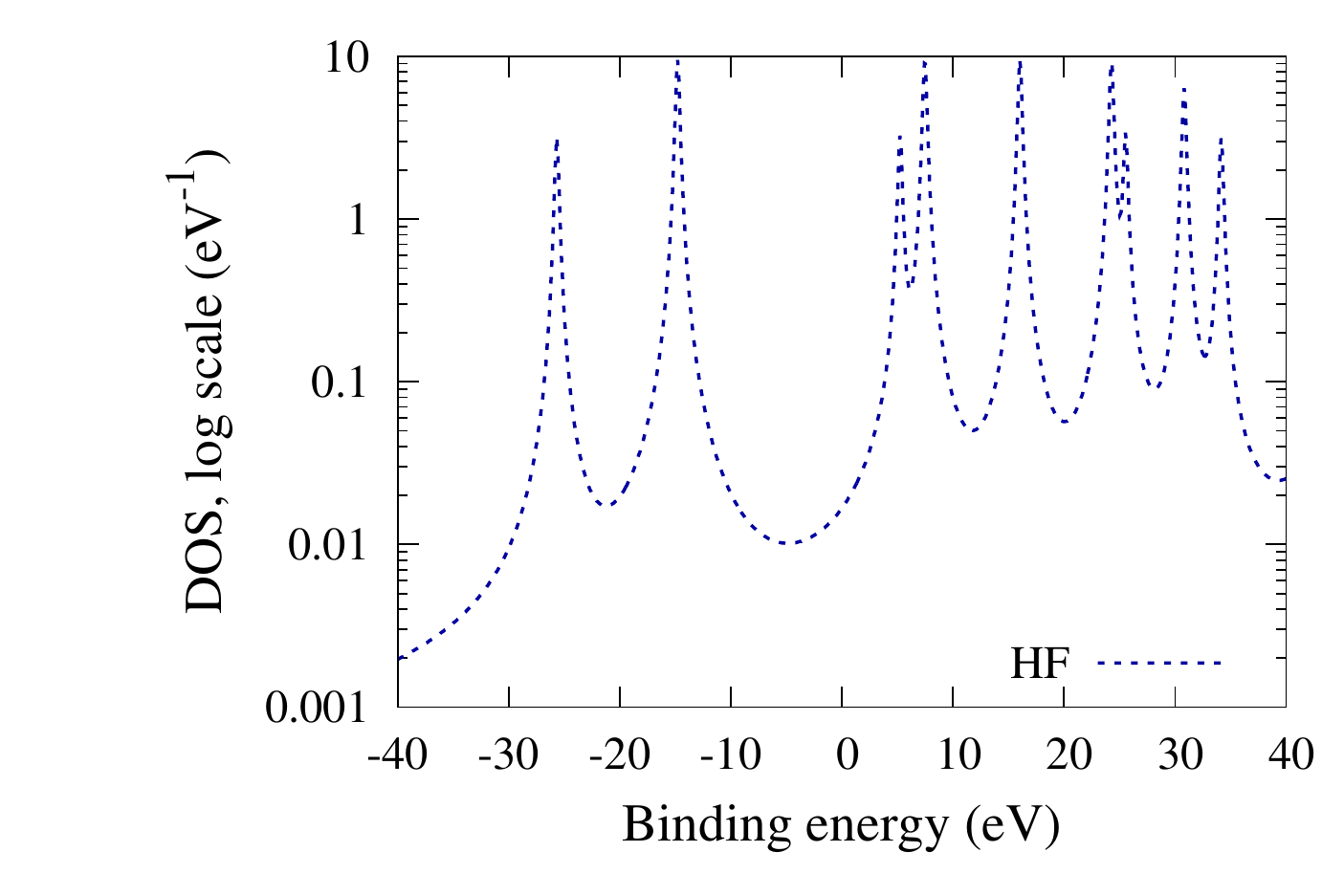} &
\includegraphics[width=7cm, angle=0,viewport=10 50 408 320,clip]{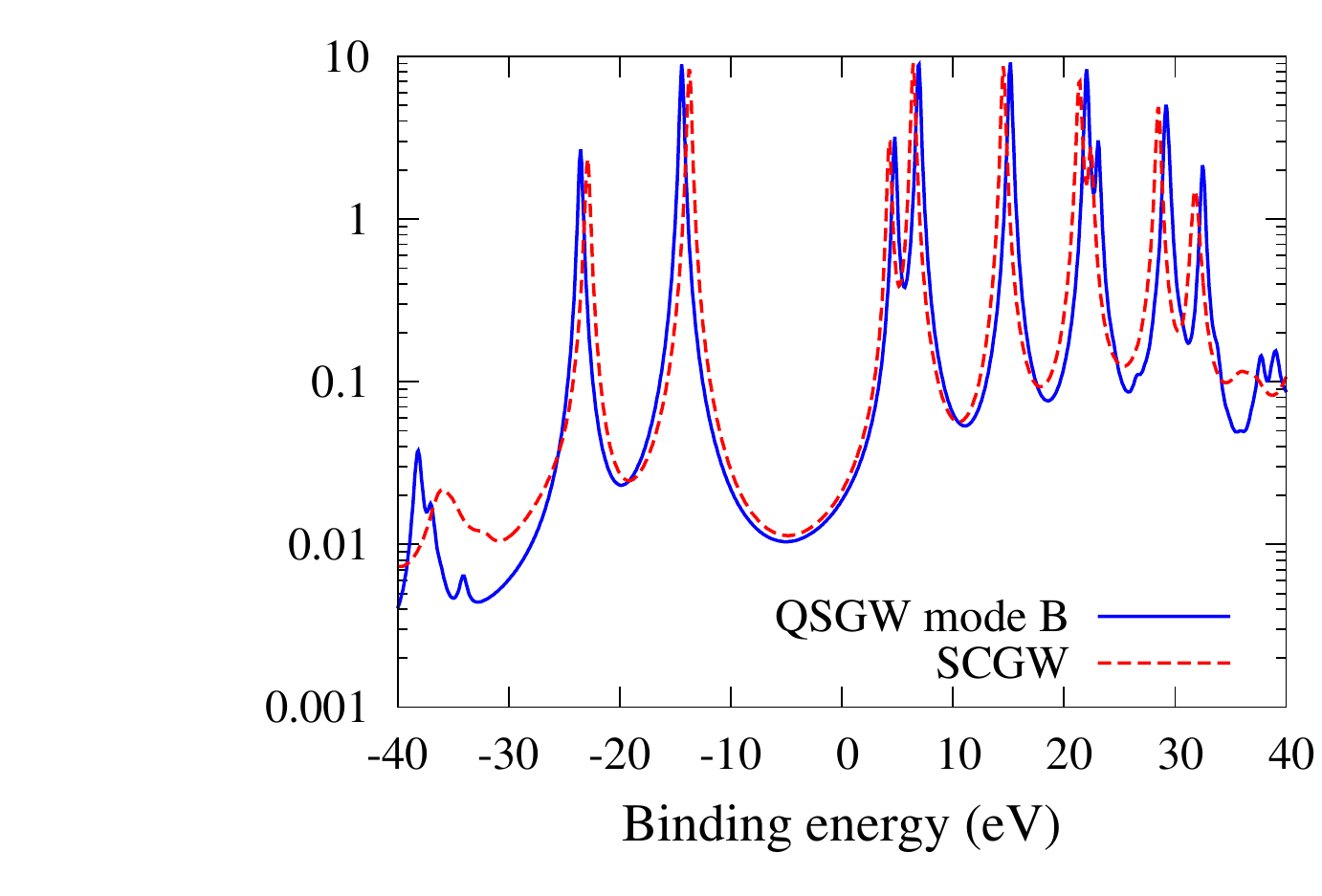} \\
\centerline{a)} & \centerline{b)} \\
\includegraphics[width=7cm, angle=0,viewport=10 50 408 311,clip]{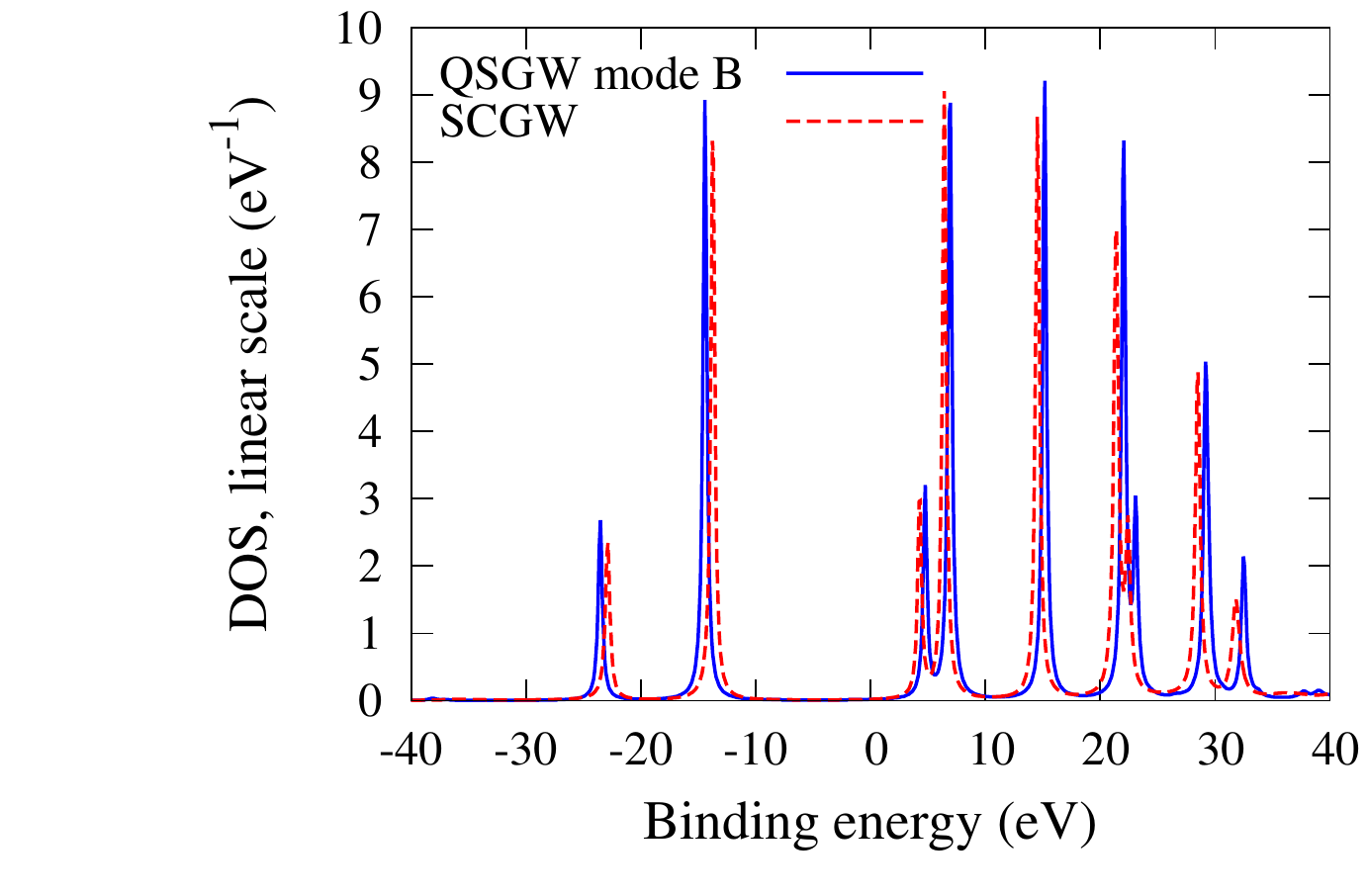} &
\includegraphics[width=7cm, angle=0,viewport=10 50 410 311,clip]{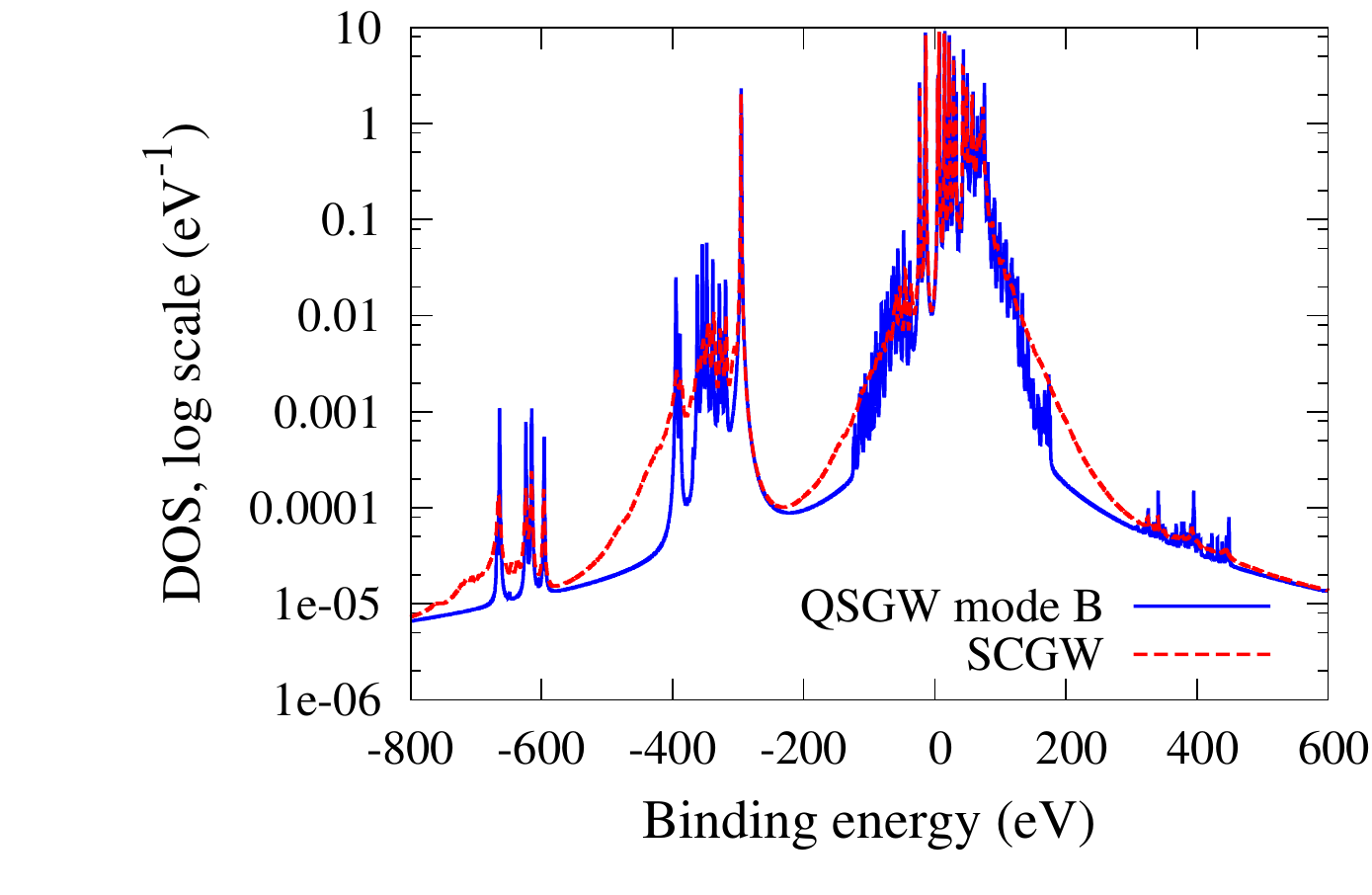} \\
\centerline{c)} & \centerline{d)}
\end{tabular}
\caption{\label{f:dos-methane}
DOS of methane computed within HF, HF-$G_0W_0$, QS$GW$ and 
SC$GW$ approximations. The 
situation is similar to the case of helium in Fig.~\ref{f:dos-helium}:
The DOS of QS$GW$ is similar
to that of HF-$G_0W_0$, while SC$GW$ gives rise to a more
complex satellite structure. However, using a linear scale to represent 
the DOS, the outcomes of QS$GW$ and SC$GW$ calculations
look qualitatively similar.
}
\end{figure}

Figure~\ref{f:dos-methane} present similar calculations for the  
methane molecule (CH$_4$). Panel (a) presents the HF DOS as a reference. 
Panel (b) demonstrates that  the main satellite peaks are somewhat less intense for the 
SC$GW$ DOS than for the QS$GW$ result.  A similar observation can 
be made when comparing SC$GW$ and HF-$G_0W_0$ results. Indeed, 
QS$GW$ and HF-$G_0W_0$ are rather similar also in the case of methane.
This observation about SC$GW$ in accord with previous observations for electron gas 
\cite{PhysRevB.54.8411}.
Panel (d) shows the DOS in a larger frequency range. We can see here 
that the secondary satellite structures  
become a smooth background with an exponential decay. This is related to the much
denser density of one-electron states in the case of methane as compared to helium.

\bibliography{scgw-article-supplementary}